\documentstyle[aps,pre,psfig]{revtex} 
 
\newcommand{\beq}{\begin{equation}} 
\newcommand{\eeq}{\end{equation}} 
\newcommand{\bea}{\begin{eqnarray}} 
\newcommand{\eea}{\end{eqnarray}} 
 
\begin{document} 
 
\draft 
 
\title{\bf Geometric approach to Hamiltonian dynamics and statistical mechanics} 
 
\author{Lapo Casetti\cite{lapo}} 
\address{Istituto Nazionale per la Fisica della Materia (INFM),  
Unit\`a di Ricerca del Politecnico di Torino, 
Dipartimento di Fisica, Politecnico di Torino, 
Corso Duca degli Abruzzi 24, I-10129 Torino, Italy}  
 
\author{Marco Pettini\cite{marco}} 
\address{Osservatorio Astrofisico di Arcetri, Largo Enrico Fermi 5, 
I-50125 Firenze, Italy, 
and Istituto Nazionale per la Fisica della Materia (INFM), 
Unit\`a di Ricerca di Firenze, 
Largo Enrico Fermi 2, I-50125 Firenze, Italy}  
 
\author{E. G. D. Cohen\cite{egd}} 
\address{The Rockefeller University, 1230 York Avenue, New York,  
New York 10021-6399} 
  
\date {December 1, 1999} 
\maketitle 
 
\begin{abstract} 
This paper is a review of results which have been recently 
obtained by applying 
mathematical concepts drawn, in particular, from differential geometry and 
topology, to 
the physics of Hamiltonian dynamical systems with many degrees of freedom of 
interest for statistical mechanics.

The first part of the paper concerns the applications of methods used in 
classical 
differential geometry to study the chaotic dynamics of Hamiltonian systems.  
Starting 
from the identity between the trajectories of a dynamical system and the 
geodesics in its 
configuration space, when equipped with a suitable metric, a geometric 
theory of chaotic 
dynamics can be developed, which sheds new light on the origin of chaos in 
Hamiltonian 
systems.  In fact, it appears that chaos can be induced not only by negative 
curvatures, as 
was originally surmised, but also by positive curvatures, provided the 
curvatures are 
fluctuating along the geodesics.  In the case of a system with a large number 
of degrees of 
freedom it is possible to approximate the chaotic instability behavior of the 
dynamics by 
means of a geometric model independent of the dynamics, which allows then an 
analytical estimate of the largest Lyapunov exponent in terms of the averages 
and 
fluctuations of the curvature of the configuration space of the system.

In the second part of the paper  the phenomenon of phase transitions is 
addressed  
and it is here that topology comes into play.  In fact, when a system 
undergoes a phase 
transition,  the fluctuations of the configuration-space curvature, when 
plotted as a 
function of either the temperature or the energy of the system, exhibit a 
singular behavior 
at the phase transition point, which can be qualitatively reproduced using 
geometric 
models.  In  these models the origin of the singular behavior of the 
curvature fluctuations 
appears to be caused by a topological transition in configuration space, 
which  
corresponds to the phase transition of the physical system.  This leads us 
to put forward a 
{\em Topological Hypothesis} (TH).  The content of the TH is that phase 
transitions would be 
related at a deeper level to a change in the topology of the configuration 
space of the 
system.  We will illustrate this on a simple model, the mean-field XY model, 
where the 
TH can be checked directly and analytically. Since this model is of a rather 
special nature, 
namely a mean-field model with infinitely-ranged interactions, we discuss 
other more 
realistic (non mean-field-like) models, which can not be solved analytically, 
but which do 
supply direct supporting evidence for the TH via numerical simulations.  
 
\end{abstract} 

\begin{flushright}

{\em To the memory of Lando Caiani}

\end{flushright}
 
\tableofcontents 
 
\section{Introduction} 
This paper deals with the application of concepts drawn from mathematics, in
particular from differential geometry and topology, 
to problems in statistical physics.  
The mathematical tools 
involved come from Riemannian geometry and from  
Morse theory, respectively. As to the physics,  
the applications of 
these concepts will be brought to bear on dynamical systems  
with many degrees of 
freedom, including eventually the thermodynamic limit. 
 
In order to contain this report to a reasonable size and yet make it 
accessible to as wide a readership as possible, and since it makes use of 
concepts which might not be known to everyone, we chose the 
following format.  
 
The first part of the main text  
is aimed at a reader who is familiar with the basics of 
Riemannian geometry, for example at the level of a course in general 
relativity. As to the second part, the knowledge of Morse theory at an  
elementary level is assumed. However, for those who are not familiar with  
these branches of mathematics, we have provided in extended Appendices 
the main points which are needed to follow the exposition. Similarly, we 
assume that the reader is familiar with the basics of dynamical systems 
theory, 
but again we summarized in an Appendix the main concepts. In all cases 
references to the literature are made for the details.  
Both the main text and the Appendices are written as a compromise 
between mathematical rigour and a physcist's accessibility; in case
of conflict, we always favoured the latter. 
 
This way we hope that a reader familiar with the basic mathematical tools will 
be able to read the report straightforwardly. We have made a special effort to 
emphasize logical structure and physical content, and we hope that the  
report will provide a clear survey of 
what has been achieved applying geometrical methods to dynamical systems and 
statistical mechanics. At the same time we would like this paper to allow the 
reader to familiarize herself or himself with this new field and to  
stimulate new developments and 
contributions to the many points which are still open.  
 
Throughout the paper we will consider classical Hamiltonian dynamical systems 
with $N$ degrees of freedom, confined in a finite volume\footnote{We will 
mainly apply our results to systems defined on a lattice, so that we will not explicitly 
refer to the volume of the system.}, whose Hamiltonian is of the form 
\beq 
{\cal H} = \frac{1}{2}\sum_{i=1}^N p_i^2 + V(q_1,\ldots,q_N)~, 
\label{H} 
\eeq 
where the $q$'s and the $p$'s are, respectively,  
the coordinates and the conjugate momenta of the system. 
Our emphasis is on systems  
with a large number of degrees of freedom. 
The dynamics of the system (\ref{H}) is defined in the  
$2N$-dimensional phase space spanned by the $q$'s and the $p$'s. 
Our aim is to relate the dynamical and the statistical properties of  
the system (\ref{H}) with the geometrical and topological properties of the  
phase space where the dynamical trajectories of the system live.  
It turns out that as long as we consider Hamiltonians of the form 
(\ref{H}) we can restrict ourselves to the study of the geometry 
and the topology of the $N$-dimensional 
configuration space\footnote{Actually we 
will also consider an enlarged configuration space with two extra dimensions.} 
without loosing information. In fact as we shall see in Sec.\ \ref{sec_geodyn} 
the dynamical trajectories can be seen as geodesics of the configuration 
space, 
provided the latter has been endowed with a suitable metric. As to the 
topology, 
later on we shall see that also all the relevant information on the 
topology of 
phase space, from the point of view of Morse theory, 
is encoded in the potential 
energy function $V(q_1,\ldots,q_N)$, 
so that also the topological investigation can be 
restricted to the configuration space. 
 
This is similar to what happens in the classical 
statistical mechanics of Hamiltonian systems of the form (\protect\ref{H}),  
where the momenta can be integrated out and the statistical measure  
can be defined on the configuration space alone. 
We remark that this is true for 
both the microcanonical and the canonical ensemble. 
 
The structure of the paper is the following: 
after a short historical introduction (Sec.\ \ref{sec_hist}),  
the main body of the paper  
is organized in two parts, which, though being tightly related, are 
also to a great extent independent from each other, so they could be also read 
separately without encountering too many difficulties. 
 
The first part (Secs.\ \ref{sec_geodyn} and \ref{sec_geochaos})  
concerns the applications of tools belonging to classical 
differential geometry to study the chaotic dynamics of Hamiltonian systems of 
the form (\ref{H}), in particular for those  
with a large number of degrees of freedom. Starting from the 
identity between the trajectories of the dynamical system (\ref{H}) and the 
geodesics of the configuration space equipped with the Jacobi or the Eisenhart 
metric, we develop a geometric theory of chaotic dynamics which sheds a new 
light on the origin of chaos in Hamiltonian systems. In fact it turns out that 
chaos can be induced not only by negative curvatures, but also by positive 
curvatures, provided the curvatures are fluctuating 
along the geodesics. In the 
case of a large number of degrees of freedom it is possible to describe the 
instability of the dynamics by means of a geometric model independent of the 
dynamics, which provides an analytical estimate of the largest Lyapunov 
exponent in terms of the averages and fluctuations of the curvature of 
the configuration space. We clarified the exposition  
with the application of the general concepts to two special examples; however, 
since some of the calculations are rather lengthy, we did not provide all the 
details, referring the reader for those to the appropriate literature. 
 
While in the first part 
we deal with the application of differential geometry to  
dynamical systems with a large number of degrees of freedom, 
we do not touch upon one of the 
most spectacular properties of large systems, namely that in the thermodynamic 
limit $N \to \infty$ they may show sharp phase transitions. In the second part 
(Secs.\ \ref{sec_geopt} and \ref{sec_topo}) we address this point,  
and it is here that topology comes into play. 
In fact, when a system undergoes a 
phase transition, the fluctuations of the configuration-space curvature, 
when plotted as a function of either the temperature or the energy,  
have a singular behavior at the transition point which 
can be qualitatively reproduced using a geometric model. In such a model  
the origin of the singular behavior of the curvature fluctuations resides 
in a topological change. This leads us to put forward a  
{\em Topological Hypothesis} (TH). The content of the TH is 
that phase transitions (at least, continuous phase transitions) would at a 
deeper level be related to a particular change in the topology  
of the configuration 
space of the system. We will illustrate this on a simple model, the 
mean-field XY model, where the TH can be checked directly and analytically. 
Since this model is of a rather special nature, namely a mean-field model 
with infinitely-ranged interactions, we discuss other more realistic (non 
mean-field-like) models, which can not be solved analytically, but which do 
supply indirect supporting evidence for the TH via numerical simulations. 

We emphasize that the purpose of the work discussed in the second part
of this Report is not to extend the existing theory of phase transitions
for systems described by a classical Hamiltonian. Rather, we try to
extend the foundation for the occurrence of phase transitions to transitions
in the topology of the configuration space of the system undergoing a
phase transition. This way we hope to make a new connection between a
branch of pure mathematics (topology) and a branch of statistical mechanics
(phase transitions). Such a connection appears to lead to a new 
approach to phase transitions.

A brief historical summary, to place the content of the present review in
its historical context, is given in the next Section.

\section{Historical remarks} 
\label{sec_hist} 
Without attempting to be exhaustive, a few historical comments might be 
helpful to place  the recent contributions about the 
geometrical approach to dynamics and statistical physics which are reviewed 
in the present article, in a more general context. This makes  the 
present Section  an exception to the self-containedness 
of this review paper, because we mention here concepts which are not necessary 
to understand the topics treated in the rest of the paper.

The idea of looking at the collection of solutions of Newton's equations
of motion from a geometric point of view dates back to H.\ Poincar\'e and to
the development of the qualitative theory of differential equations.
Tackling the famous problem of the integrability of the three-body problem,
Poincar\'e also discovered that generic classical Hamiltonian systems, in 
spite of their deterministic nature, lack predictability, i.e. were unstable, 
because of their 
extreme sensitivity to the initial conditions. Such an instability of 
classical dynamics originates in homoclinic intersections, which Poincar\'e
described in his {\it M\'ethodes Nouvelles de la M\'ecanique C\'eleste}
\cite{Poincare} 
without ``even attempting to draw'' them (see Appendix \ref{app_chaos}).
 However his geometric treatment of  
dynamics, later developed by E.\ Cartan among others, involves 
submanifolds of phase space using what is now called symplectic geometry
\cite{Arnold}. Although 
of undeniable elegance, symplectic geometry is not very helpful to advance 
our knowledge about those regions in phase space where the dynamics 
is unstable, knowledge which is relevant for statistical mechanics. 

The name of Poincar\'e, together with that of E.\ Fermi, is also 
associated with an important theorem about the non-existence of analytic
integrals of motion, besides energy, for generic nonlinear Hamiltonian systems
describing at least three interacting bodies \cite{Poincare,Fermi}; 
this is the origin of the 
accessibility of the whole constant energy hypersurface of phase space which 
is determined by the initial conditions. The Poincar\'e-Fermi theorem has been 
generally considered by the physics 
community as sufficient to legitimate classical statistical mechanics from a  
dynamical viewpoint\footnote{Kol'mogorov-Arnol'd-Moser(KAM) theory 
\protect\cite{KAM} might seem 
capable of explaining how, in spite of this ``no-go'' theorem due to 
Poincar\'e and Fermi,
ergodicity could fail. However, the exceedingly tiny -- and fastly vanishing
with the number $N$ of degrees of freedom -- nonintegrable perturbations that 
are required to keep a positive 
measure of regular regions in phase space, do, in general, not have any 
physically appreciable effect even when only a few bodies are considered.}.

One had to wait until the 1940s when  a qualitatively new attempt emerged to
make use of geometric concepts in the investigation of newtonian dynamics and
its connection with statistical mechanics. Though very different from the clear 
mathematical expositions of Poincar\'e, and much more in the spirit of a 
physicist, it was N.\ S.\ Krylov \cite{Krylov} in his doctoral dissertation, 
who showed for the first time the existence of a close relationship between
dynamical instability (seen as the exponential 
amplification of small deviations in the initial conditions of a collection 
of colliding objects representing idealized atoms in a gas) and phase space
mixing. Phase mixing is a stronger property than ergodicity and is far
more relevant to physics than ergodicity. In fact, 
while ergodicity assures the equality of time and phase space averages of
physical quantities, phase mixing addresses the rate of approach of time to
ensemble averages. Phase mixing entails thus the
convergence of time averages to ensemble averages in a finite time. 
In modern terms,
Krylov realized the necessity of chaotic dynamics to obtain  
phase mixing and to make the connections between dynamics and statistical 
mechanics stronger. Moreover Krylov, in view of what we are going
to discuss in the next Sections, also has the great historical merit of having
attempted for the first time to bridge the dynamical foundations of 
statistical mechanics with
a widely developed and powerful field of mathematics: Riemannian differential
geometry.
Krylov knew certain mathematical results concerning the properties of
geodesic flows on compact negatively curved manifolds
by Hadamard, Hedlund and Hopf \cite{HHH},    
and he realized their potential interest to physics, once newtonian
dynamics is rephrased in terms of Riemannian geometric language. 
Such a possibility
was well known since the beginning of the century, mainly due to the work of  
T.\ Levi-Civita, in particular that 
the principle of stationary action entails the identity of a classical 
mechanical flow with a geodesic flow in a configuration space, endowed with 
a suitable metric.
Krylov's efforts concentrated on the analysis of the properties of physical
systems which move in negatively curved regions in configuration space.
For example, he discussed how the presence of an inflection point in the 
Lennard-Jones potential could influence the dynamics of a dilute gas (through 
the appearance of regions of negative scalar curvature in configuration space)
and lead to a strong instability of the dynamics.
These attempts have been very influential on the development of the so-called
abstract ergodic theory \cite{ergodic_theory}, where Anosov flows 
\cite{Anosov} (e.g., geodesic flows on compact manifolds
with negative curvature) play a prominent role.
Ergodicity and mixing of these flows have been thoroughly investigated.
To give an example, Sinai proved ergodicity and mixing for two hard spheres
by just showing that such a system is equivalent to a 
geodesic flow on a negatively-curved compact manifold \cite{Sinai}

From time to time, Krylov's intuitions have
been worked out further by several physicists for whom we refer to Refs.
\cite{Aizawa,Ong,Gutzwiller,vanVelsen,Gurzadyan,Knauf,Szydlowski,AizawaII,Nobbe,Heller,Kandrup}.
They invariably discovered,
much to their surprise, that geodesic flows associated with physical
Hamiltonians do not live on negatively curved manifolds, despite their 
chaoticity even if the latter is well developed; only very few exceptions 
to this are known, in fact the two low dimensional models discussed in Refs.
\cite{Gutzwiller,Knauf}, where chaos is actually associated 
with negative curvature. 
Worse, for certain models the regions of negative
curvature of the mechanical manifolds apparently shrink by increasing the
number $N$ of degrees of freedom.

This somewhat biased search for negative curvatures has been the main
obstacle to an effective use of the geometric framework originated by Krylov 
to explain the source of chaos in Hamiltonian systems. 
On the other hand, it is true that 
the Jacobi equation, which describes the (in)stability of a geodesic flow,
is in practice only tractable on negatively curved manifolds,
formidable mathematical difficulties are encountered in treating 
the (in)stability of geodesic flows on manifolds of 
non-constant and not everywhere negative curvature.
Moreover, for this kind of problems, intuition can hardly help. However,
the advent of computers has been here of invaluable help. In this connection, 
it may not be out of place to quote here some of 
the sentences which S.\ Ulam remembered from the far-looking 
conversations he had with E.\ Fermi and J.\ von Neumann \cite{Ulam}: 

\begin{quotation}
``After the war, during one of his frequent summer visits to Los Alamos, Fermi
became interested in the development and potentialities of the electronic
computing machines. He held many discussions with me on the kind of future
problems which could be studied through the use of such machines. We decided
to try a selection of problems for heuristic work where in the absence of 
closed analytic solutions experimental work on a computing machine would 
perhaps contribute to the understanding of properties of solutions.

(...) Fermi expressed often the belief that future fundamental 
theories in physics may involve nonlinear operators and equations, and that 
it would be useful to attempt practice in the mathematics needed for the 
understanding of nonlinear systems (...).''
\end{quotation}

As a matter of fact, only during the last few years an interplay 
between analytic 
methods and numerical simulations has made possible to overcome the mentioned
difficulties, proving the effectiveness of the Riemannian 
geometric approach 
to dynamical systems of  interest to statistical mechanics, field 
theory, and condensed matter physics 
\cite{Pettini,tesi_lapo,pre93,prl95,CerrutiPettiniII,PettiniValdettaro,tesi_cecilia,tesi_lando,CerrutiPettini,pre96,cecilia_master,lapo_thesis,gasrari,CerrutiFranzosiPettini,cccp,pre98,jpa98,jpa99}.
As we shall see in the following, this has extended the domain of 
application of geometric techniques, and has also introduced a new point
of view about the origin of chaos in Hamiltonian systems, as well as new 
methods to describe and understand it.  

The use of Finsler manifolds (generalizations of Riemannian manifolds that
allow the geometrization of velocity dependent potentials as well)
has also been proposed in 
Refs.\ \cite{Dryuma,Cipriani}. An analysis of dynamics based on the geometry
of trajectories, rather than of the manifolds on which they move, has 
been proposed in Ref.\ \cite{Casartelli}.  

For what concerns the use of geometric and topologic concepts in statistical
mechanics, we must distinguish between macroscopic and microscopic phase 
spaces. A macroscopic phase space is a low-dimensional space spanned by 
macroscopic variables, like temperature, pressure, volume, chemical potential 
etc, in other words it is in general a parameter space.
In the '70s some applications to the study of phase transitions of the 
theory of singularities of differentiable maps 
(popularly known as Catastrophe theory), which includes Morse 
theory, were proposed. These followed R.\ Thom's remark that  the simplest 
example of the classical critical point as it appears in the van der Waals
equation corresponds to the Riemann-Hugoniot catastrophe \cite{Thom,Poston}.

An elegant and deep formulation of phase transitions related to structural 
instability and using one of the most beautiful theorems in differential
topology, the Atiyah-Singer index theorem, was proposed by M.\ Rasetti in
Ref.\ \cite{Mario}.

Other very recent proposals of geometric and topologic methods in macroscopic
phase spaces have been put forward in Refs.\ \cite{Ruppeiner,Votyakov}. 

In recent papers some elements of the geometry of constant energy
hypersurfaces $\Sigma_E$ of phase space have been used for the microcanonical 
definitions of the temperature and the specific heat \cite{Rugh,Giardina}, 
in Ref.\ \cite{CSCP} a topological property of the $\Sigma_E$ has been related 
to their mean curvature from which a relationship between 
thermodynamics and topology emerged. 

The description of phase transitions through geometric and topologic changes
in the microscopic phase space has never been considered until very recently.
It appeared as a natural consequence of the above mentioned new 
developments in the Riemannian theory of Hamiltonian chaos. 
These newly proposed ideas, as well as the conceptual path that led from 
the geometry of dynamics to topology and 
phase transitions, are reviewed in this paper.

\section {Riemannian geometry and Hamiltonian dynamics} 
\label{sec_geodyn} 
A Hamiltonian system whose kinetic energy is a quadratic form  
in the velocities  
is referred to as a {\em natural} Hamiltonian system. Every  
Newtonian system, 
that is a system of particles interacting through forces derived 
from a potential, i.e. of the form (\ref{H}), belongs to this  
class. 
The trajectories of a natural system can be seen  
as geodesics  
of a suitable Riemannian manifold. This classical result is  
based on a 
variational formulation of dynamics. In fact Hamilton's  
principle states 
that the natural motions of a Hamiltonian system are the  
extrema of 
the functional (Hamiltonian action ${\cal S}$) 
\beq 
{\cal S} = \int {\cal L} \, dt~, 
\eeq 
where ${\cal L}$ is the Lagrangian function of the system,  
and the  
geodesics of a Riemannian manifold are the extrema of the  
length 
functional  
\beq 
\ell = \int ds~,
\eeq 
where $s$ is the arc-length parameter. 
Once a connection between length and action is established,  
by means 
of a suitable choice of the metric, it will be possible to  
identify 
the geodesics with the physical trajectories. 
 
\subsection{Geometric formulation of the dynamics} 
 
The Riemannian formulation of classical dynamics is far from  
unique, 
even if we restrict oureselves to the case of natural systems. There  
are many 
possible choices for the ambient space and its metric.  
The most commonly known choice --- dating back to the nineteenth  
century --- 
is the so-called Jacobi metric on the configuration space 
of the system. Actually 
this was the geometric framework of Krylov's work. 
Among other possibilities, we will also consider a metric  
originally 
introduced by Eisenhart on an enlarged configuration space-time.  
The choice of the metric to be used will be dictated mainly by convenience. 
 
These choices certainly do not contain all the possibilities of  
geometrizing 
conservative dynamics. For instance, with regards to  
systems whose 
kinetic energy is not quadratic in the velocities --- the  
classical example is 
a particle subject to conservative as well as velocity-dependent  
forces, such as the  
Lorentz force --- it is impossible to give a Riemannian  
geometrization, 
but becomes possible in the more general  
framework of a Finsler 
geometry \cite{Finsler}. 
However, we will not consider this here, and restrict ourselves 
to standard Hamiltonian systems.   
 
For a summary of the notation and the concepts of differential geometry that will be used in the following we refer the reader to Appendix \ref{app_geo}.  
The summation convention over repeated indices will be always used throghout  
the paper, if not explicity stated otherwise. 
 
\subsubsection{The Jacobi metric} 
 
Let us consider an autonomous   
dynamical system, i.e., a system with interactions which do not 
explicitly depend on time, whose  
Lagrangian 
can be written as 
\beq 
{\cal L} = T - V =  
\frac{1}{2} a_{ij} \dot q^i \dot q^j - V(q)~,  
\label{L} 
\eeq 
where the dot stands for a derivative with respect to the parameter on
which the $q$'s depend\footnote{Such a parameter is the time $t$ here, but 
could also be the arc-length $s$ in the following.}, and $q$ is a shorthand
notation for all the coordinates $q_1,\ldots,q_N$. Both these conventions 
will be used throughout the paper, when there is no possibility of confusion.

The Hamiltonian ${\cal H} = T + V$ is an integral of motion, 
whose value, the energy $E$, is a conserved quantity.   
Hence 
Hamilton's priciple can be cast in Maupertuis' form \cite{Arnold}: 
the natural motions of the system are the stationary paths in the  
configuration space $M$ 
for the functional 
\beq 
{\cal A} = \int_{\gamma(t)} p_i \, dq^i  
         = \int_{\gamma(t)} \frac{\partial {\cal  
L}}{\partial \dot q^i} 
          \, \dot q^i\, dt  
\label{action_M} 
\eeq 
among all the isoenergetic curves, i.e. the curves  
$\gamma(t)$ connecting 
the initial and final points parametrized so that 
the Hamiltonian ${\cal H}({p},{q})$ is a constant equal  
to the energy 
$E$. The fact that the curves must be isoenergetic with energy $E$ 
implies that the accessible part of the configuration space is not 
the whole $M$, but only the subspace $M_E \subset M$ defined by  
\beq 
M_E = \{ q \in M : V(q) \leq E \}~. 
\eeq 
In fact a curve $\gamma'$ which lies outside $M_E$ will never be 
parametrizable in such a way that the energy is $E$, because $\gamma'$ 
will then pass through points where $V > E$ and the kinetic 
energy is positive\footnote{The accessible configuration space 
$M_E$ can then be seen as the union 
of all the ``sub-configuration spaces'' $\{ q \in M : V(q) = E - T \}$
that one gets for all the possible values of $T$, $0 \leq T \leq E$.}.
 
The kinetic energy $T$ is a homogeneous function of  
degree two 
in the velocities, hence Euler's theorem implies that 
\beq 
2T = \dot q^i \frac{\partial {\cal L}}{\partial \dot q^i}~, 
\eeq 
and Maupertuis' principle reads as 
\beq 
\delta{\cal A} = \delta \int 2T \, dt = 0~. 
\label{M_principle} 
\eeq 
The configuration space $M$ of a dynamical system with $N$  
degrees of 
freedom has a differentiable manifold structure, and  
the 
Lagrangian coordinates $(q_1,\ldots,q_N)$ can be regarded as  
local 
coordinates on $M$. The latter becomes a Riemannian manifold 
once a proper metric is defined. For the sake of simplicity, let us  
consider systems of the form (\ref{H}), so that 
the kinetic energy matrix is diagonal  
and the masses are all equal to one, i.e., $a_{ij} = \delta_{ij}$. 
If we write 
\beq 
g_{ij} = 2[E - V(q)]\, \delta_{ij}~; 
\label{g_J_def} 
\eeq 
then Eq.\ (\ref{M_principle}) becomes 
\beq 
0 = \delta \int 2T \, dt =  
\delta \int \left( g_{ij} \dot q^i \dot q^j \right)^{1/2} \,  
dt = 
\delta \int ds ~,    
\eeq 
so that the natural motions are the geodesics of $M$  
provided  
$ds$ is the arc-length element, i.e., the metric  
on $M$ 
is given by the tensor whose components are just the $g_{ij}$ defined 
in Eq.\ (\ref{g_J_def}). This metric is referred to as  
the 
{\em Jacobi metric}, and its arc-length element is 
\beq 
ds^2 \equiv g_{ij}dq^i dq^j = 2[E - V(q)]\, \frac{dq^i}{dt}  
\frac{dq_i}{dt} \, dt^2 = 4[E - V(q)]^2\,  dt^2 ~. 
\label{ds_J} 
\eeq 
The geodesic equations written in the local coordinate frame  
$(q^1,\ldots,q^N)$ are (see Appendix \ref{app_geo}, Eq.\ (\ref{eqs_geodet}))
\beq 
\frac{D \dot\gamma}{ds} \equiv \frac{d^2 q^i}{ds^2} + 
\Gamma^i_{jk} \frac{dq^j}{ds}  
\frac{dq^k}{ds} = 0~, 
\label{eq_geodesics_loc} 
\eeq 
where $D/ds$ is the covariant derivative along the curve $\gamma(s)$
(see Appendix \ref{app_geo}, Eqs.\ (\ref{def_D}) and (\ref{D_local})), 
$\dot\gamma = dq/ds$ is the velocity vector of the geodesic and  
the $\Gamma$ are the Christoffel symbols. Using the definition of the 
Christoffel symbols (see Appendix \ref{app_geo}, Eq.\ (\ref{Gamma})) 
and Eq.\ (\ref{g_J_def}) it is  straightforward
to show that the Eqs.\ (\ref{eq_geodesics_loc}) become 
\beq 
\frac{d^2 q^i}{ds^2} + \frac{1}{2(E - V)} 
\left[2 \frac{\partial (E-V)}{\partial q_j} \frac{dq^j}{ds}  
\frac{dq^i}{ds}  
- g^{ij} \frac{\partial (E-V)}{\partial q_j}  
g_{km}\frac{dq^k}{ds} \frac{dq^m}{ds} \right] = 0~, 
\eeq 
whence, using Eq.\ (\ref{ds_J}),  
Newton's equations are recovered, 
\beq 
\frac{d^2 q^i}{dt^2} = - \frac{\partial V}{\partial q_i}~. 
\label{eqsNewton}
\eeq  
 
Note that the Jacobi metric is obtained by a conformal change of the  
kinetic energy 
metric $a_{ij}$ --- see Eq.\ (\ref{g_J_def}) and Appendix \ref{app_geo},
\S \ref{app_sec_curvature}.  
In fact the general result 
for the Riemannian geometrization of natural 
Hamiltonian dynamics is the following 
 
\smallskip 
 
\noindent{\bf Theorem.} 
{\em Given a dynamical system on a Riemannian manifold $(M,a)$, i.e., a  
dynamical system whose Lagrangian is} 
\[ 
{\cal L} = \frac{1}{2} a_{ij} \dot q^i \dot q^j - V(q)~, 
\] 
{\em then it is always possible to find a conformal transformation  
of the metric,} 
\[ 
g_{ij} = e^{\varphi(q)} a_{ij} 
\] 
{\em such that the geodesics of $(M,g)$ are the trajectories of the  
original dynamical system; this transformation is defined by} 
\[ 
\varphi(q) = \log [E - V(q)]~. 
\] 
\smallskip 
 
\noindent The proof proceeds as above, 
using Eqs.\ (\ref{g_J_def})-(\ref{eqsNewton}) 
and simply replacing all the $\delta_{ij}$ 
matrices with the kinetic energy matrix $a_{ij}$;  
for details, see e.g. Ref.\ \cite{Ong}. 
 
\subsubsection{The Eisenhart metric} 
\label{Eisenhart_metric} 
We could try to consider the configuration spacetime $M  
\times {\bf R}$ as an alternative 
ambient space for the geometrization of dynamics, with local  
coordinates 
$(q^0 = t, q^1,\ldots,q^N)$, and to define a metric starting  
from 
Hamilton's principle $\delta \int {\cal L} \, dt = 0$. 
We could try to define a metric tensor by multiplying the Lagrangian  
(\ref{L}) 
by $2(dq^0)^2$:  
\beq 
ds^2 = 2 {\cal L} \, (dq^0)^2 =   
\left(g_{\cal L}\right)_{\mu\nu} dq^\mu dq^\nu 
 = a_{ij} \, dq^i dq^j -2V(q)\, (dq^0)^2~,  
\eeq 
where $\mu$ and $\nu$ run from $0$ to $N$ and  $i$ and $j$ run from 1 to $N$; however, one can easily verify that  
the geodesics of the manifold $(M\times{\bf  
R}, g_{\cal L})$  
are then not the natural motions of the systems, since 
the Lagrangian is not an integral of the motion.  
 
However, we can consider an ambient space  
with an 
extra dimension, $M\times {\bf R}^2$, with local coordinates 
$(q^0,q^1,\ldots,q^i,\ldots,q^N,q^{N+1})$. This space can be  
endowed  
with a non-degenerate pseudo-Riemannian metric (see Appendix \ref{app_geo},
\S \ref{sec_app_metrics})), first  
introduced by 
Eisenhart \cite{Eisenhart}, whose arc-length is  
\beq 
ds^2 = \left( g_E \right)_{\mu\nu}\, dq^{\mu}dq^{\nu} =  
a_{ij} \, dq^i dq^j -2V(q)(dq^0)^2  
+ 2\, dq^0 dq^{N+1} ~, 
\label{g_E} 
\eeq 
where $\mu$ and $\nu$ run from $0$ to $N+1$ and  $i$ and $j$ run from 1 to $N$, and which, from now on, will be referred to as the {\em  
Eisenhart metric}, and whose metric tensor will be denoted as $g_E$.  
The relation between the geodesics of 
this manifold and the natural motions of the dynamical system  
is contained in the following \cite{Lichnerowicz} 
\smallskip 
 
\noindent{\bf Theorem (Eisenhart).} 
{\em The natural motions of a Hamiltonian dynamical system  
are obtained as the canonical projection  
of the geodesics 
of $(M\times {\bf R}^2,g_E)$ on the configuration  
space-time,  
$\pi : M\times {\bf R}^2 \mapsto M\times {\bf R}$.  
Among the totality of geodesics, only those whose  
arclengths are
positive-definite and are given by} 
\beq 
ds^2 = c_1^2 dt^2 
\label{ds_Eisenhart} 
\eeq 
{\em correspond to natural motions; the  
condition (\ref{ds_Eisenhart}) can be equivalently cast in the following  
integral form 
as a condition on the extra-coordinate $q^{N+1}$:} 
\beq 
q^{N+1} = \frac{c_1^2}{2} t + c^2_2 - \int_0^t {\cal L}\,  
d\tau~, 
\label{qN+1} 
\eeq 
{\em where $c_1$ and $c_2$ are given real constants. 
Conversely, given a point $P \in M \times {\bf R}$ belonging  
to a trajectory 
of the system, and given two constants $c_1$ and $c_2$, the  
point 
$P' = \pi^{-1} (P) \in M \times {\bf R}^2$, with   
$q^{N+1}$ given by  
(\ref{qN+1}), describes a geodesic curve in $(M\times {\bf  
R}^2,g_E)$ 
such that $ds^2 = c_1^2 dt^2$.} 
 
\smallskip 
 
\noindent For the full proof, see Ref.\ \cite{Lichnerowicz}. 
Since the constant $c_1$ is arbitrary, we will always set  
$c_1^2 = 1$ in order 
that $ds^2 = dt^2$ on the physical geodesics. 
 
From Eq.\ (\ref{g_E}) follows that  
the explicit table of the components of the Eisenhart metric is 
\beq 
g_E = \left( 
\begin{array}{ccccc} 
-2V(q)& 0       & \cdots        & 0     & 1     \\ 
0       & a_{11}& \cdots        & a_{1N}& 0     \\ 
\vdots  & \vdots& \ddots        & \vdots& \vdots\\ 
0       & a_{N1}& \cdots        & a_{NN}& 0     \\ 
1       & 0     & \cdots        & 0     & 0     \\ 
\end{array} \right) \label{gE} 
\eeq 
where $a_{ij}$ is the kinetic energy metric. The  
non-vanishing Christoffel symbols, in the case $a_{ij} =  
\delta_{ij}$, are only 
\beq 
\Gamma^i_{00} = - \Gamma^{N+1}_{0i} = \partial_i V~,  
\label{Gamma_E} 
\eeq 
so that the geodesic equations (\ref{eq_geodesics_loc}) read  
\begin{mathletters} 
\beq 
\frac{d^2q^0}{ds^2}  =  0 ~, \label{eqgeo0}  
\eeq 
\beq 
\frac{d^2q^i}{ds^2} +\Gamma^i_{00}  
\frac{dq^0}{ds}\frac{dq^0}{ds}  =  0~, 
\label{eqgeoi}  
\eeq 
\beq 
\frac{d^2q^{N+1}}{ds^2} +\Gamma^{N+1}_{0i} \frac{dq^0}{ds} 
\frac{dq^i}{ds}  =  0 ~; \label{eqgeoN+1} 
\eeq 
\end{mathletters} 
using $ds = dt$ one  
obtains 
\begin{mathletters} 
\beq 
\frac{d^2q^0}{dt^2}  =  0~, \label{eqgeo0t}  
\eeq 
\beq 
\frac{d^2q^i}{dt^2}  =  - \frac{\partial V}{\partial q_i}  
~,\label{eqgeoit}  
\eeq 
\beq 
\frac{d^2q^{N+1}}{dt^2}  =   - \frac{d{\cal L}}{dt} ~.  
\label{eqgeoN+1t} 
\eeq 
\label{eqs_geo_eisenhart}
\end{mathletters} 
Eq. (\ref{eqgeo0t}) only states that $q^0=t$,  
The $N$ equations (\ref{eqgeoit}) 
are Newton's equations, and Eq.\ (\ref{eqgeoN+1t}) is 
the differential version of Eq.\ (\ref{qN+1}). 
 
The fact that in the framework of the Eisenhart metric the dynamics can be 
geometrized with an affine parametrization of the arc-length, i.e.,  
$ds = dt$, will be extremely useful in the following, together 
with the remarkably simple curvature properties of the Eisenhart metric (see 
\S \ref{curv_mech}). 
  
\subsection{Curvature and stability} 
 
The geometrization of the dynamics is a natural framework  
for the study 
of the stability of the trajectories of a dynamical system,  
for it  
links the latter with the stability of the geodesics; the latter 
is {\em completely 
determined} by the curvature of the manifold, as shown below.  
 
Studying the stability of the dynamics means determining the evolution  
of perturbations of a given trajectory. This implies that one should follow 
the evolution of the linearized (tangent) flow along the reference trajectory. 
For a Newtonian system, writing the perturbed trajectory as 
\beq 
\tilde{q}^i(t) = q^i(t) + \xi^i(t) ~, 
\eeq 
substituting this expression in the equations of motion 
\beq 
\ddot q^i = -\frac{\partial V(q)}{\partial q^i}~, 
\eeq 
and retaining terms up to first order in the $\xi$'s,  
one finds that the perturbation obeys the so-called {\em tangent dynamics 
equation} which reads as 
\beq 
\ddot\xi^i = - \left( \frac{\partial^2 V(q)}{\partial q^i \partial q^j} 
\right)_{q^i = q^i(t)} \xi^j~. 
\label{eq_dintang} 
\eeq 
This equation should be solved together with the dynamics in order to 
determine the stability or instability of the trajectory: when the norm  
of the  perturbations 
grows exponentially, the trajectory is unstable, otherwise it is stable.  
 
Let us now translate the stability problem into geometric language. 
By writing, in close analogy to what has been done above in the case of 
dynamical systems, a perturbed geodesic as 
\beq 
\tilde{q}^i(s) = q^i(s) + J^i(s) ~, 
\eeq 
and then inserting this expression in the equation for the geodesics 
(\ref{eq_geodesics_loc}), one finds 
that the evolution of the perturbation vector $J$ is given by the following 
equation: 
\beq 
\frac{D^2 J^i}{ds^2} + R^i_{~jkl} \frac{dq^j}{ds} J^k \frac{dq^l}{ds} = 0~, 
\label{eq_jacobi_geo} 
\eeq 
where $R^i_{~jkl}$ are the components of the Riemann curvature tensor (see  
Appendix \ref{app_geo}, Eq.\ (\ref{curv_components})).  
Equation 
(\ref{eq_jacobi_geo}) is referred to as the Jacobi equation, and the 
tangent vector field $J$ as the Jacobi field. This equation was first studied by 
Levi-Civita and is also often referred to as the equation of Jacobi and 
Levi-Civita. For a derivation we refer to Appendix \ref{app_geo},  
\S \ref{sec_app_jacobi}, where it is also shown that one can 
always assume that $J$ is orthogonal to the velocity vector along the
geodesic, $\dot\gamma$, i.e.,
\beq
\langle J, \dot\gamma \rangle = 0~,
\label{J_ort}
\eeq
where $\langle \bullet , \bullet \rangle$ stands for the scalar product
induced by the metric (see Appendix \ref{app_geo}, Eq.\ (\ref{scalprod})).
The remarkable 
fact is that the evolution of $J$ --- and then the stability or instability of 
the geodesic --- is completely determined by the {\em curvature} of 
the manifold. 
Therefore, if the metric is induced by a physical system, as  
in the case 
of Jacobi or Eisenhart metrics, such an  
equation links  
the stability or instability of the trajectories to the  
curvature of the ambient manifold.  
 
The subject of the next sections is  
precisely to exploit 
such a link in order to describe and understand the  
behaviour of those physical 
systems whose trajectories are mainly unstable.  
 
However, before that, we have to give explicit expressions for  
the curvature of the mechanical manifolds, 
i.e., of those manifolds whose Riemannian structure is induced by the dynamics 
via the Jacobi or the Eisenhart metric. 
 
\subsection{Curvature of the mechanical manifolds} 
 
\label{curv_mech} 
 
\label{par_jacobi} 
We already observed that the Jacobi metric is a conformal  
deformation 
of the kinetic-energy metric, whose components are given by the kinetic energy 
matrix $a_{ij}$. In the case of systems whose kinetic energy 
matrix is diagonal, 
this means that the Jacobi metric is conformally flat (see  
Appendix \ref{app_geo}, \S \ref{app_sec_curvature}).  
This greatly simplifies the  
computation of 
curvatures. It is convenient to define then a  
symmetric tensor 
$C$ whose 
components are \cite{Ong} 
\beq 
C_{ij} = \frac{N - 2}{4(E-V)^2} \left[ 2(E-V)  
\partial_i\partial_j V +  
3 \partial_i V \partial_j V  - \frac{\delta_{ij}}{2} |\nabla  
V|^2 
\right]~, 
\eeq 
where $V$ is the potential, $E$ is the energy, and $\nabla$  
and $| \cdot |$  
stand for the Euclidean gradient and norm, respectively.  
The curvature of $(M_E,g_J)$ can be expressed through  
$C$. In fact,  
the components of the Riemann tensor are 
\beq 
R_{ijkm} = \frac{1}{N - 2} \left[ C_{jk} \delta_{im} - C_{jm}  
\delta_{ik} + 
C_{im} \delta_{jk}  - C_{ik} \delta_{jm}   
\right]~. 
\eeq 
By contraction of the first and third indices,  
we obtain the Ricci tensor, whose components  
are (see Appendix \ref{app_geo}, Eq.\ (\ref{ricci_tensor}))
\beq 
R_{ij} =  \frac{N - 2}{4(E-V)^2} \left[ 2(E-V)  
\partial_i\partial_j V +  
3 \partial_i V \partial_j V  \right]  
+ \frac{\delta_{ij}}{4(E-V)^2} \left[ 2(E-V) \triangle V - 
(N-4)|\nabla V|^2 \right]~, 
\eeq 
and by a further contraction we obtain the scalar curvature 
(see Appendix \ref{app_geo}, Eq.\ (\ref{scalar_trace}))
\beq 
{\cal R} = \frac{N - 1}{4(E-V)^2} \left[ 2(E-V) \triangle V  
- (N-6)|\nabla V|^2 \right]~. 
\eeq 
 
The curvature properties of the Eisenhart metric are much  
simpler than those 
of the Jacobi metric, and this is obviously a great  
advantage from a  
computational point of view. The only non-vanishing  
components of the 
curvature tensor are  
\beq 
R_{0i0j} = \partial_i\partial_j V~, 
\eeq 
hence the Ricci tensor has only one nonzero component 
\beq 
R_{00} = \triangle V~, 
\label{ricci_eisenhart} 
\eeq 
and the scalar curvature is identically vanishing, 
\beq 
{\cal R} = 0~. 
\eeq 

To summarize, we have shown that the dynamical trajectories of a Hamiltonian
system of the form (\ref{H}) can be seen as geodesics of the configuration 
space, or of an enlargement of it, once a suitable metric is 
defined\footnote{As already stated at the beginning of this Section, there
are many other possible choices for the ambient manifold and it metric:
some other possible choices are described 
in Ref.\ \protect\cite{lapo_thesis}.}.
The general relationship which holds between dynamical and geometrical 
quantities regardless of the precise choice of the metric can be
sketched as follows:
\beq
\begin{array}{ccc}
\text{\sf dynamics} & & \text{\sf geometry}  \\
 & & \\
\text{\sf (time)}~~t & \sim & s~~ \text{\sf (arc-length)} \\ 
\text{\sf (potential energy)}~~V & \sim & g~~\text{(\sf metric)} \\
\text{\sf (forces)}~~ \partial V & \sim & \Gamma ~~\text{(\sf Christoffel 
symbols)} \\
\text{\sf (``curvature'' of the potential)}~~ \partial^2 V, (\partial V)^2 
& \sim & R~~\text{\sf (curvature of the manifold)} 
\end{array}
\eeq
In the case of the Eisenhart metric, all these relations are 
extremely simple (maybe as simple as possible). In fact the physical time
$t$ can be chosen as equal to the arc-length $s$, the metric tensor
$g_E$ contains only the potential energy $V$, the non-vanishing
Christoffel symbols $\Gamma$ are equal to the forces $\partial V$,
and the components of the Riemann curvature tensor $R$ contain
only the second derivatives of the potential energy, 
$\partial_i\partial_j V$.  

We have also shown that the stability of the dynamical 
trajectories can be mapped onto the stability of the geodesics, which
is completely determined by the curvature of the manifold.
We will show in \S \ref{geom_formula} that, 
in the case of the Eisenhart metric, 
as a consequence of its remarkably 
simple properties,  also the relationship
between the stability of the trajectories and the stability of the
geodesics becomes as simple as possible, i.e., the Jacobi equation
(\ref{eq_jacobi_geo}) becomes identical to tangent dynamics equation
(\ref{eq_dintang}).

\section {Geometry and chaos} 
\label{sec_geochaos} 
The purpose of the present section is to describe in some  
detail how it is 
possibile, using the Jacobi equation as the main tool, to reach  
a twofold objective: 
first, to obtain a deeper understanding of the origin of  
chaos in Hamiltonian 
systems, and second, to obtain quantitative informations on  
the ``strength'' of 
chaos in these systems. Some basic concepts about Hamiltonian chaos and the 
definition of Lyapunov exponents are 
summarized in Appendix \ref{app_chaos}.   
 
\subsection{Geometric approach to chaotic dynamics} 
 
A physical theory should provide a conceptual framework for  
modeling 
and understanding --- at least at a qualitative level ---  
the observed features 
of the system which is the object of the theory, and should  
also have 
a predictive content, i.e. should provide quantitative tools apt to  
compute, at least 
approximately, the outcomes of the experiments (no matter if it concerns 
laboratory experiments or numerical experiments performed on a 
computer). According 
to these requirements, a satisfactory theory of  
deterministic chaos is  
certainly still lacking. In fact in both aspects the current  
theoretical  
approaches to chaos have some problems, especially if we  
consider the case 
of conservative flows, i.e., of the dynamics of conservative systems of 
ordinary differential equations. 
 
To explain the origin of chaos in  
conservative  
dynamics   
one usually invokes the existence of invariant  
hyperbolic sets  
--- or horseshoes --- in phase space, like those generated  
by homoclinic  
intersections of perturbed separatrices 
(see Appendix \ref{app_chaos}). In order  
to quantify the degree of instability of a trajectory or of  
a system 
we must instead resort to the notion of Lyapunov exponents.  
The Lyapunov exponents are 
asymptotic quantities and their relation with  
local properties  
of phase space, like horseshoes, is far from evident;  
nonetheless they 
provide the natural measure of the degree of chaos, measuring the typical 
time scales over which a trajectory looses the memory of its initial 
conditions. 
A rigorous definition of the existence of  
chaotic 
regions in the phase space of a system, based on the  
detection of horseshoes, 
does not provide any quantitative tool to 
measure chaos; 
on the other hand, Lyapunov exponents allow a very precise  
measure of 
chaos but give no information at all on the origin of such a  
chaotic  
behaviour. From a conceptual point of view this situation is  
far from 
being satisfactory, not to speak of the fact that the  
practical 
application of the methods to search horseshoes becomes  
extremely difficult 
as the number of degrees of freedom is large \cite{MarsdenRatiu}. 
From the predictive point of view the situation is even 
worse, for  
no analytic method at all exists to compute Lyapunov  
exponents, 
at least in the case of flows of physical relevance. It is  
worth 
noticing that in a recent paper \cite{Gozzi}, Gozzi and Reuter  
have shown that one could build, in principle, a  
field-theoretic 
framework to compute Lyapunov exponents, but the practical  
application 
of such methods is still unclear. Needless to say, all the  
tools 
belonging to canonical perturbation theory, which have  
undergone remarkable  
developments in the last years \cite{can_perturb_theory}, 
can hardly be used to  
compute quantities like Lyapunov exponents since in this  
framework one 
can only describe the regular, i.e., nonchaotic, features of phase space. 
 
The geometric approach to dynamical instability allows a  
unification  
of the method to measure chaos with the explanation of its 
origin. In fact the evolution of the field $J$ given by the  
Jacobi equation (\ref{eq_jacobi_geo}) contains all the   
information  
needed to compute Lyapunov exponents, and  makes us also recognize  
in the curvature 
properties of the ambient manifold the origin of chaotic  
dynamics. 
 
Obviously, also this approach is far from being free of problems. 
For 
instance, the only case in which it is possible to rigorously 
prove  
that some definite curvature properties imply chaos in the  
geodesic flow,  
is the case of compact manifolds whose curvature is  
everywhere negative. 
In this case every point of the manifold is hyperbolic: in a  
sense this is 
the opposite limit to the integrable case. 
Though abstract and unphysical, such systems can help  
intuition. 
In a geodesic flow on a compact negatively-curved manifold,  
the negative curvature forces nearby geodesics to separate  
exponentially, 
while the compactness ensures that such a separation does  
not reduce to 
a trivial ``explosion'' of the system and obliges the  
geodesics to 
fold. The joint action of stretching and folding is the  
essential 
ingredient of chaos. 
 
Krylov tried to apply this framework to explain the origin  
of 
mixing in physical dynamical systems. Unfortunately for many  
systems  
in which chaos is detected the curvatures are found mainly  
positive, 
and there are examples, for instance the H\'enon-Heiles  
system --- see Eq.\ (\ref{Henon-Heiles}) ---  
geometrized with the Jacobi metric and  
the  
Fermi-Pasta-Ulam model --- see Eq.\ (\ref{v_fpu_beta}) --- geometrized with  
the Eisenhart metric, where curvatures are 
always positive even in the presence of  
fully developed chaos. Hence even positive curvature must be  
able to produce chaos. 
 
Only recently an example has been found of a compact surface with positive  
curvature, 
where the presence of chaotic regions coexisting with  
regular ones 
can be rigorously proved \cite{K>0_chaos},  
and this provides mathematical support for the available numerical  
evidence that  
negative curvature is not necessary at all to have chaos in  
a geodesic flow \cite{Pettini,pre93,pre96}. 
What then {\em is} the crucial feature of the curvature which is  
required to produce 
chaos? There is not yet a definite answer --- at least on  
rigorous grounds --- 
to this question. Nevertheless it is sure that, if positive,  
curvature 
must be non-constant in order to originate instability, and  
we shall see 
that the {\em curvature fluctuations} along the geodesics 
can be responsible for the insurgence of an instability  
through a mechanism  
very close 
to parametric instability.  
 
The advantages of the geometric approach to chaos are not  
only  
conceptual: also on predictive grounds this framework proves very  
useful. 
For, starting from the Jacobi equation, it  
is  
possible to obtain an effective stability equation which  
allows one to 
obtain an analytic estimate of the largest Lyapunov exponent in the  
thermodynamic limit \cite{prl95,pre96}. Such an estimate turns out to be  
in very  
good agreement with the results of numerical simulations  
for a number of 
systems (see \S \ref{applications}). In order to understand  
the derivation 
of such an effective stability equation, let us investigate in  
greater detail 
the relation between stability and curvature which was  
introduced  
in the last section. 
 
\subsubsection{Geometric origin of Hamiltonian chaos} 
 
Let us consider an $N$-dimensional Riemannian (or  
pseudo-Riemannian)  
manifold $(M,g)$ and a local coordinate frame with  
coordinates $(q^1,\ldots,q^N)$.  
 
We already observed that the evolution of the Jacobi field $J$,  
which contains the  
whole information on  
the stability of the geodesic flow, is completely determined by the curvature 
tensor $R$ 
through the Jacobi equation (\ref{eq_jacobi_geo}).  
Unfortunately the number  
of independent 
components of the tensor $R$ is ${\cal O}(N^4)$ --- even if  
this number can 
be considerably reduced by symmetry  considerations --- 
so that Eq.\  (\ref{eq_jacobi_geo}) becomes rather untractable already at  
fairly small 
dimensions.  
 
Nevertheless there is a particular case in which the Jacobi  
equation 
has a remarkably simple form: the case of {\em isotropic}  
manifolds (see Appendix \ref{app_geo}, \S \ref{app_sec_curvature}), where 
Eq.\ (\ref{eq_jacobi_geo}) becomes 
\beq 
\frac{D^2 J^i}{ds^2} + K\, J^i = 0~, 
\label{eq_jacobi_cost} 
\eeq 
where $K$ is the constant sectional curvature of the  
manifold (see Appendix \ref{app_geo}, Eq.\ (\ref{K})). 
Choosing a geodesic frame, i.e. an orthonormal frame  
transported along the geodesic, covariant derivatives become ordinary 
derivatives, i.e., $D/ds \equiv d/ds$, so that the  
solution of Eq. (\ref{eq_jacobi_cost}), with initial  
condition 
$J(0) = 0$ and $dJ(0)/ds = w(0)$, is 
\beq 
J(s) = \left\{ \begin{array}{lr} 
		 \frac{w(s)}{\sqrt{K}} \sin \left(\sqrt{K}\, s  
\right) 
		& (K > 0)~; \\ 
		& \\ 
		 s\, w(s) 
		& (K = 0)~; \\ 
		& \\ 
		 \frac{w(s)}{\sqrt{- K}} \sinh \left(\sqrt{ - K}\,  
s \right) 
		& (K < 0)~.  
		\end{array} \right. 
\label{sol_cost} 
\eeq 
The geodesic flow is unstable only if $K < 0$, and in this  
case the 
instability exponent is just $\sqrt{-K}$. 
 
As long as the curvatures are negative, the geodesic flow is  
unstable even 
if the manifold is no longer isotropic, and by means of the  
so-called 
``comparison theorems'' (mainly Rauch's theorem, see e.g.  
\cite{doCarmo}) 
it is possible to prove that the instability exponent is  
greater or 
equal to $(- \max_M (K))^{1/2}$ \cite{ergodic_theory}.  
On the contrary, no exact  results  
of general validity have yet 
been found for the dynamics of geodesic flows on manifolds whose curvature 
is neither constant nor everywhere negative. 
 
Equation (\ref{eq_jacobi_cost}) is valid only if $K$ is  
constant. Nevertheless 
in the case in which $\dim M = 2$ (surfaces), the Jacobi  
equation 
--- again written in a geodesic reference frame for the sake  
of simplicity --- 
takes a form very close to that for isotropic manifolds, 
\beq 
\frac{d^2 J}{ds^2} + K(s) \, J = 0~, 
\label{eq_jacobi_2} 
\eeq  
where 
\beq 
K(s) = \frac{1}{2} {\cal R}(s) 
\eeq 
and, contrary to Eq.\ (\ref{eq_jacobi_cost}), it is no longer a constant.  
With ${\cal R}(s)$ we denote the scalar curvature of the  
manifold at the  
point $P = \gamma(s)$ (see Appendix \ref{app_geo}, 
Eq.\ (\ref{scalar_curvature})).  
If the geodesics are 
unstable, Equation (\ref{eq_jacobi_2}) has exponentially growing solutions.  
As far as we know, the solutions of   
(\ref{eq_jacobi_2}) can  
exhibit an exponentially growing envelope in two cases:  
\begin{itemize} 
\item[{\bf a.}] the curvature $K(s)$ takes negative values;  
\item[{\bf b.}] the curvature $K(s)$, though mainly or even  
exclusively positive, 
fluctuates in such a way that it triggers a sort of {\em  
parametric instability} 
mechanism. 
\end{itemize} 
In the first case, the mechanism that is at the origin of the instability of the 
geodesics is the one usually considered in ergodic theory \cite{ergodic_theory}.  
But in the second 
case a new mechanism of instability,  
which does not require the presence of negatively curved 
regions on the manifold, shows up: the fluctuations of the curvature along the 
geodesic make the geodesic unstable. 
 
Let us now turn to physics, i.e., to the case of a mechanical manifold:  
in the case of the Jacobi metric with $N=2$ the scalar  
curvature written 
in standard (Lagrangian) coordinates reads as  
\beq 
{\cal R} = \frac{(\nabla V)^2}{(E-V)^3} + \frac{\triangle  
V}{(E-V)^2}~, 
\label{curv_scal_J} 
\eeq 
where $\nabla$ and $\triangle$ stand respectively for the  
euclidean gradient 
and Laplacian operators. Hence we can have ${\cal R} < 0$  
only if 
$\triangle V < 0$, i.e., for stable physical potentials, 
when the potential has inflection  
points. In these 
cases Krylov's idea can work --- even if in the  
high-dimensional case 
this becomes very complicated --- and we may have  
dynamical chaos induced 
by negative curvatures of the manifold. Indeed Krylov was  
mainly concerned 
with weakly non-ideal gases, or in general dilute systems, where  
for typical interatomic interactions $\triangle V < 0$  
so that the curvatures can be  
negative (see Krylov's PhD thesis in Ref.\ \cite{Krylov}). 
 
We will now show one example in which, though chaos is present, 
curvatures are positive. 
Let us consider the H\'enon-Heiles model  
\cite{HenonHeiles},  
whose Hamiltonian is \label{Henon_page} 
\beq 
{\cal H} = \frac{1}{2} \left( p_x^2 + p_y^2 \right) +  
\frac{1}{2} \left( x^2 + y^2 \right) + x^2 y - \frac{1}{3}  
y^3 ~. 
\label{Henon-Heiles} 
\eeq 
This model was introduced in an astrophysical framework to  
study the 
motion of a star in an axially symmetric galaxy, but it can  
also be regarded 
as a model of a triatomic molecule (after having used  
translational symmetry to  
eliminate the center-of-mass coordinate) \cite{Gutzwiller_book}.   
The H\'enon-Heiles model is a cornerstone in the study of  
Hamiltonian chaos: 
it was the first physical model for which chaos was found and  
where a transition 
from a mainly regular to a mainly chaotic phase space was  
identified under a variation of the energy. 
In this model, Eq. (\ref{eq_jacobi_2}) is exact, but ${\cal  
R} > 0$ everywhere. 
Hence chaos in this system cannot come from any negative  
curvature in 
the associated mechanical (Jacobi) manifold. As we shall see  
later on (see e.g.\ \S \ref{applications}), 
the absence of negative curvatures in the associated  
mechanical manifolds 
is not a peculiarity of this model, for it is shared with  
many 
systems of interest for field theory and condensed matter  
physics which have chaotic trajectories. In particular, 
all the systems that in the low-energy limit behave as a 
collection of harmonic 
oscillators do 
belong to this class. 
 
In these cases the second of the previously discussed  
instability mechanisms, the one mentioned in item {\bf b}, comes in:  
curvature  
fluctuations may induce chaos through parametric  
instability. 
The latter is a well-known feature of differential equations  
whose 
parameters are time-dependent. The classical example (see  
e.g.\  Arnol'd's book \cite{Arnold}) is the mathematical swing,  
i.e. a pendulum, initially at rest, whose length is modulated in  
time. 
If the modulation contains frequencies resonating with the  
free pendulum's fundamental frequency, the stable equilibrium  
position gets unstable and the swing starts to oscillate with growing  
amplitude. 
In Eq.\ (\ref{eq_jacobi_2}), 
$\sqrt{K(s)}$ and $s$ play the role of a frequency  
and of time, respectively, so that this equation   
can be thought of as the  
equation of motion 
of a harmonic oscillator with time-dependent frequency, often  
referred to as 
a (generalized) Hill's equation \cite{Abramowitz}.  
By expanding $K(s)$ in a  
Fourier series  
we get  
\beq 
K(s) = K_0 + \sum_{n=1}^\infty \left[a_n \cos(n \omega s) +  
b_n \sin  
(n \omega s) \right]~, 
\label{K_fourier} 
\eeq 
where $\omega = 2\pi/L$ and $L$ is the length of the geodesic. 
The presence of resonances between the average frequency 
$\sqrt{K_0}$ and the frequency in  
some term in the expansion (\ref{K_fourier}) eventually forces  
an exponential growth of the solutions of the equation.  
In the simplest case, in which only one coefficient of the 
series (\ref{K_fourier}), say $a_1$, is non-vanishing, 
the equation is called  
the Mathieu 
equation and it is possible to compute analytically both  
the bounds of the instability regions in the parameter  
space and 
the actual value of the instability exponent 
\cite{Abramowitz}. At variance with 
the Mathieu case, in the 
general case, where a large number of coefficients 
of the Fourier decomposition 
of $K(s)$ is nonzero, it is much more difficult to do something  
similar.  
Hence there is not yet any rigorous proof of the fact that this kind of 
parametric instability  
is the mechanism that produces chaos in Hamiltonian  
dynamical 
systems --- in the two-degrees-of-freedom case or in the  
general case --- 
and this still remains a conjecture.  
Nevertheless such a conjecture is strongly supported by at  
least two 
facts. First of all, in recent papers  
\cite{CerrutiPettini,PettiniValdettaro} it 
has been shown that the solutions of the Jacobi equation  
(\ref{eq_jacobi_2}) 
for the H\'enon-Heiles model and for a model of quartic  
coupled oscillators 
show an oscillatory behaviour with an exponentially growing  
envelope --- 
which is precisely what one expects from parametric  
instability --- in the 
chaotic regions, while the oscillations are bounded in the  
regular regions. 
Second, also in high-dimensional flows the  
components of the Jacobi field $J$ oscillate with an  
exponentially 
growing amplitude as long as the system is non-integrable,  
whereas they  
exhibit only bounded oscillations for integrable systems.  
Moreover,  
in the high-dimensional case (i.e., for systems with a large number of degrees 
of freedom) it is possible to establish a  
{\em quantitative} 
link between the largest Lyapunov exponent and the curvature  
fluctuations. In fact,   
as we shall see in the following, in the high-dimensional  
case it is  
possibile to write down, under suitable approximations, an  
effective stability  
equation which looks very similar to Eq.\ (\ref{eq_jacobi_2}),  
but where the squared  
frequency $K(s)$ is  
a stochastic process, and, through this equation, it 
is possible to give an analytical estimate of the largest Lyapunov exponent.
Since, from now on, we are going to consider only the largest 
Lyapunov exponent, the latter will be referred to as just the Lyapunov 
exponent.  
  
\subsubsection{Effective stability equation in the 
high-dimensional case} 
 
Let us now study the problem of the stability of the  
geodesics in manifolds whose dimension $N$ is large: according to  
the correspondence between geometry and dynamics 
introduced in Sec.\  \ref{sec_geodyn},  
we are considering a system with a large number $N$ of degrees 
of freedom.  
 
Our starting point is the Jacobi equation (\ref{eq_jacobi_geo}). Our aim is to 
derive from it an effective stability equation which no longer depends on the 
dynamics, i.e., on the evolution of the particular geodesic that we are 
following, but only on the average curvature properties of the manifold. 
To do that, we need some assumptions and approximations which are not valid in 
general but which are very reasonable in the case of large-$N$ 
mechanical manifolds. For the sake of clarity, we first summarize the 
assumptions and approximations leading to our final result, and later on we 
discuss them more thoroughly. Further details can be found in the papers where 
this approach was originally put forward \cite{prl95,pre96}. 
 
\begin{itemize} 
 
\item[{\bf 0.}] We assume that the evolution of a generic geodesic is chaotic. 
This assumption is reasonable in the case of a manifold whose 
geodesics are the 
trajectories of a generic Hamiltonian system with a large number of degrees of 
freedom $N$, for in this case the overwhelming majority of the trajectories  
will be chaotic. This bears a certain similarity to Gallavotti and Cohen's 
``Chaotic Hypothesis'' \cite{GallavottiCohen}.

\item[{\bf 1.}] We assume that the manifold is {\em quasi-isotropic}. Loosely 
speaking, this assumption means that 
the manifold can be regarded somehow as a locally deformed constant-curvature  
manifold. However, we will give this assumption a precise formulation 
later, in 
Eqs.\ (\ref{qh}). This approximation allows us 
to get rid of the dependence of the Jacobi equation 
(\ref{eq_jacobi_geo}) on the 
full Riemann curvature tensor by replacing it with an effective  
sectional curvature ${\cal K}(s)$ along the geodesic; moreover,  
the Jacobi equation becomes diagonal. 
\item[{\bf 2.}] To get rid of the dependence of the effective  
sectional curvature ${\cal K}(s)$ 
on the dynamics, i.e., on the evolution of the geodesic, we model  
${\cal K}(s)$ with a 
stochastic process. This assumption is motivated by Assumption 0 above. 
Moreover, as long as we consider 
a high-dimensional mechanical manifold associated to a Hamiltonian flow  
with $N$ degrees of freedom and we are eventually interested  
in taking the thermodynamic limit $N \to \infty$, the sectional curvature  
is formed by adding up many independent terms,  
so that invoking a central-limit-theorem-like argument, 
${\cal K}(s)$ is expected to behave, in first approximation, as a gaussian 
stochastic process.  
\item[{\bf 3.}] We assume that the statistics of the effective  
sectional curvature ${\cal K}$  
is the same as that of the Ricci curvature $K_R$, which is a suitably averaged 
sectional curvature (see Appendix \ref{app_geo}, 
Eq.\ (\ref{ricci_curvature})). 
Such an assumption is consistent 
with Assumption 1 above, for in a constant-curvature manifold the sectional 
curvature equals the Ricci curvature times a constant,  
and allows us to compute the mean and the variance of the stochastic 
process introduced in Assumption 2 in terms of the average and the variance of 
$K_R$ along a generic geodesic. 
\item[{\bf 4.}] The last step, which completely decouples the problem of the 
stability of the geodesics from the evolution of the geodesics themselves, 
consists in replacing the (proper) time averages of the 
Ricci curvature with static 
averages computed with a suitable probability measure $\mu$. 
If the manifold is  
a mechanical manifold, the natural choice for $\mu$ is  
the microcanonical measure. Again this assumption is reasonable if 
Assumption 0 is valid.  
\end{itemize} 
After these steps, we end up with an effective stability 
equation which no longer depends on the evolution of the geodesics, 
but only on 
the average and fluctuations of the Ricci curvature of the manifold.  
  
Let us now discuss more thoroughly the above-sketched procedure. For that,  
it is convenient to introduce the Weyl projective tensor $W$, 
whose components are given by \cite{Goldberg} 
\begin{equation} 
W^i_{~jkl} = R^i_{~jkl} - \frac{1}{N-1} (R_{jl} \delta^i_k -  
R_{jk}  
\delta^i_l)~, 
\label{weyl} 
\end{equation} 
where $R_{ij} = R^m_{~imj}$ are the components of the Ricci curvature tensor 
(see Appendix \ref{app_geo}, Eq.\ \ref{ricci_tensor}). 
Weyl's projective tensor measures the deviation from isotropy of a  
given manifold, since it vanishes identically for an isotropic manifold. 
Then we can reformulate the Jacobi equation  
(\ref{eq_jacobi_geo}) 
in the following form \cite{pre96}:  
\beq 
\frac{D^2 J^i}{ds^2}  +   
\frac{1}{N-1} R_{jk} \frac{dq^j}{ds} \frac{dq^k}{ds} J^i  - 
\frac{1}{N-1} R_{jk} \frac{dq^j}{ds} J^k \frac{dq^i}{ds}   
+ W^i_{~jkl} \frac{dq^j}{ds} J^k \frac{dq^l}{ds} = 0~.  
\label{eq_jacobi_weyl} 
\eeq 
For an isotropic manifold 
the third term in Eq.\ (\ref{eq_jacobi_weyl}) vanishes because $R_{jk} =  
K\, g_{jk}$ (see Appendix A, Eq.\ \ref{Ricci_comp_const}) so that 
$R_{jk} \dot q^j J^k = K \, 
\langle \dot \gamma,J\rangle $,
and $\langle \dot\gamma,J \rangle = 0$ (see Eq.\ \ref{J_ort}).  
Thus, for an isotropic manifold  
Eq.\ (\ref{eq_jacobi_weyl}) collapses to Eq.\  (\ref{eq_jacobi_cost}), 
in fact the second term is nothing but $K \, J^i$. 
When the manifold is not isotropic,  
we see that Eq.\ (\ref{eq_jacobi_weyl}) retains  
the structure of  
Eq.\ (\ref{eq_jacobi_cost}) up to its second term, since the coefficient of 
$J^i$ is still a scalar. This coefficient has now the meaning of a 
sectional curvature averaged, at any given point, over the $N-1$  
independent directions orthogonal to $\dot\gamma$, the velocity vector of the 
geodesic. However, such a mean sectional curvature is no longer  
constant along the geodesic $\gamma(s)$, and is nothing but that the Ricci 
curvature $K_R$ divided by $N-1$ (see Appendix \ref{app_geo}, 
Eq.\ \ref{ricci_curv_comp}).  
The fourth term 
of (\ref{eq_jacobi_weyl}) accounts for the local degree of  
anisotropy of the ambient manifold. 
 
Let us now consider a geodesic frame: in this case Eq.\ (\ref{eq_jacobi_weyl}) 
can be rewritten as  
\begin{equation} 
\frac{d^2 J^i}{ds^2}+ k_R(s)\, J^i - r^i_j(s)\,  
J^j + w^i_j (s)\, J^j = 0 
\label{sistemaHill} 
\end{equation} 
where, to ease the notation, we have put 
\begin{mathletters} 
\beq 
k_R (s) = \frac{K_R}{N-1} 
= \frac{1}{N-1} R_{jk} \frac{dq^j}{ds} \frac{dq^k}{ds}~;  
\label{k_R_eqj} 
\eeq 
\beq 
r^i_j (s) = \frac{1}{N-1} R_{jk} \frac{dq^k}{ds} \frac{dq^i}{ds} ~;  
\label{r} 
\eeq 
\beq 
w^i_j (s) = W^i_{~kjl} \frac{dq^k}{ds} \frac{dq^l}{ds} ~. 
\label{w} 
\eeq 
\end{mathletters} 
Being a scalar quantity, the value of $k_R$ is independent of the  
coordinate system. Now let us formulate our Assumption 1, namely,  
that the manifold is quasi-isotropic, in a more precise way. 
To do that, we recall
that (see Appendix \ref{app_geo}, Eqs.\ (\ref{R_comp_const}) and 
(\ref{Ricci_comp_const}))  
if and only if the manifold is isotropic, 
i.e., has constant curvature, 
the Riemann curvature tensor and the Ricci tensor can be written in   
the remarkably simple forms, i.e., 
\begin{mathletters} 
\beq 
R_{ijkl} = K \, (g_{ik}g_{jl} - g_{il}g_{jk})~, 
\label{R_const} 
\eeq 
and 
\beq 
R_{ij} = K \, g_{ij}~, 
\label{Ricci_const} 
\eeq 
\label{tensors_const} 
\end{mathletters} 
where $K$ is a scalar constant, the sectional curvature of the manifold. The 
precise formulation of Assumption 1 is now that along a generic geodesic  
the Riemann curvature tensor and the Ricci tensor retain 
the same functional form as in the case (\ref{tensors_const}), i.e., that  
\begin{mathletters} 
\beq 
R_{ijkl}  \approx  {\cal K}(s)\, (g_{ik}g_{jl} - g_{il}g_{jk})~, \label{qh1} 
\eeq 
and 
\beq  
R_{ij}  \approx  {\cal K}(s)\, g_{ij}~, \label{qh2} 
\eeq 
\label{qh} 
\end{mathletters} 
where ${\cal K}(s)$, which is no longer a constant, is an effective sectional 
curvature. In the general case we are not able to give a rigorous explicit  
expression for ${\cal K}(s)$, because the functional dependence postulated in 
Eqs.\ (\ref{qh}) holds true only for constant-curvature manifolds. However, 
the effective curvature ${\cal K}(s)$ is expected to be 
essentially the sectional curvature $K(\dot\gamma,J)$ (see Appendix 
\ref{app_geo}, Eq.\ (\ref{ricci_curvature}))  
measured along the geodesic in the directions 
of the velocity vector $\dot\gamma = dq/ds$ 
and of the Jacobi vector $J$. 
 
Combining Eqs.\ (\ref{r}) and (\ref{qh2}), 
and recalling that the vector $J$ is 
orthogonal to the velocity of the geodesic, i.e., $g_{ij}\frac{dq^i}{ds} J^j = 
0$, we find that the third term in Eq.\ (\ref{sistemaHill}), 
$-r^i_j J^j$, vanishes as in the isotropic case.  
Now we combine Eqs.\ (\ref{weyl}) and (\ref{qh1}) to obtain  
\beq 
W^i_{~jkl} \approx {\cal K}(s) (\delta^i_j g_{kl} - \delta^i_l g_{kj}) - 
\frac{1}{N-1} (R_{jl}\delta^i_k - R_{jk} \delta^i_l) 
\eeq 
so that Eq.\ (\ref{w}) can be rewritten as 
\beq 
w^i_j \approx {\cal K}(s)\delta^i_j - k_R(s) \delta^i_j - {\cal K}(s) 
\frac{N - 2}{N - 1} 
\frac{dq^i}{ds} g_{kj}  
\frac{dq^k}{ds}~, 
\label{w_appr} 
\eeq 
where we have used the definition of $k_R$ given in Eq.\ (\ref{k_R_eqj}) and  
the approximation (\ref{qh2}) for the Ricci tensor. Let us 
now insert Eq.\ (\ref{w_appr}) 
into Eq.\ (\ref{sistemaHill}): the last term of Eq.\ (\ref{w_appr}) vanishes 
after having been multiplied by $J^j$ and summed over $j$, 
because $J$ and $dq/ds$ are orthogonal, and the term 
$k_R(s)J^i$ is canceled the term $-k_R(s)J^i$ 
coming from Eq.\ (\ref{w_appr}), so that   
Eq.\ (\ref{sistemaHill}) is finally rewritten as 
\begin{equation} 
\frac{d^2 J^i}{ds^2}+  {\cal K}(s)\,J^i = 0~. 
\label{eq_Hill} 
\end{equation} 
Equation (\ref{eq_Hill}) is now diagonal. 
However, in order to use it, we should 
know the values of  
${\cal K}(s)$ along the geodesic. Here, Assumptions 2 and 3 come 
into play: we replace ${\cal K}(s)$ with a stochastic gaussian 
process, and we assume that its probability 
distribution is the same as that of 
the Ricci curvature, 
\begin{equation} 
{\cal P}({\cal K})\simeq{\cal P}(K_R)~. 
\label{probabilita} 
\end{equation} 
Such an assumption is consistent with our Assumption 1, because for 
an isotropic manifold the sectional curvature is identically equal to
the Ricci curvature divided by $N-1$, so that, if the manifold is 
quasi-isotropic, it is natural to assume that the probability distributions
of the sectional curvature and of the Ricci curvature are similar.
Moreover, such an assumption is also the only easy one, because we are
able to compute, under some further assumptions, the probability
distribution of $K_R$, but we do not know anything about $\cal K$.

To be consistent with the definition of the sectional and the Ricci 
curvatures (see Appendix \ref{app_geo}, Eq.\ (\ref{curv_relations})),  
the following relations are assumed to hold for the first two cumulants
of (\ref{probabilita}): 
\begin{mathletters}  
\beq 
\langle {\cal K}(s) \rangle_s \simeq 
\frac{1}{N-1} \langle K_R(s)   
\rangle_s \equiv \langle k_R(s) \rangle_s~, 
\label{momento1} 
\eeq 
\beq 
\langle [{\cal K}(s) - {\overline {\cal K}}]^2\rangle_s \simeq 
\frac{1}{N-1} \langle [K_R(s) - \langle K_R \rangle_s]^2  
\rangle_s \equiv \langle \delta^2 k_R \rangle_s~, 
\label{momento2} 
\eeq 
\end{mathletters} 
where $\langle \cdot \rangle_s$ stands for a proper-time  
average along a  
geodesic $\gamma(s)$.  
In general, the probability distributions (\ref{probabilita})  
will not be Gaussian, i.e., other 
cumulants in addition to the first two will be nonvanishing.  
However, we already observed that since for a large system  
$K_R$ is 
obtained by summing a large number of independent components,  
it is reasonable to assume 
that a sort of central limit theorem holds 
and that a gaussian approximation is 
sufficient.  
 
Our approximation for the effective sectional curvature ${\cal K}(s)$ is then 
the stochastic process  
\begin{equation} 
{\cal K}(s) \simeq \langle k_R(s) \rangle_s +  
\langle \delta^2 k_R \rangle^{1/2}_s 
\, \eta(s)~, 
\label{K_stocastica} 
\end{equation} 
where $\eta(s)$ is a random gaussian process with zero mean  
and unit variance. 
 
Finally, in order to completely decouple the stability equation from  
the dynamics, we use Assumption 4 and we 
replace time averages with static averages computed with a  
suitable measure $\mu$. If the manifold is a mechanical  
manifold 
the geodesics are the natural motions of the systems,  
and a natural choice for $\mu$ is then  the microcanonical  
ensemble, so that Eq. (\ref{K_stocastica}) becomes 
\begin{equation} 
{\cal K}(s) \simeq \langle k_R(s) \rangle_\mu + \langle \delta^2 k_R 
\rangle^{1/2}_\mu 
\, \eta(s)~. 
\label{K_stoc_mu} 
\end{equation} 
Our final effective stability equation is then  
\begin{equation} 
{{d^2\psi}\over{ds^2}}+ \langle k_R\rangle_\mu\,\psi +  
\langle\delta^2 k_R\rangle_\mu^{1/2}\,\eta (s)\,\psi = 0~, 
\label{eq_Hill_psi} 
\end{equation} 
where $\psi$ stands for {\em any} of the components of $J$, since  
all of them 
now obey the {\em same} effective equation of motion. 
 
Equation (\ref{eq_Hill_psi}) implies that, if the  
manifold is 
a mechanical manifold, the growth-rate 
of $\psi$ gives the dynamical instability exponent in our  
Riemannian framework. 
Equation (\ref{eq_Hill_psi}) is a scalar equation which,  
{\em independently of the 
knowledge of the dynamics}, provides a measure of the   
degree of instability 
of the dynamics itself through the behavior of $\psi (s)$.  
The peculiar 
properties of a given Hamiltonian system enter 
Eq.\  (\ref{eq_Hill_psi}) only through 
the global geometric properties $\langle k_R\rangle_\mu$ and  
$\langle\delta^2 k_R\rangle_\mu$ of the ambient Riemannian  
manifold (whose 
geodesics are natural motions) and are sufficient, as long  
as our Assumptions 1-4 hold, to determine the average 
degree of chaoticity of the dynamics. 
Moreover, $\langle  
k_R\rangle_\mu$ and  
$\langle\delta^2 k_R\rangle_\mu$ are microcanonical averages, so that they are 
functions of the energy $E$ of the system,  
or of the energy per degree of freedom $\varepsilon = E/N$ which is  
the relevant parameter as $N \to \infty$.  
Thus from (\ref{eq_Hill_psi}) we can obtain the energy  
dependence of the  
geometric instability exponent. 
 
Within the validity of our Assumptions 1-4, transforming 
the Jacobi equation  
(\ref{eq_jacobi_geo}) into Eq.\  (\ref{eq_Hill_psi}), its original  
complexity of the Jacobi equation has been considerably  
reduced: from a tensor  
equation we have obtained an effective scalar equation  
formally representing 
the equation of motion of a stochastic oscillator.  
In fact, Eq.\ (\ref{eq_Hill_psi}), with a self-evident notation,  
is of the form 
\begin{equation} 
{{d^2\psi}\over{ds^2}}+ k(s)\, \psi =0 
\label{eq_stoc_osc} 
\end{equation} 
where $k(s)$, the squared frequency, is a gaussian  
stochastic process. 
 
Moreover, such an equation admits a very suggestive  
geometric interpretation, 
since it is a scalar equation, i.e., it is formally the Jacobi  
equation  
on a 2-dimensional manifold whose Gaussian curvature is  
given, along a geodesic, 
by the random process $k(s)$, which can be regarded as an  
``effective''  
low-dimensional manifold approximatng the ``true'' high-dimensional  
manifold where the dynamics of the geodesic flow takes  
place. This is 
the real geometrical content of our quasi-isotropy  
hypothesis. 
Hence the average global curvature properties  
$\langle k_R\rangle_\mu$ and  
$\langle\delta^2 k_R\rangle_\mu$, in addition 
to being the ingredients for a geometric computation of the  
instability 
exponent, convey also information on the geometric  
structure of this  
effective manifold. Thus we expect that it will be possible  
to gain 
some insight in the global properties of the dynamics by  
simply studying 
the behaviour of these average curvature properties as the  
energy is varied. 
 
\subsubsection{A geometric formula for the Lyapunov exponent  
\label{geom_formula}} 
 
Let us now study the properties of the solutions 
of Eq.\ (\ref{eq_stoc_osc}) in order to obtain an analytic  
estimate for   
the Lyapunov exponent. 
The derivation of the stochastic  
oscillator  
equation does not depend on a particular choice of the  
metric; within the approximations discussed above,  
Equation (\ref{eq_stoc_osc}) holds regardless of the choice of the metric. 
However, to make explicit the connection between the solutions 
of Eq.\ (\ref{eq_stoc_osc}) and the stability of a dynamical system, 
one has to choose a particular metric;   
in the case of Hamiltonian systems of the form (\ref{H}), the choice of  
the Eisenhart metric is the simplest one.  
 
For this reason, we shall from now on restrict ourselves to standard  
Hamiltonian systems  
with a diagonal kinetic energy matrix, i.e., $a_{ij} =  
\delta_{ij}$, 
choosing as ambient manifold for the geometrization of the  
dynamics 
the enlarged configuration space-time 
equipped with the Eisenhart metric (\ref{g_E}). 
The case of the Jacobi metric is discussed in  Ref.\ \cite{pre96}. 
 
We will proceed as follows. $(i)$ We will show that in the present case
the Jacobi equation (\ref{eq_jacobi_geo}) is {\em equal} to the tangent
dynamics equation (\ref{eq_dintang}). $(ii)$ We will replace
the arc-length $s$ with the time $t$ and we will explicitly compute 
the average and the fluctuations of the Ricci curvature along
a geodesics in terms of dynamical observables, so that the (static)
probability distribution of the stochastic process $k(t)$ which
models the effective sectional curvature is defined. $(iii)$ We will
give an estimate for the time correlation function of the process $k(t)$.
$(iv)$ We will solve the stochastic oscillator equation, obtaining
an analytical formula for the Lyapunov exponent.

Let us now consider item $(i)$. 
As a consequence of the simple structure of the curvature  
tensor for the
Eisenhart metric (see \S \ref{curv_mech}),
the Jacobi equation (\ref{eq_jacobi_geo}) takes the form (we recall that the 
manifold has now dimension $N+2$; all the indices 
$i,j,k,\ldots$ run from $1$ to 
$N$) 
\begin{mathletters} 
\beq 
\frac{D^2 J^0}{ds^2} +R^0_{i0j}{{dq^i}\over{ds}}J^0 
{{dq^j}\over{ds}}+R^0_{0ij}{{dq^0}\over{ds}}J^i{{dq^j}\over{ 
ds}}= 0 ~, 
\label{JLC_gE_1} 
\eeq 
\beq 
\frac{D^2 J^i}{ds^2} +R^i_{0j0}\left( {{dq^0} 
\over{ds}}\right)  
^2J^j+R^i_{00j}{{dq^0}\over{ds}}J^0{{dq^j}\over{ds}}+ 
R^i_{j00}{{dq^j}\over{ds}}J^0{{dq^0}\over{ds}}=0 ~,  
\label{JLC_gE_2} 
\eeq 
\beq 
\frac{D^2 J^{N+1}}{ds^2} 
+R^{N+1}_{i0j}{{dq^i}\over{ds}}J^0 
{{dq^j}\over{ds}}+R^{N+1}_{ij0}{{dq^i}\over{ds}}J^j{{dq^0} 
\over{ds}}=0~,  
\label{JLC_gE_3} 
\eeq 
\label{eq_JLC_gE}
\end{mathletters}
where, for the sake of clarity, we have written out Eq.\ (\ref{eq_JLC_gE})
separately for the $0$, the $ i = 1,\ldots,N$, and the $N+1$ components,
respectively.  
As $\Gamma^0_{ij}=0$ (see Eq.\ \ref{Gamma_E}) 
we obtain, from the definition of covariant derivative (see 
Appendix \ref{app_geo}, Eq.\ \ref{D_local}), $D J^0/ds = dJ^0/ds$, and, as 
$R^0_{~ijk}=0$ (see \S \ref{curv_mech}), we find that Eq.\ (\ref{JLC_gE_1}) 
becomes 
\begin{equation} 
{{d^2J^0}\over{ds^2}}=0~, 
\end{equation} 
so that $J^0$ does not accelerate and, without loss of  
generality, we can 
set $\left. \frac{dJ^0}{ds}\right|_{s = 0} = J^0(0)=0$. 
Combining the latter result with the  
definition of covariant derivative we obtain 
\beq 
\frac{D J^i}{ds} = \frac{d J^i}{ds} + 
\Gamma^i_{0k} \frac{dq^0}{ds} J^k  
\eeq  
and using $dq^0/ds = 0$ we finally get 
\begin{equation} 
{{D^2J^i}\over{ds^2}}={{d^2J^i}\over{ds^2}} 
\end{equation} 
so that Eq.\ (\ref{JLC_gE_2}) gives,  
for the projection in configuration space of the separation  
vector, 
\begin{equation} 
\frac{d^2 J^i}{ds^2} + \frac{\partial^2 V}{\partial q_i  
\partial q_k} 
\left(\frac{dq^0}{ds}\right)^2 J_k = 0 ~. 
\label{eq_jacobi_E} 
\end{equation} 
Equation (\ref{JLC_gE_3}) describes the 
evolution of 
$J^{N+1}$, which, however, does not contribute to the norm of $J$ because  
$g_{N+1N+1}=0$,  
so we can disregard it. 
 
Along the physical geodesics of 
$g_E$, $ds^2 = (dq^0)^2 = dt^2$, so that Eq.\ (\ref{eq_jacobi_E}) 
is exactly the usual tangent dynamics equation  
(\ref{eq_dintang}), 
provided the identification  
$\xi = J$ is made. 
This clarifies then the relationship between the geometric  
description of the 
instability of a geodesic flow and the conventional  
description of dynamical  
instability. We stress that from a formal viewpoint this is  
a peculiarity 
of the Eisenhart metric; nevertheless the physical content  
of this result 
is valid independently of the metric used, as long as the  
identification between 
trajectories and geodesics holds true. For, in recent  
papers  
\cite{CerrutiPettini,PettiniValdettaro} it has been found  
that using the Jacobi 
metric the solutions of the Jacobi equation and of those of  
the tangent 
dynamics equation --- which in this case are two well-distinct equations --- 
look strikingly similar. 
 
We now turn to item $(ii)$. 
The Ricci curvature is obtained saturating the Ricci tensor 
with the components 
of the velocity vector $dq/ds$ (see Appendix \ref{app_geo},  
Eq.\ (\ref{ricci_curv_comp})). 
In the present case, the only non-vanishing component of the Ricci tensor is  
$R_{00} = \triangle V$ (see Eq.\ (\ref{ricci_eisenhart})),  
so that the dynamical observable which  
corresponds to the Ricci 
curvature along a geodesic depends only on the coordinates  
and not 
on the velocities and reads 
\beq 
K_R(q) = {\triangle V}~, 
\label{k_R_obs} 
\eeq 
where we have used that, along a physical geodesic, 
$(dq^0)^2 = dt^2 = ds^2$. Using again this result  
we replace the arc-length $s$ along  
the geodesic with the physical time $t$, 
and the stochastic oscillator equation  
(\ref{eq_stoc_osc}) can be written  
\beq 
\frac{d^2 \psi}{dt^2} + k(t)\, \psi = 0~, 
\label{stoc_osc_t} 
\eeq 
where mean and variance of $k(t)$ are given by  
\begin{mathletters} 
\beq 
k_0  \equiv  \langle k_R \rangle_\mu  =   
\frac{1}{N}\langle \triangle V \rangle_\mu ~, \label{k_0}  
\label{mean_k}
\eeq 
\beq 
\sigma^2_k  \equiv  \langle \delta^2 k_R \rangle_\mu  =  
\frac{1}{N}\left( \langle (\triangle V)^2 \rangle_\mu -  
\langle \triangle V \rangle^2_\mu \right) ~. \label{sigma_k} 
\label{variance_k}
\eeq 
\label{stat_k}   
\end{mathletters} 
Since we are considering systems with large $N$ ---  
eventually 
taking the limit $N \to \infty$ --- we replaced $N-1$  
with $N$ in Eqs. (\ref{stat_k}). 
 
We consider now item $(iii)$. 
The process $k(t)$ is not completely defined unless its time  
correlation  
function,  
\beq 
\Gamma_k(t_1,t_2) = \langle k(t_1)k(t_2) \rangle - \langle k(t_1) 
\rangle \langle k(t_2)\rangle~,  
\eeq 
is given. The simplest choice is to assume that 
$k(t)$ is a stationary and $\delta$-correlated process, so that 
\begin{equation} 
\Gamma_k(t_1,t_2) = \Gamma_k(\vert t_2 - t_1\vert )  
= \Gamma_k(t) = \tau \, \sigma_k^2 \, \delta(t)~,  
\label{corr_k} 
\end{equation} 
where $\tau$ is the characteristic correlation time scale of the process. 
 
Before we can actually solve Eq.\ (\ref{stoc_osc_t}), we have then to give 
an explicit expression for $\tau$.  
To do that, first we will show how two independent characteristic  
correlation time scales, which will be referred to as $\tau_1$ 
and $\tau_2$, respectively, can be defined, then we will estimate $\tau$ 
by combining these two time scales. 
 
A first time scale, which we will refer to as $\tau_1$, is  
associated  to 
the time needed to cover the average distance between 
two successive {\em conjugate points} along a geodesic. Conjugate points  
\cite{doCarmo}   
are the points where the Jacobi vector field vanishes. As long as the curvature 
is positive and its fluctuations are small, compared to the average, two nearby 
geodesics will remain close to each other until a conjugate point is reached. 
At each crossing of a conjugate  
point the Jacobi 
vector field increases as if the geodesics  
were kicked (this is what happens when parametric  
instability is active). Thus the average distance between conjugate points 
provides a relevant correlation time scale.  It can be proved that 
\cite{doCarmo,Cheeger} if the sectional  
curvature $K$ is bounded as  $0< L\leq K \leq H$,  
then the distance $d$ between two 
successive conjugate points is bounded by  
$d > \frac{\pi}{2\sqrt{H}}$. The upper bound $H$ of the curvature can then be 
approximated in our framework by 
\beq 
H \simeq k_0 +\sigma_k~, 
\eeq 
so that we can define $\tau_1$ as (remember that $dt = ds$) 
\begin{equation} 
\tau_1 = d_1 = 
\frac{\pi}{2 
\sqrt{k_0 +\sigma_k}}~. 
\label{dstar} 
\end{equation} 
This time scale is expected to be the most relevant  
only as long as the curvature is positive and the fluctuations are small,  
compared to the average. 
 
Another time scale, referred to as $\tau_2$, is related to the  
local curvature fluctuations. These will be felt on a length  
scale of  
the order of, at least, $l =1/\sqrt{\sigma_k}$  
(the average fluctuation of curvature 
radius). The scale $l$ is expected to be the relevant one when  
the fluctuations  
are of the same order of magnitude as the average curvature. 
Locally, the metric of a manifold can be approximated by \cite{doCarmo} 
\beq 
g_{ik}\simeq\delta_{ik} - \frac{1}{6}R_{ikjl}u^iu^k ~, 
\eeq 
where the $u^i$ are the components of the displacements from the point around  
which we are approximating the metric.   
When the sectional curvature is positive (resp.\ negative),  
lengths and time intervals --- on a scale $l$ --- are enlarged 
(resp.\ shortened) by a  
factor $(l^2 K/6)$,  
so that the period $\frac{2\pi}{\sqrt{k_0}}$  
has a fluctuation amplitude $d_2$ given by  
$d_2 =\frac{l^2 K}{6}\frac{2\pi}{\sqrt{k_0}}$; replacing $K$  
by the most probable value $k_0$ one gets 
\begin{equation} 
\tau_2 = d_2 = 
 \frac{l^2 k_0}{6}\frac{2\pi}{\sqrt{k_0}} 
        \simeq \frac{k_0^{1/2}} 
         {\sigma_k}~. 
\label{tau2star} 
\end{equation} 
Finally $\tau$ in Eq.\ (\ref{corr_k}) is obtained by  
combining $\tau_1$ 
with $\tau_2$ as follows 
\begin{equation} 
\tau^{-1} =  \tau_1^{-1} + \tau_2^{- 
1}~. 
\label{taufinale} 
\end{equation} 
The present definition of $\tau$ is obviously by no means  
a direct consequence of any theoretical result, but only a reasonable  
estimate. Such an estimate might well be improved  
independently of the general geometric framework.  
 
Now that all the quantities entering Eq.\ (\ref{stoc_osc_t}) have 
been fully defined, we can turn to item $(iv)$, i.e., to the solution 
of Eq.\ (\ref{stoc_osc_t}). 
Whenever $k (t)$ has a non-vanishing stochastic 
component, any solution $\psi (t)$ has an exponentially  
growing envelope 
\cite{VanKampen} whose growth-rate provides a measure of the  
degree of 
instability. How can one relate such a growth-rate with the Lyapunov 
exponent of the physical system? Let us recall that, for a standard  
Hamiltonian system of the form (\ref{H}), the  
Lyapunov exponent can be computed as the following limit 
(see Appendix \ref{app_chaos}, Eq.\ \ref{def_lambda_standard}): 
\beq 
\lambda = \lim_{t \to \infty} \frac{1}{2t} \log  
\frac{\xi_1^2(t) +  
\cdots + \xi_N^2(t) + \dot\xi_1^2(t) +  
\cdots + \dot\xi_N^2(t) }{\xi_1^2(0) +  
\cdots + \xi_N^2(0) + \dot\xi_1^2(0) +  
\cdots + \dot\xi_N^2(0) } 
\label{def_lambda_std} 
\eeq 
where the $\xi$'s are the components of the tangent vector, i.e., 
of the perturbation of a reference trajectory, which obey
the tangent dynamics equation (\ref{eq_dintang}). 
In the case of Eisenhart metric, each component  
of the Jacobi vector field $J$ can be identified with the 
corresponding component of the tangent vector $\xi$; moreover, $\psi$ in  
Eq.\ (\ref{stoc_osc_t}) stands for {\em any} of the components of $J$, 
which obey the {\em same} effective equation. Thus,   
Eq.\ (\ref{def_lambda_std}) becomes
\begin{equation} 
\lambda = \lim_{t\to\infty} \frac{1}{2t} \log  
\frac{\psi^2(t) + \dot\psi^2(t)}{\psi^2(0) +  
\dot\psi^2(0)}~, 
\label{def_lambda_gauss_prel} 
\end{equation} 
where $\psi(t)$ is solution of Eq.\ (\ref{stoc_osc_t}). Equation  
(\ref{def_lambda_gauss_prel}) is our estimate for the (largest) Lyapunov  
exponent.  
 
As a stochastic differential 
equation, the solutions of Eq.\ (\ref{stoc_osc_t}) 
are properly defined after an averaging over 
the realizations of the stochastic process: 
referring to such an averaging as $\langle \bullet \rangle$, we rewrite 
Eq.\ (\ref{def_lambda_gauss_prel}) as 
\begin{equation} 
\lambda = \lim_{t\to\infty} \frac{1}{2t} \log  
\frac{\langle \psi^2(t) \rangle + \langle \dot\psi^2(t) \rangle} 
{\langle \psi^2(0) \rangle +  
\langle \dot\psi^2(0) \rangle}~. 
\label{def_lambda_gauss} 
\end{equation} 
The evolution of $\langle \psi^2 \rangle$, $\langle \dot\psi^2 \rangle$  and  
$\langle \psi \dot\psi  \rangle$, i.e., of the vector of the 
second moments of $\psi$, obeys the following equation, which can 
be derived by means of a technique, developed by Van Kampen  
and sketched in 
Appendix \ref{app_sto}:  
\begin{equation} 
\frac{d}{dt}\left( 
\begin{array}{c} 
\langle\psi^2\rangle \\ 
\langle\dot{\psi}^2\rangle \\ 
\langle\psi\dot{\psi}\rangle 
\end{array} \right) = \left( 
\begin{array}{ccc} 
0 & 0 & 2 \\ 
\sigma^2_k\tau & 0 & -2k_0 \\ 
-k_0 & 1 & 0 
\end{array} \right)\left( 
\begin{array}{c} 
\langle\psi^2\rangle \\ 
\langle\dot{\psi}^2\rangle \\ 
\langle\psi\dot{\psi}\rangle 
\end{array} \right) 
\label{vankamp} 
\end{equation} 
where $k_0$ and $\sigma_k$ are the mean and the  
variance of $k (t)$, defined in Eqs.\ (\ref{mean_k}) and (\ref{variance_k}),
respectively. 
Equation (\ref{vankamp}) can be solved by diagonalizing the matrix on  
the r.h.s.\ of (\ref{vankamp}). The result for the evolution of   
$\langle \psi^2\rangle  
+\langle\dot\psi^2\rangle\ $ is 
\beq 
\langle \psi^2(t)\rangle  
+\langle\dot\psi^2(t)\rangle\ = \left( \langle \psi^2(0)\rangle  
+\langle\dot\psi^2(0)\rangle\right) \, \exp(\alpha t)~,  
\eeq 
where $\alpha$ is the only real eigenvalue of the matrix. According to  
Eq.\ (\ref{def_lambda_gauss}), the Lyapunov exponent is given by $\lambda 
= \alpha/2$, so that, by computing explicitly $\alpha$, one then finds 
the final  
expression 
\begin{mathletters} 
\beq 
\lambda(k_0,\sigma_k,\tau)   =  \frac{1}{2} 
\left(\Lambda-\frac{4k_0} 
{3 \Lambda}\right)~,  
\eeq 
\beq 
\Lambda   =   
\left(\sigma^2_k\tau+\sqrt{\left(\frac{4k_0}{3} 
\right)^3+\sigma^4_k\tau^2}\,\right)^{1/3}~. 
\eeq 
\label{lambda_gauss} 
\end{mathletters} 
All the quantities $k_0$, $\sigma_k$ and $\tau(k_0,\sigma_k)$ can be  
computed 
as {\it static} averages, as functions of the energy per degree of
freedom, $\varepsilon$ (see Eqs.\ (\ref{mean_k}) and (\ref{variance_k})).
Therefore --- within the limits of validity  of the  
assumptions made above ---  Eqs.\ (\ref{lambda_gauss})  
provide an approximate analytic  
formula to compute 
the largest Lyapunov exponent independently of the numerical  
integration of  
the dynamics and of the tangent dynamics.  
 
Let us remark that expanding Eqs.\ (\ref{lambda_gauss}) in the  
limit $\sigma_k \ll k_0$ one finds that  
\beq 
\lambda \propto \sigma_k^2 
\label{lyap_fluct} 
\eeq   
which shows how close the relation is between curvature  
fluctuations and dynamical instability. 
 
\subsection{Some applications} 
\label{applications} 
Let us now discuss briefly the results of the application of the 
geometric techniques described up to this point to some Hamiltonian models.  
In particular, we shall consider two cases: 
a chain of coupled nonlinear oscillators (the so-called FPU 
$\beta$ model, first introduced by Fermi, Pasta and Ulam in Ref.\ \cite{FPU}) 
and 
a chain of coupled rotators (the 1-d $XY$ model). The reason of the choice of 
these two particular models  
is that they allow fully analytic calculations and are 
well-suited to show advantages and limitations of the theory.  
The geometric theory developed 
above has already been applied to many other cases, some of which 
will be addressed  
in Sec.\ \ref{sec_geopt}. For other applications we refer to the 
literature: in particular, a model of a homopolymer chain has been  
studied in Ref.\ \cite{cecilia_master},  
a model of a three-dimensional 
Lennard-Jones crystal has been studied in Ref.\ \cite{gasrari}, and a 
classical lattice gauge theory has been considered in  
Ref.\ \cite{jpa99}.  
Without entering into the details, 
we would like to single out a result which is 
shared by all the models considered until now. In all these models the 
functional dependence of the largest Lyapunov exponent on the
energy per degree of freedom $\varepsilon$, in the low-$\varepsilon$ limit,  
is numerically found to be  
\beq 
\lambda (\varepsilon) \propto \varepsilon^2~. 
\eeq 
No explanation of this ``universal'' behavior is yet at hand, and for some 
cases doubts about the validity of such a scaling with energy 
have been raised, 
because the numerical determination of Lyapunov exponents at low 
$\varepsilon$ is difficult.
However, the application of the geometric theory has 
provided a theoretical confirmation 
of this behavior in all the cases considered.  
 
The systems we now consider are 1-$d$ models with nearest-neighbor  
interactions 
whose Hamiltonians $\cal H$ have the standard form (\ref{H}) with 
\beq 
V = \sum_{i = 1}^{N} v (q_{i} - q_{i-1})~. 
\eeq 
The interaction potentials are, respectively,   
\begin{mathletters} 
\beq 
v (x)  =  \frac{1}{2} x^2 + \frac{u}{4} x^4 \qquad  
\mbox{(FPU-$\beta$ model)}~, \label{v_fpu_beta}  
\eeq 
or 
\beq 
v (x)  =  - J \cos x  \qquad ~~ 
\mbox{(1-d $XY$ model)}~. \label{v_rotatori}  
\eeq 
\label{1d_models} 
\end{mathletters} 
In Eq.\ (\ref{v_fpu_beta}) we used $u$ instead of the customary $\beta$ in 
order to avoid confusion with the inverse temperature $\beta$. We assume  
$u > 0$. 
 
The geometric quantities --- in the  
framework of the Eisenhart metric --- which are relevant in the  
quasi-isotropy 
approximation to describe the ``effective'' structure of the mechanical 
manifold,  
and which enter the geometric formula for the Lyapunov  
exponent, are the average and the root mean square (r.m.s.) 
fluctuations of the 
Ricci curvature of the mechanical manifold. They are defined as statistical 
averages computed in the {\em microcanonical} ensemble  
(see Eqs.\ (\ref{stat_k})). First, we will show how these microcanonical  
quantities 
can be computed starting from the {\em canonical} partition 
function, which  can be calculated exactly for an  
infinite chain, i.e., $N \to \infty$, for both models (\ref{1d_models}). 
Then, we will apply this procedure to each of the two  
models  (\ref{1d_models}). 
 
The average and fluctuations, within  
the microcanonical ensemble, of any observable function $f(q)$, 
can be computed as follows, in terms of the 
corresponding quantities in the canonical ensemble. 
The canonical configurational partition function $Z(\beta )$ is  
given by 
\begin{equation} 
Z(\beta ) =\int d q\,e^{-\beta\,V(q)} 
\end{equation} 
where $d q=\prod_{i=1}^{N}dq_i$.  
The canonical average $\langle f\rangle_{\rm can}$ of the  
observable $f$ can be computed as  
\begin{equation} 
\langle f\rangle_{\rm can} = [Z(\beta)]^{-1} \int d q\,  
f(q)\,e^ 
{- \beta V(q)}~~. 
\end{equation} 
From this average, we can obtain the microcanonical average  
of $f$, $\langle f\rangle_{\mu}$,  
in the following (implicit) parametric form \cite{LPV}  
\begin{equation} 
\left.\begin{array}{l} 
\langle f\rangle_{\mu}(\beta) = \langle f  
\rangle_{\rm can}(\beta)\\ 
\\ 
\varepsilon(\beta) = {\displaystyle 
\frac{1}{2\beta} - \frac{1}{N}\frac{\partial}{\partial\beta} 
[\log Z(\beta)]} 
\end{array}\right\} \, \rightarrow  
\langle f\rangle_{\mu}(\varepsilon)  
\label{mm} 
\end{equation} 
Note that Eq.\ (\ref{mm}) is strictly valid only in the  
thermodynamic limit;  
at finite $N$, $\langle f\rangle_{\mu}(\beta) = \langle  
f \rangle_{\rm can} 
(\beta )+ {\cal O}(\frac{1}{N})$. 
 
Contrary to the computation of $\langle f\rangle$,  
which is 
insensitive to the choice of the probability measure in the  
$N\rightarrow\infty$ limit,  
computing the fluctuations of $f$, i.e., of  
$\langle \delta^{2}f\rangle=\frac{1}{N}\langle \left(f- 
\langle f\rangle \right) 
^{2}\rangle$, by means of the canonical or microcanonical  
ensembles yields  
different results. The relationship between the canonical --- i.e.  
computed with the Gibbsian weight $e^{-\beta{\cal H}}$ ---  
and the  
microcanonical fluctuations,  
is given by the Lebowitz-Percus-Verlet formula \cite{LPV} 
\begin{equation} 
\langle 
\delta^{2}f\rangle_\mu  
(\varepsilon)=\langle\delta^{2}f\rangle_{\rm can}(\beta)- 
\frac{\beta^{2}}{c_{V}} 
\left[\frac{\partial\langle  
f\rangle_{\rm can}(\beta)}{\partial\beta}\right]^{2}, 
\label{corr2} 
\end{equation} 
where 
\begin{equation} 
c_{V}=-\frac{\beta^{2}}{N}\frac{\partial\langle  
{\cal H}\rangle_{\rm can}}{\partial\beta} 
\end{equation} 
is the specific heat at constant volume and $\beta =\beta  
(\varepsilon )$ is 
given in implicit form by the second equation in (\ref{mm}). 

The average $k_0$ and the fluctuations $\sigma_k$ of the Ricci curvature
per degree of freedom are then obtained by replacing $f$ with the 
explicit expression for Ricci curvature, which, 
according to the definition given in Eq.\ (\ref{k_R_obs}), is 
\beq 
K_R (q) = \sum_{i=1}^N \frac{\partial^2}{\partial q_i^2}\, 
v (q_{i} - q_{i-1})~, 
\label{k_R_obs_chain} 
\eeq  
in Eqs.\ (\ref{mm}) and (\ref{corr2}), respectively.

We now turn to the two applications mentioned above. 
 
\subsubsection{FPU $\beta$ model} 
 
For the FPU $\beta$ model the dynamical observable which  
corresponds 
to the Ricci curvature reads, according to Eq.\ (\ref{k_R_obs_chain}),  
\begin{equation} 
K_R = 2N + 6 u \sum_{i=1}^N \left( q_{i+1} - q_i  
\right)^2~. 
\label{kRFPU} 
\end{equation} 
Note that $K_R$ is always positive and that this is also true  
for the 
sectional curvature along a physical geodesic. 
Computing the microcanonical average of $K_R$ according to Eq.\ (\ref{mm})
we find that in the  
thermodynamic limit $k_0 (\varepsilon)$ is implicitly given by 
(the details are reported in Ref.\ \cite{pre96}) 
\begin{equation} 
\left. \begin{array}{rcl} 
\langle k_R \rangle_{\rm can} (\theta) & = & {\displaystyle  
2 + \frac{3}{\theta}\, \frac{D_{-3/2}(\theta)}{D_{- 
1/2}(\theta)} } \\ 
& & \\ 
\varepsilon(\theta) & = & {\displaystyle 
\frac{1}{8\sigma}\left[\frac{3}{\theta^2}+\frac{1}{\theta}\, 
\frac{D_{-3/2}(\theta)}{D_{-1/2}(\theta)}\right] } 
\end{array} \right\} \, \rightarrow k_0 (\varepsilon)~, 
\label{kR(eps)FPU} 
\end{equation} 
where the $D_\nu$ are parabolic cylinder functions \cite{Abramowitz} and 
$\theta$ is a parameter proportional to $\beta$, so that $\theta \in  
[0,+\infty]$.  
 
\begin{figure} 
\centerline{\psfig{file=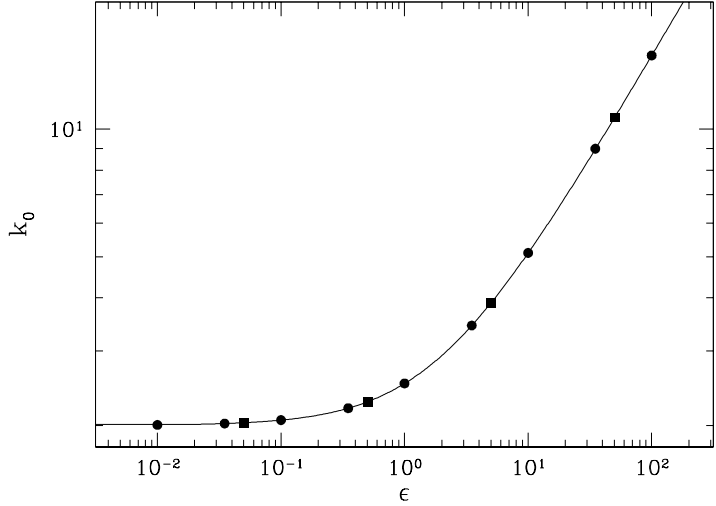,height=8cm,clip=true}} 
\caption{Average Ricci curvature (Eisenhart metric)  
per degree of freedom, $k_0$, vs.\ energy 
density $\varepsilon$ for the FPU-$\beta$ model. The  
continuous line 
is the analytic computation according to Eq.  
(\protect\ref{kR(eps)FPU}); 
circles and squares are time averages obtained by numerical  
simulations 
with $N = 128$ and $N = 512$ respectively; $u = 0.1$. 
From Ref.\ \protect\cite{pre93}.} 
\label{fig_curv_fpu} 
\end{figure}

\begin{figure} 
\centerline{\psfig{file=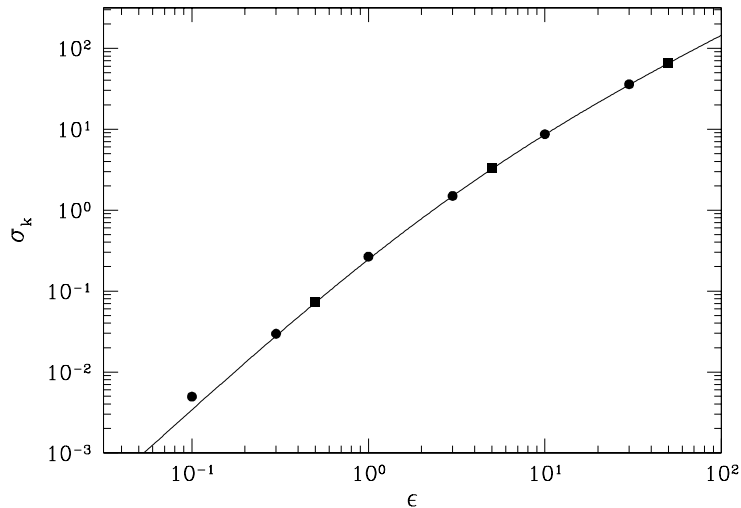,height=8cm,clip=true}} 
\caption{Fluctuations of the Ricci curvature  
(Eisenhart metric), $\sigma_k$ vs.\ energy density $\varepsilon$ for the  
FPU-$\beta$ model. 
Symbols and parameters as in Fig.\  
\protect\ref{fig_curv_fpu};  
the continuous line now refers to Eq.\  
(\protect\ref{fluttkR(eps)FPU}). 
From Ref.\ \protect\cite{pre93}.} 
\label{fig_flutt_fpu} 
\end{figure}

Let us now compute the fluctuations
\begin{equation} 
\sigma^2_k (\varepsilon ) =  
\frac{1}{N}\langle \delta^2 K_R \rangle^{\,}_\mu  
(\varepsilon ) = 
\frac{1}{N} \langle \left( K_R - \langle K_R \rangle  
\right)^2  
\rangle^{\,}_\mu~. 
\end{equation} 
According to Eq.\  (\ref{corr2}), first the canonical fluctuation, $\langle \delta^2 k_R \rangle_{\rm can} (\beta)= 
\frac{1}{N} \langle \left( K_R - \langle K_R \rangle  
\right)^2 \rangle_{\rm can} 
(\beta)$, has to be computed and then a correction term  
must be added. 
For the canonical fluctuation we obtain \cite{pre96}  
\begin{equation} 
\langle \delta^2 k_R \rangle_{\rm can} (\theta) = 
\frac{9}{\theta^2} \, \left\{ 2 - 2\theta \,  
\frac{D_{-3/2}(\theta)}{D_{-1/2}(\theta)} - 
\left[\frac{D_{-3/2}(\theta)}{D_{-1/2}(\theta)}\right]^2  
\right\}~, 
\label{fluttkcan(theta)} 
\end{equation} 
and the final result for the 
fluctuations of the Ricci curvature is  
\begin{equation} 
\left. \begin{array}{rcl} 
\langle \delta^2 k_R \rangle^{\,}_\mu (\theta) &  
= & {\displaystyle 
\langle \delta^2 k_R \rangle_{\rm can} (\theta)-  
\frac{\beta^2}{c_V(\theta)} 
\left( \frac{\partial \langle k_R \rangle_{\text{can}}  
(\theta)}{\partial\beta} 
\right)^2 } \\ 
 & & \\ 
\varepsilon(\theta) & = & {\displaystyle 
\frac{1}{8\mu}\left[\frac{3}{\theta^2}+\frac{1}{\theta}\, 
\frac{D_{-3/2}(\theta)}{D_{-1/2}(\theta)}\right] } 
\end{array} \right\} \, \rightarrow \sigma^2_k (\varepsilon )~, 
\label{fluttkR(eps)FPU} 
\end{equation} 
where $\langle \delta^2 k_R \rangle_{\rm can} (\theta)$ is given by 
(\ref{fluttkcan(theta)}), $\partial \langle k_R \rangle (\theta)/ 
\partial\beta$ is given by 
\begin{equation} 
\frac{\partial \langle k_R \rangle (\theta)}{\partial\beta}  
= 
\frac{3}{8\mu\theta^3}\, \frac{\theta \, D^2_{-3/2}(\theta)+  
2(\theta^2 -1)D_{-1/2}(\theta) D_{-3/2}(\theta) -2\theta  
D^2_{-1/2}(\theta)} 
{D^2_{-1/2}(\theta)} ~, \label{dkRdbeta} 
\end{equation} 
and the specific heat per particle $c_V$ is found to be 
\cite{LiviPettiniRuffoVulpiani,pre96} 
\begin{eqnarray} 
c_V(\theta )&=& 
{\displaystyle \frac{1}{16 D^2_{-1/2}(\theta )} }\left\{ (12  
+ 2\theta^2) 
D^2_{-1/2}(\theta ) + 2\theta D_{-1/2}(\theta )D_{- 
3/2}(\theta )\right. 
 \nonumber \\ 
&-&\left. \theta^2 D_{-3/2}(\theta ) \left[ 2\theta D_{- 
1/2}(\theta ) + 
D_{-3/2}(\theta )\right] \right\}~. 
\label{calspec} 
\end{eqnarray} 
The microcanonical averages and fluctuations
computed in Eqs.\ (\ref{kR(eps)FPU}) and  
(\ref{fluttkR(eps)FPU}) are compared in Figs.\ \ref{fig_curv_fpu}  
and \ref{fig_flutt_fpu}  
with their corresponding time averages computed along  
numerically simulated trajectories  
of the FPU $\beta$-model with the potential (\ref{v_fpu_beta})  
for $N=128$ and $N=512$ with $u = 0.1$.  
Though the microcanonical averages have to be computed in the 
thermodynamic limit, the agreement between time and ensemble  
averages  
is excellent already at $N=128$.  
 
\begin{figure} 
\centerline{\psfig{file=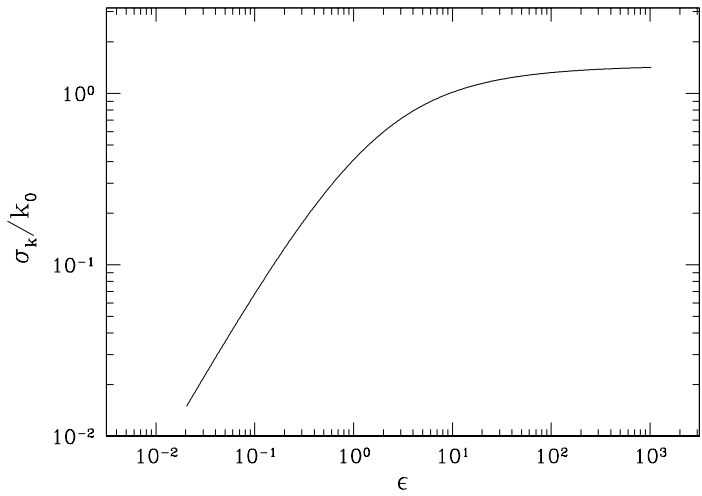,height=8cm,clip=true}} 
\caption{Fluctuations of the Ricci curvature  
(Eisenhart metric) divided by the average curvature, 
$\sigma_k/k_0$, vs.\ energy density $\varepsilon$ for the  
FPU-$\beta$ model.} 
\label{fig_fluttratio_fpu} 
\end{figure}

Before we comment on these results,  
we remark here that in many Hamiltonian dynamical systems 
different dynamical regimes can be found as the energy per degree of freedom 
$\varepsilon$ is varied (see, for a review, Ref.\ \cite{rnc} and 
references quoted therein). In particular,  
in the FPU-$\beta$ model, a weakly chaotic regime is 
found for specific energies 
smaller than $\varepsilon_c \approx 0.1/u$ 
\cite{PettiniLandolfi,PettiniCerruti,pre93}.  
Although in the weakly chaotic regime the 
dynamics is chaotic (i.e., the Lyapunov exponent is positive, though small), 
mixing is very slow, as witnessed by the existence of a rather long memory of 
the initial conditions, i.e., of long relaxation 
times if the initial conditions are far from equilibrium. For $\varepsilon$ 
larger than $\varepsilon_c$ 
the dynamics is strongly chaotic and relaxations are 
fast. The precise  
origin of these phenomena is still to be understood. However, the 
geometric approach described here is able to provide a suggestive 
interpretation \cite{pre93,lapo_thesis}. Let us consider 
Fig.\  \ref{fig_fluttratio_fpu}, 
where the ratio of the fluctuations and the average curvature  
$\sigma_k/k_0$ is reported. As  
$\varepsilon \to 0$, 
$\sigma_k \ll k_0$, so that the manifold looks essentially like  
a constant 
curvature manifold with only small curvature fluctuations. This situation 
corresponds to the weakly chaotic dynamical regime. On the 
contrary, as $\varepsilon$ is larger than $\varepsilon_c$, 
$\sigma_k /k_0$ tends to saturate towards a value of order  
unity, 
thus indicating that in the high-energy (strongly chaotic)  
regime 
the curvature fluctuations are of the same order of 
magnitude as the average 
curvature, so that the system no longer ``feels'' the  
isotropic (and integrable) 
limit. Hence the geometric approach can give a hint for  
understanding, at least  
qualitatively, the origin of weak and strong chaos in the  
Fermi-Pasta-Ulam model. 
 
\begin{figure} 
\centerline{\psfig{file=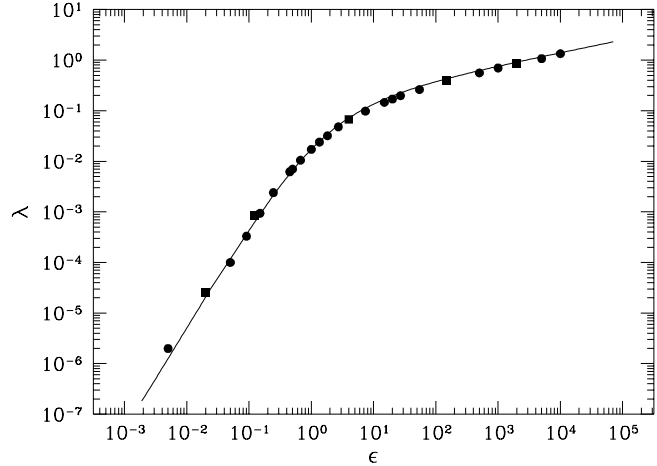,height=8cm,clip=true}} 
\caption{Lyapunov exponent $\lambda$ vs. energy density  
$\varepsilon$  
for the FPU-$\beta$ model with $u = 0.1$.  
The continuous line is the theoretical computation  
according to Eq.\ (\protect\ref{lambda_gauss}), while the  
circles and  squares 
are the results of numerical simulations with $N$  
respectively 
equal to 256 and 2000. 
From Ref.\ \protect\cite{pre96}.} 
\label{fig_lyap_fpu} 
\end{figure}

The geometric theory also allows us to make a quantitative  
prediction for the Lyapunov 
exponent as a function of $k_0$ and $\sigma_k$ via Eq.\  
(\ref{lambda_gauss})~, 
which turns out to be 
extremely accurate. 
The analytic result is shown in Fig.\ \ref{fig_lyap_fpu} and is 
compared with numerical simulations made for different values of $N$,  
for the FPU-$\beta$ case in  
a wide range 
of energy densities --- more than six orders of magnitude  
\cite{prl95,pre96}.  
The agreement 
between theory and simulations is remarkably good. 
 
\subsubsection{1-d $XY$ model} 
 
\label{sec_XY1d} 
If the canonical coordinates $q_i$ and $p_i$ are given the  
meaning of angular 
coordinates and momenta, the 1-d $XY$ model,  
whose potential energy is given in 
Eq.\ (\ref{v_rotatori}), describes a linear chain 
of $N$ rotators constrained to rotate on a plane and coupled  
by a 
nearest-neighbor interaction. 
This model can be formally obtained by restricting  
the classical Heisenberg model with $O(2)$ symmetry to one  
spatial dimension. 
The potential energy of the $O(2)$ Heisenberg model is  
$V= -J\sum_{\langle i,j \rangle}{\bf s}_{i}\cdot{\bf  
s}_{j}$, where the sum is 
extended only over nearest-neighbor pairs, $J$ is the  
coupling constant and 
each ${\bf s}_{i}$ has unit modulus and rotates in the plane. 
To each ``spin'' ${\bf s}_{i}= (\cos q_{i},\sin q_{i})$, the  
velocity  
${\bf \dot s}_{i}=(-\dot q_{i} \sin q_{i}, 
\dot q_{i} \cos q_{i})$ is associated, so that ${\cal  
H}=\sum_{i=1}^{N} 
\frac{1}{2}\dot{\bf s}_{i}^{2} -J \sum_{\langle i,j \rangle}  
{\bf s}_{i}\cdot {\bf s}_{j}$. 
 
This Hamiltonian system has two integrable limits. In the  
small energy limit it represents a chain of harmonic  
oscillators, as can be seen by expanding the potential energy in a power 
series, 
\begin{equation} 
{\cal H}(p,q) 
\simeq\sum_{i=1}^{N}\left\{\frac{p_{i}^{2}}{2}+J 
(q_{i+1}-q_{i})^2 - 1 \right\}~, 
\label{hamrotE0} 
\end{equation} 
where $p_i = \dot q_i$, whereas in the high-energy limit a system  
of freely rotating 
objects is found, 
because the kinetic energy becomes much larger than the bounded 
potential energy. 
 
The dynamics of this system has been extensively studied  
recently 
\cite{LiviPettiniRuffoVulpiani,EscandeLiviRuffo,tesi_cecilia}.  
Numerical simulations and theoretical arguments independent  
of 
the geometric approach (see in particular Ref.\  
\cite{EscandeLiviRuffo}) 
have shown that also in this system there exist weakly and strongly chaotic 
dynamical regimes. It has been found that  
the dynamics is weakly chaotic in the low- and high-energy 
density regions, close to the two integrable limits. On the  
contrary, 
fully developed chaos is found in the intermediate-energy  
region. 
 
According to Eq.\ (\ref{k_R_obs_chain}),  
the expression of the Ricci curvature $K_{R}$, computed with  
the Eisenhart metric, is 
\begin{equation} 
K_{R}(q) 
=2J\sum_{i=1}^{N}\cos(q_{i+1}-q_{i}). 
\end{equation} 
We note that for this model a relation exists between  
the potential energy 
$V$ and Ricci curvature $K_R$: 
\begin{equation} 
V({q})= JN - \frac{K_{R}(q)}{2}~. 
\label{vincolo} 
\end{equation} 
 
\begin{figure} 
\centerline{\psfig{file=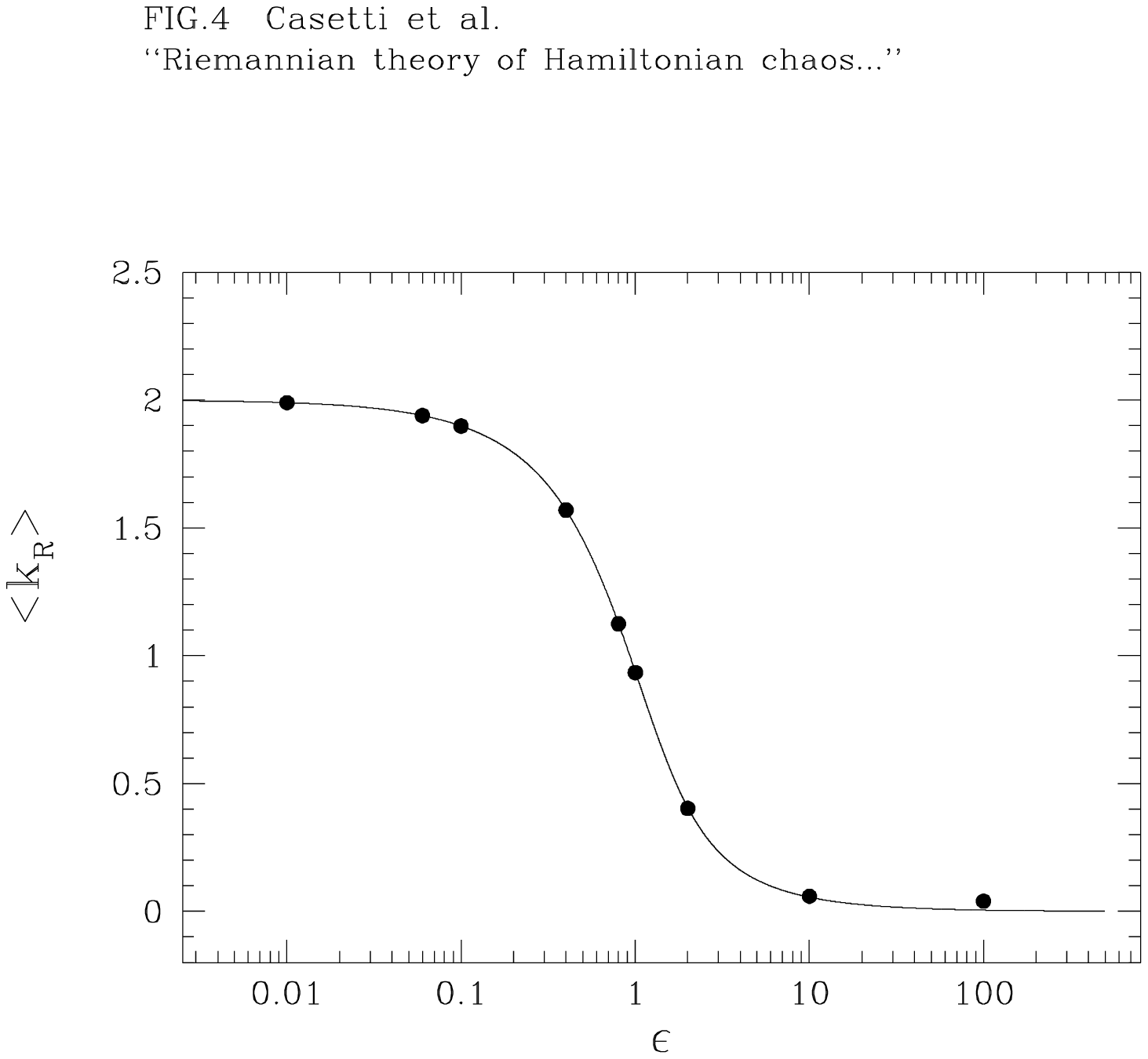,height=8cm,clip=true}} 
\caption{Average Ricci curvature (Eisenhart metric)  
per degree of freedom $k_0$ vs.\ specific energy 
$\varepsilon$ for the coupled rotators model. The  
continuous line 
is the result of an analytic computation according to  
Eq.\ (\protect\ref{media-rot}); the full
circles are time averages obtained by numerical simulations 
with $N = 150$; $J = 1$. 
From Ref.\ \protect\cite{pre96}.} 
\label{fig_curv_rot} 
\end{figure}

\begin{figure} 
\centerline{\psfig{file=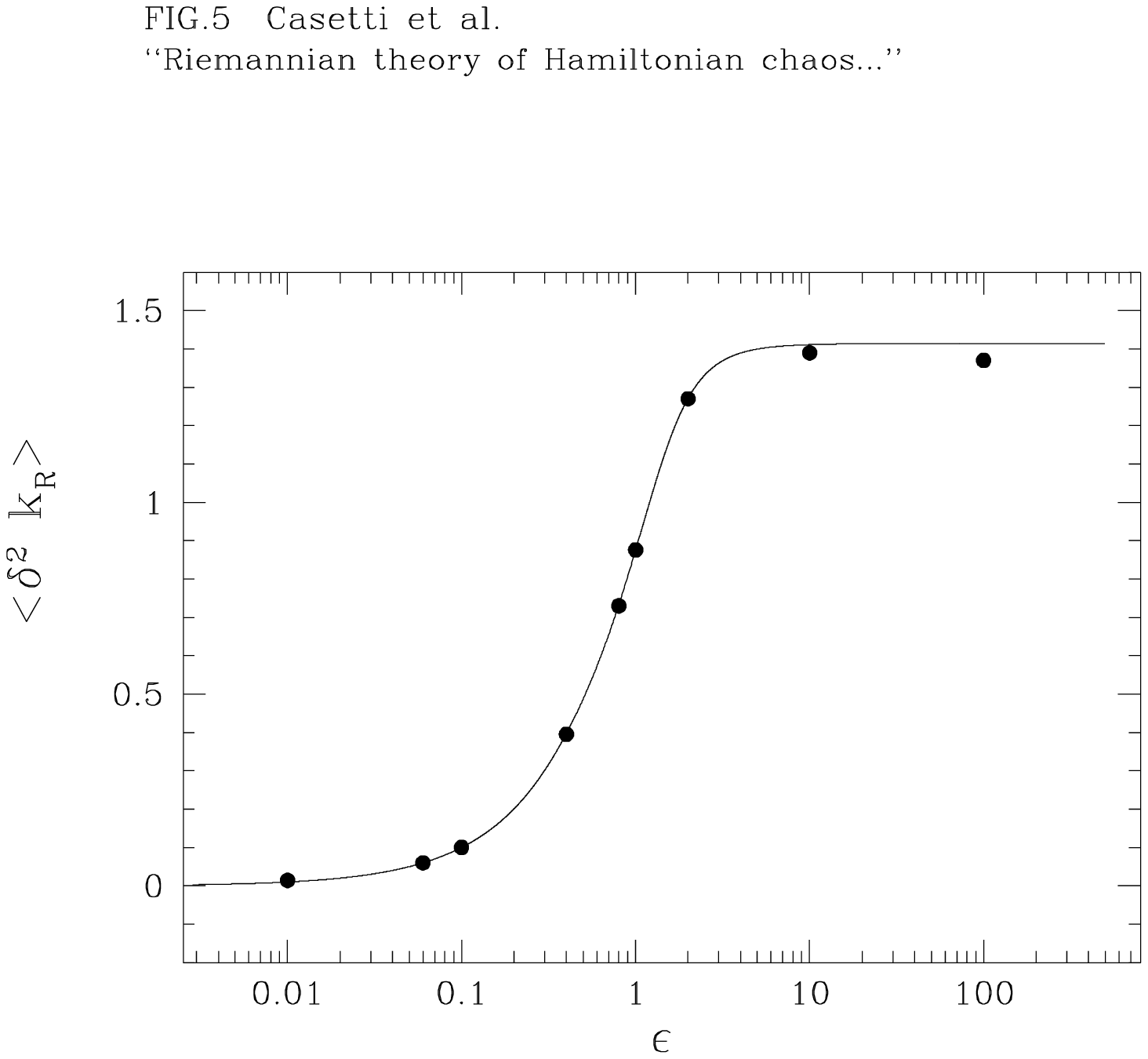,height=8cm,clip=true}} 
\caption{Fluctuation of the Ricci curvature (Eisenhart  
metric) 
$\sigma^2_k$ vs.\ specific energy $\varepsilon$ for the coupled  
rotators model. 
Symbols and parameters as in Fig.\ \protect\ref{fig_curv_rot};  
the continuous line now refers to  
Eq.\ (\protect\ref{flutt-rot}). 
From Ref.\ \protect\cite{pre96}.} 
\label{fig_flutt_rot} 
\end{figure}

The average Ricci curvature can be again expressed by implicit  
formul{\ae} (see Ref.\ \cite{pre96} for details) 
\begin{equation} 
\left. \begin{array}{l} 
\langle k_{R}\rangle_{{}_\mu}(\beta)=  
2J\,{\displaystyle\frac{I_{0}(\beta J)} 
{I_{1}(\beta J)}} \\ 
\\ 
\varepsilon(\beta)={\displaystyle\frac{1}{2\beta} }+ 
J\left(1- \,{\displaystyle\frac{I_{1}(\beta J)}{I_{0}(\beta  
J)} }\right) 
\end{array}\right\} \, \rightarrow k_0(\varepsilon)~, 
\label{media-rot} 
\end{equation} 
where the $I_\nu$'s are modified Bessel functions of index $\nu$ 
\cite{Abramowitz}.  
The fluctuations are given by the implicit equations 
\begin{equation} 
\left. \begin{array}{l} 
\langle\delta^{2}k_{R}\rangle(\beta)= 
{\displaystyle  
\frac{ 4 J}{\beta}\frac{\beta J I_{0}^{2}(\beta 
J)-I_{0}(\beta J)I_{1}(\beta J)-\beta J I_{1}^{2}(\beta J)} 
{I_{0}^{2}(\beta 
J)\left[1+2\left(\beta J\right)^{2}\right]-2\beta J  
I_{1}(\beta J)I_{0}(\beta 
J)-2\left[\beta J I_{1}(\beta J)\right]^{2}} }\\ 
\\ 
\varepsilon(\beta)={\displaystyle \frac{1}{2\beta}+J\left[1- 
\frac{I_{1}(\beta J)}{I_{0}(\beta J)}\right]} 
\end{array}\right\} \, \rightarrow \sigma^2_k (\varepsilon)~. 
\label{flutt-rot} 
\end{equation} 
In Figs.\ \ref{fig_curv_rot} and \ref{fig_flutt_rot}  
a comparison between analytical and numerical results is 
provided for the average Ricci curvature and its  
fluctuations. The agreement  
between ensemble and time averages is again very good.  
 
\begin{figure} 
\centerline{\psfig{file=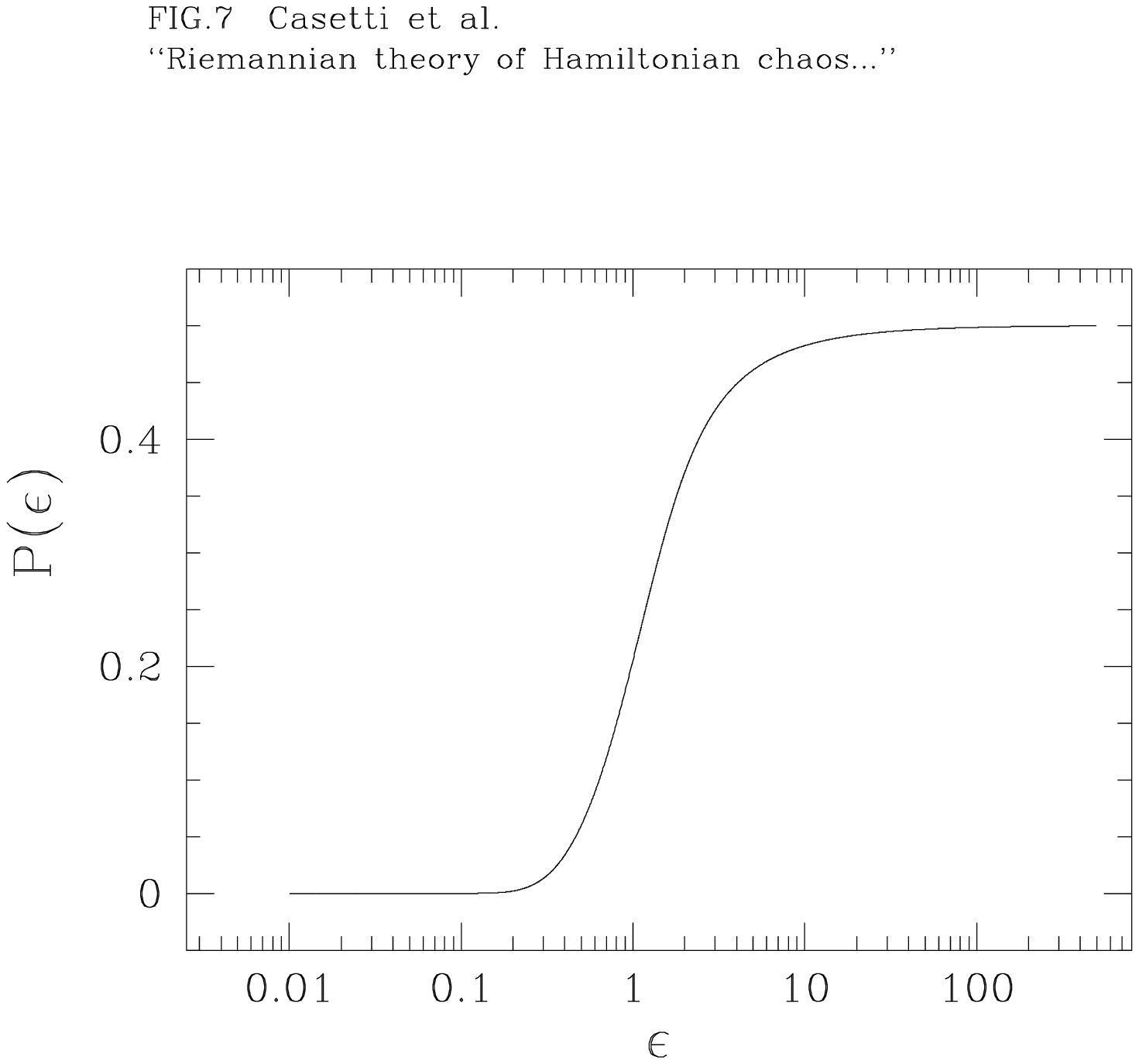,height=8cm,clip=true}} 
\caption{Estimate of the probability $P(\varepsilon)$  
of occurrence of negative sectional curvatures in the  
1-d $XY$ model 
according to Eq.\ (\protect\ref{prob.cos.neg}); $J = 1$.  
From Ref.\ \protect\cite{pre96}.
\label{fig_pro_neg}} 
\end{figure}

Looking at Figs.\ \ref{fig_curv_rot} and \ref{fig_flutt_rot}  
we realize that 
the low-energy weakly chaotic region has the same geometric  
properties as 
the corresponding region of the FPU model, as expected, since  
the two low-energy integrable 
limits are the same. On the contrary, in the high-energy  
weakly chaotic region 
the fluctuations are far from being small with respect to  
the average curvature. 
The average curvature $k_0(\varepsilon)$ vanishes as $\varepsilon \to 
\infty$. In this case the weakly chaotic dynamics seems  
related to the fact 
that the manifold $(M\times {\bf R}^2,g_E)$ looks almost  
flat along the 
physical geodesics. The bounds of the two weakly chaotic  
regions, as estimated  
in Ref.\ \cite{EscandeLiviRuffo}, coincide  
with the 
values of $\varepsilon$ where the asymptotic behaviours of  
$k$ (low-energy 
region) and $\sigma_k$ (high-energy region) set in, respectively.  
Moreover, the case of the coupled rotators  
is very different from the FPU case, since the  
sectional curvature 
$K(s)$ along a geodesic can take negative values. The  
probability  
$P(\varepsilon)$ that $K(s) < 0$ can be analytically  
estimated in the following 
simple way. 
The explicit 
expression of the sectional curvature $K({\dot\gamma},{\xi})$, 
relative to  
the plane spanned by 
the velocity vector ${\dot\gamma}= dq/dt$ and a generic vector ${\xi} \bot  
{\dot\gamma}$, is (see Appendix \ref{app_geo}, Eq.\ \ref{sec_curv_comp}) 
\begin{equation} 
K({\dot\gamma},  
\xi)=R_{0i0k}\frac{dq^0}{dt}\frac{\xi^i}{\Vert{\xi}\Vert} 
\frac{dq^0}{dt}\frac{\xi^k}{\Vert{\xi}\Vert} \equiv  
\frac{\partial^{2}V}{\partial q^{i}\partial  
q^{k}}\frac{\xi^i \xi^k} 
{\Vert{\xi}\Vert^{2}}~, 
\end{equation} 
so that, computing $\partial^2 V/\partial q^i \partial q^k$ using the 
explicit form of $V(q)$ given in Eq.\ (\ref{v_rotatori}), 
we get  
\begin{equation} 
K({\dot\gamma},{\xi})=\frac{J}{\Vert{\xi}\Vert^{2}}\sum_{i=1}^{N} 
\cos(q_{i+1}-q_{i})\left[\xi^{i+1}-\xi^{i}\right]^{2} 
\label{csezrot} 
\end{equation} 
for the 1-d $XY$ model. 
We realize, by simple inspection of Eq.\  (\ref{csezrot}),  
that 
the probability of finding $K<0$ along a geodesic must be  
related to the 
probability of finding an angular difference larger than  
$\frac{\pi}{2}$  
between two nearest-neighboring rotators. 
From Eq.\ (\ref{csezrot}) we see that $N$ orthogonal directions of the  
vector $\xi$ exist 
such that the sectional curvatures --- relative to the $N$  
planes spanned by  
these vectors together with $\dot\gamma$ --- are just $\cos (q_{i+1}-  
q_i)$. These directions are defined by the unit vectors of components 
$(1,0,\ldots,0),(0,1,0,\ldots,0),\ldots,(0,\ldots,0,1)$. Hence 
the probability $P(\varepsilon )$ of occurrence of a  
negative value of a
cosine is used to estimate the probability of occurrence of  
negative sectional 
curvatures along the geodesics. This probability function, 
calculated using the Boltzmann weight,  
has the following 
simple expression \cite{tesi_cecilia,pre96} 
\begin{equation} 
P(\varepsilon) = \frac{\int_{-\pi}^{\pi}\Theta(-\cos  
x)e^{\beta J \cos 
x}dx}{\int_{-\pi}^{\pi}e^{\beta J \cos x}dx}= 
\frac{\int_{\frac{\pi}{2}}^{\frac{3\pi}{2}}e^{\beta J \cos 
x}dx}{2\pi I_{0}(\beta J)}, 
\label{prob.cos.neg} 
\end{equation} 
where $\Theta(x)$ is the Heavyside unit step function 
and $I_0$ the modified Bessel function of index $0$.  
$P(\varepsilon)$ 
is plotted in 
Fig.\ \ref{fig_pro_neg}. We see 
that in the strongly chaotic region such  a probability  
starts to increase 
rapidly from a very small value, while it approaches an  
asymptotic value 
$P(\varepsilon) \simeq 1/2$ when the system enters  
its high-energy weakly  
chaotic region. 
 
When the sectional curvatures are positive\footnote{The sectional curvature 
is always strictly positive  
in the FPU $\beta$ model; in the 1-d $XY$ model, in the low energy 
region, negative 
sectional curvatures can occur, 
but have a very small probability.} chaos is produced by  
curvature  
fluctuations, hence we expect chaos to be weak as long as  
$\sigma_k/k_0 \ll 1$,  
and to become strong when $\sigma_k \approx k_0$. On the  
contrary, when  
$K(s)$ can assume both positive and negative values, the  
situation is much 
more complicated, for there are now two different and  
independent sources 
of chaos: negative curvature which directly induces a  
divergence of nearby 
geodesics, and the bumpiness of the ambient manifold which  
induces such 
a divergence via parametric instability. The results for the  
coupled rotators 
model suggest that as long as the negative curvatures are  
``few'' they do 
not dramatically change the picture, and may strengthen the  
parametrically 
generated chaos, while when their occurrence is equally likely  
as the occurrence 
of positive curvatures, the two mechanisms of chaos seem to  
inhibit each 
other and chaos becomes weak\footnote{The fact that the two  
mechanisms, when 
comparable, can inhibit rather than strengthen each other can be  
considered 
a ``proof'' of the fact that their nature is intrinsically  
different. A similar 
situation is found also in some billiard systems, where there  
are two mechanisms 
which can produce chaos: $(i)$ defocusing, due to positively curved boundaries, and $(ii)$ divergence of the trajectories due to scatterings with 
negatively curved boundaries 
\protect\cite{BunimovichCasatiGuarneri}.}.  
 
\begin{figure} 
\centerline{\psfig{file=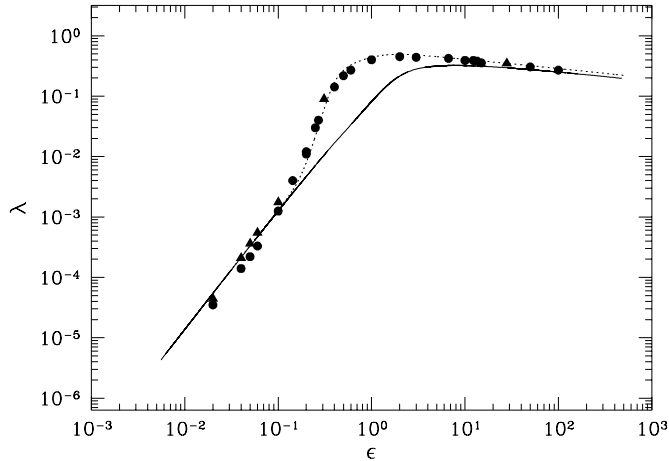,height=8cm,clip=true}} 
\caption{Lyapunov exponent $\lambda$ vs. energy density  
$\varepsilon$  
for the 1-d $XY$ model with $J = 1$.  
The continuous line is the theoretical  
computation according to Eq. (\protect\ref{lambda_gauss}),  
while full circles, squares and triangles  
are the results of numerical simulations with $N$,  
respectively, 
equal to 150, 1000, and 1500. The dotted line is the  
theoretical result 
where the value of $k$ entering Eq.  
(\protect\ref{lambda_gauss}) has been 
corrected according to Eq. (\protect\ref{Kcorr}) with  
$\alpha = 150$. 
From Ref.\ \protect\cite{pre96}.} 
\label{fig_lyap_rot} 
\end{figure} 
 
Such a qualitative picture is consistent with the result of the  
geometric computation of $\lambda$ for the coupled rotator 
model. The result of the application of Eq.\  
(\ref{lambda_gauss}) to this 
model is plotted in Fig.\ \ref{fig_lyap_rot} (solid line). 
There is agreement between analytic  
and numeric values of the Lyapunov exponent only at  
low and high 
energy densities. Like in the FPU case, at low energy, in  
the quasi-harmonic 
limit, we find $\lambda(\varepsilon)\propto\varepsilon^2$.  
At high energy 
$\lambda(\varepsilon)\propto\varepsilon^{-1/6}$; here  
$\lambda(\varepsilon)$ is a  
decreasing function of $\varepsilon$ because for  
$\varepsilon\rightarrow\infty$ the 
systems is integrable. 
 
However, in the intermediate energy range our theoretical prediction 
underestimates the actual degree of chaos of the dynamics. 
This energy range coincides with the region of fully 
developed (strong) chaos. According to the above discussion  
the origin  
of the underestimation can be found in the fact that the  
role of the  
negative curvatures, which appears to strengthen chaos in this  
energy 
range, is not correctly taken into account.  The sectional  
curvature $K(s)$, whose expression is given by Eq.\ (\ref{csezrot}),  
can take negative values with non-vanishing  
probability  
regardless of the value of $\varepsilon$,  
whereas, as long as $\varepsilon < J$, 
this possibility is lost in the replacement of $K$ by the Ricci  
curvature, due to   
the constraint 
(\ref{vincolo}), which implies that at each point of the manifold 
\begin{equation} 
k_R(\varepsilon )\geq 2(J-\varepsilon)~. 
\end{equation} 
Thus our approximation fails to account for the 
presence of negative sectional curvatures at   
values of $\varepsilon$ smaller than $J$.  
In Eq.\ (\ref{csezrot}) the cosines have different and  
variable weights,  
$(\xi^{i+1}-\xi^i)^2$, which make it in principle possible to  
find somewhere 
along a geodesic a $K<0$ even with only one negative cosine.  
This is not the 
case for $k_R$ where all the cosines have the same weight.  
 
Let us now show how the theoretical results can be improved. 
Our strategy is to modify the model for $K(s)$ in some {\it  
effective}  
way which takes into account the just mentioned difficulty of  
$k_R(s)$ to 
adequately model $K(s)$. This will be achieved by suitably  
``renormalizing''  
$k_0$ or $\sigma_k$ to obtain an ``improved'' gaussian  
process which can better model the behavior of the sectional curvature. 
Since our ``bare'' gaussian model underestimates negative  
sectional 
curvatures in the strongly chaotic region, the simplest way  
to  
renormalize the gaussian process is to shift the peak of the  
distribution ${\cal P}(K_R)$ toward the negative  
axis to make the average smaller. This can easily  be done 
by the following rescaling of the average curvature $k_0$:   
\begin{equation} 
k_0 = \langle k_{R}(\varepsilon)\rangle\rightarrow  
\frac{\langle k_{R}(\varepsilon)\rangle} {1+\alpha  
P(\varepsilon)}. 
\label{Kcorr} 
\end{equation} 
This correction has no influence either  when  
$P(\varepsilon)\simeq 0$ (below  
$\varepsilon\simeq 0.2$) or when $P(\varepsilon)\simeq 1/2$  
(because in that case 
$\langle k_{R}(\varepsilon)\rangle\rightarrow 0$). 
The simple  
correction (\ref{Kcorr}) makes use of the information we have obtained 
analytically, i.e., 
of the $P(\varepsilon)$ given in Eq.\ (\ref{prob.cos.neg}), 
and is sufficient 
to obtain an excellent agreement of the theoretical prediction 
with the numerical data over the whole range of energies, as shown in  
Fig.\ \ref{fig_lyap_rot}. 
The parameter $\alpha$ in (\ref{Kcorr}) is a free parameter, and its value
is determined so as to obtain the best agreement  
between numerical and theoretical data. The result shown 
in Fig.\ \ref{fig_lyap_rot} (dotted line) is obtained with  
$\alpha = 150$, but also very different values of $\alpha$, up to  
$\alpha \simeq 1000$,  
yield a good result, i.e.,  
no particularly fine tuning of $\alpha$ is necessary to obtain a  
very good 
agreement between theory and numerical experiment. 

\subsection{Some remarks} 

Before moving to the second Part of the paper, 
let us now comment about some points of the material presented 
in the first Part of this Report. 
In particular, we would like to clarify the meaning of 
some of the approximations made and to draw the attention of the reader to
some of the points which are still open. A better understanding of these points
could lead, in our opinion, to a considerable improvement of the theory, which 
is still developing and can by no means be considered as a ``closed'' subject.

What has been presented in this  Section has a conceptual implication 
that goes far beyond the development of a method to analytically compute
Lyapunov exponents. Rather, the strikingly good agreement between 
analytic and numeric Lyapunov exponents -- obtained at the price of a 
restriction of the domain of applicability of the analytic expression worked
out for $\lambda$ -- has three main implications: 
\begin{itemize}
\item[{$(i)$}] the {\it local}
geometry of mechanical manifolds contains all the relevant information about
(in)stability of Hamiltonian dynamics; 
\item[{$(ii)$}] once a good model for the
local source of instability of the dynamics is provided, then a 
statistical-mechanical-like treatment of the average degree of instability 
of the dynamics can be worked out, in the sense that we do not need a
detailed knowledge of the dynamics but, by computing {\it global} geometric
quantities, obtain a very good estimate of the average strength of chaos;
\item[{$(iii)$}] due to  the variational formulation of newtonian dynamics,
the Riemannian-geometric framework a-priori seems -- and actually seemed in
the past (as we have pointed out in Section \ref{sec_hist}) -- the natural 
framework to investigate the instability of Hamiltonian dynamics, however no
evidence was available at all to confirm such an idea until the above
mentioned development took place. It is now evident that the efforts to
improve the theory by expanding its domain of applicability are worthwhile.
\end{itemize}
We must warn the reader though against ``blind'' applications of formula
(\ref{lambda_gauss}), i.e. without any idea about the fulfilment, 
by the Hamiltonian model under investigation, of the conditions under which 
it has been derived.

From a more technical point of view, one of the central results we have 
presented so far is the possibility of deriving, from the Jacobi equation, 
a {\em scalar} equation (Eq.\ (\ref{eq_Hill})) describing the evolution of 
the Jacobi field $J$ for a geodesic spread on a manifold. We would like to 
stress that such a result, though approximate, applies to a wide class of
Hamiltonian systems. In fact, the only hypothesis needed to get such an 
equation is the quasi-isotropy hypothesis, i.e., the assumption that 
$R_{ijkl} \approx {\cal K}(s) (g_{ik}g_{jl} - g_{il}g_{jk})$. Loosely speaking,
such an assumption means that, locally, the manifold can be regarded as 
isotropic, i.e., there is a neighborhood 
of each point where the curvature can be considered constant. This does not 
imply at all that  there are only small-scale  
fluctuations.  There can be fluctuations of curvature on many scales, provided 
that they are finite and there is a cutoff at a certain point. 
The only case in which such an assumption will surely {\em not} hold is when 
there are curvature fluctuations  
over {\em all} scales.  As  will become clear in the following, this might 
happen when the manifold undergoes a topological change, and for 
``mechanical manifolds'' this might happen at a phase transition. 
 
Other approximations come into play when one actually wants to model 
${\cal K}(s)$ 
along a geodesic with a stochastic process. 
It is true that replacing the sectional curvature 
by the Ricci curvature requires that the fluctuations are not only finite, 
but also small. Moreover, 
we use global averages to define the stochastic process, and here it is 
crucial that the fluctuations do not extend over too large scales. Thus 
Eq.\ (\ref{eq_stoc_osc}) has a less general validity than Eq.\ (\ref{eq_Hill}).
A way to improve the theory might be to try to 
replace the sectional curvature with some quantity related also to the 
gradient of the Ricci curvature, in order to make the replacement of sectional
curvature less sensitive to the large scale variations of the Ricci curvature.

To get an explicit solution of Eq.\ (\ref{eq_Hill}), an even less general  
situation must be considered, through the following steps:

\begin{itemize}

\item[{$(i)$}] using the Eisenhart metric;

\item[{$(ii)$}] considering standard systems where the kinetic energy does 
not depend on the $q$'s;

\item[{$(iii)$}] estimating the characteristic correlation time $\tau$ of the 
curvature fluctuations. 

\end{itemize}

As to item $(iii)$, we have already remarked that our estimate given in 
Eq.\ (\ref{taufinale}) is by no means a consequence of any theoretical result,
but only a reasonable estimate which could surely be improved.

As to item $(ii)$, the case of a more general kinetic  
energy matrix $a_{ij} \not = \delta_{ij}$, though not conceptually different, 
is indeed different in practice and the final result is not expected to hold 
in the same form for that case. 

Finally, (item $(i)$) should not reduce significantly the generality of the 
result.
In fact, considering the Eisenhart metric only makes 
the calculations feasible, and in principle nothing should change, if one 
were able to solve Eq.\ (\ref{eq_stoc_osc}) in the case of the Jacobi metric
(see the discussion in Ref.\ \cite{CerrutiFranzosiPettini}).
However, Eisenhart and Jacobi metrics are {\it equivalent} for what
concerns the computation of the average instability of the dynamics
\cite{CerrutiFranzosiPettini}, but they might {\it not} be {\it equivalent}
for other developments of the theory.
This in view of the fact that $(M_E,g_J)$ is a manifold which has
better mathematical properties than $(M\times{\bf R}^2,g_E)$:
$(M_E,g_J)$ is a proper Riemannian manifold, it is compact, all of its
geodesics are in one-to-one correspondence with mechanical trajectories, its
scalar curvature does not identically vanish as is the case of
$(M\times{\bf R}^2,g_E)$, it can be naturally lifted to the tangent bundle
where the associated geodesic flow on the submanifolds of constant energy
coincides with the phase space trajectories. 

Let us finally add a comment on the application of the theory to the 
calculation
of the Lyapunov exponent for the one-dimensional $XY$ model. We have seen that 
although the predictions of the theory compare reasonably well with the
numerical simulations, there is an intermediate energy range in which a
correction must be added.
As will become clear in the second Part, the very first assumption 
(quasi-isotropy) should not be satisfied for this model, due to the presence 
of topology changes in the mechanical manifolds, in fact the difficulties of 
the theory begin just at the energy density which corresponds to the appearance
of a large number of critical points of the potential energy 
(see next Sections).

\section {Geometry and phase transitions} 
\label{sec_geopt} 
In the previous Sections we have shown how simple concepts belonging to 
classical differential geometry can be successfully used as tools to build a 
geometric theory of chaotic Hamiltonian dynamics. Such a theory is able to  
describe the instability of the dynamics in classical  
systems consisting of a large number 
$N$ of mutually interacting particles, by relating these properties to the 
average and the fluctuations of the curvature of the configuration space.  
Such a relation is made quantitative through Eq.\ (\ref{lambda_gauss}), which 
provides an approximate analytical estimate of the largest 
Lyapunov exponent in 
terms of the above-mentioned geometric quantities, and 
which compares very well 
with the outcome of numerical simulations in a number of 
cases, two of which have been 
discussed in detail at the end of Sec.\ \ref{sec_geochaos}. 
 
The macroscopic properties of large-$N$ Hamiltonian systems 
can be understood by means of the traditional methods of statistical 
mechanics. One of the most striking phenomena that may happen in such systems 
is that when the external parameters (e.g., either the temperature or the  
energy) are varied until some critical value is reached,  
the macroscopic thermodynamical quantities may suddenly and even 
discontinuously 
change, so that, though the microscopic interactions are the same above 
and below the 
critical value of the parameters, its macroscopic properties may be completely 
different. Such phenomena are referred to as phase transitions. In statistical 
mechanics, phase transitions are explained as true mathematical singularities 
that occur in the thermodynamic functions in 
the limit $N \to \infty$, the so-called thermodynamic  
limit\footnote{According to G.\ E.\ Uhlenbeck \protect\cite{Cohen_Uhlenbeck},  
the use of the thermodynamic limit as an explanation of the 
singularities of the partition function was 
suggested for the first time by Kramers in the 1938 Leiden 
conference on Statistical Mechanics.}\cite{Ruelle}.  
Such singularities come from the fact 
that the equilibrium probability distribution 
in configuration space, which in the 
canonical ensemble is the Boltzmann weight  
\beq 
\varrho_{\text{can}}(q_1,\ldots,q_N) = \frac{1}{Z}\exp\left[ - \beta  
V(q_1,\ldots,q_N)\right]~, 
\eeq 
where $\beta = 1/k_B T$, $V$ is the potential energy, and $Z = \int dq  
\, e^{-\beta V(q)}$ is the configurational partition function,  
can itself develop singularities in the thermodynamic limit. 
 
The statistical-mechanical theory of phase transitions is one of the most 
elaborate and successful physical theories now at hand, and at least as 
continuous phase transitions are concerned, also quantitative  
predictions can
be made, with the aid of renormalization-group techniques, which
are in very good 
agreement with laboratory experiments and numerical simulations. We are not 
going to discuss this here, referring the reader to the vast literature 
on the subject  
\cite{Fisher,KogutWilson,Ma,Parisi,ZinnJustin,LeBellac,Goldenfeld}.  
 
However, the origin of the possibility of describing Hamiltonian systems 
via equilibrium statistical mechanics 
are the chaotic properties underlying the 
dynamics. In fact, though it is not possible to prove that generic Hamiltonian 
systems of the form (\ref{H}) are ergodic and mixing, the fact that  
the trajectories are mostly chaotic  
(i.e., for the overwhelmingly majority of the trajectories 
positive Lyapunov exponents are found) 
means that such systems can be considered 
ergodic and mixing for all practical purposes\footnote{This has been 
recently extended to non-equilibrium statistical mechanics
via the ``Chaotic Hypothesis'' \protect\cite{GallavottiCohen}.}.  

The observation that chaos is at the origin of the statistical behaviour of 
Hamiltonian systems and that chaotic dynamics can be described by means of the 
geometric methods described above, 
naturally leads to the follwing two questions: 
\begin{itemize} 
 
\item[{\bf 1.}] Is there any specific behaviour of the 
Lyapunov exponent\footnote{As in the first part of the paper, we consider 
only the largest Lyapunov exponent, which is referred to as just the Lyapunov  
exponent.}  
when the system undergoes a phase transition? 
 
\item[{\bf 2.}] What are the geometric properties of the configuration space 
manifold in the presence of 
a phase transition? 
 
\end{itemize} 
The aim of the present Section is to discuss these two questions.  
We shall first give a phenomenological description which follows 
from numerical experiments, and then we shall 
concentrate on the particular case of the mean-field $XY$ model 
where the geometrical quantities can be analytically calculated.  
From the discussion of these questions and from the (at least partial) 
answers that we find, we are lead to put forward a Topological 
Hypothesis about phase transitions,  
which will be discussed in Sec.\ \ref{sec_topo}. 
 
\subsection{Chaotic dynamics and phase transitions} 
 
In order to look for an answer to question {\bf 1} above,  
we now review the numerical results that have been obtained until now for 
various Hamiltonian dynamical systems which show a phase transition when 
considered as statistical-mechanical models for macroscopic 
systems in thermal equilibrium. 
 
The first attempt to look for a chaotic-dynamic counterpart 
of an equilibrium phase transition is in the work by Butera and 
Caravati (BC) \cite{Butera}. 
BC considered a two-dimensional $XY$ model, i.e., a  
Hamiltonian dynamical system of the form   
(\ref{H}) with the potential energy 
\beq 
V = 1 - \sum_{\langle i,j\rangle} \cos(\varphi_i - \varphi_j)~, 
\label{V_XY} 
\eeq 
where the $\varphi$'s are angles, $i$ and $j$ label the sites 
of a square lattice and the sum runs over all the nearest-neighbor sites. 
This model is the two-dimensional version of the one studied in 
Sec.\ \ref{sec_XY1d}. 
As the temperature is decreased,  
such a system undergoes a peculiar phase transition 
(referred to as the Bere\v{z}inskij-Kosterlitz-Thouless, or BKT, transition) 
from a disordered phase to a quasi-ordered phase where, though no true 
long-range order is present, correlation functions decay as power laws, 
as occurs at a critical point 
\cite{Goldenfeld}. Since there are no singularities in the finite-order 
derivatives of the free energy, the BKT transition is sometimes classified 
as an ``infinite-order'' phase transition. 
BC computed the Lyapunov exponent $\lambda$ as a function of the temperature, 
and found that $\lambda(T)$ followed a rather smooth pattern; however, in a 
region around the transition, the dependence of $\lambda$ on $T$ changed from 
a steeply increasing function to a much less steep one.  
 
\begin{figure} 
\centerline{\psfig{file=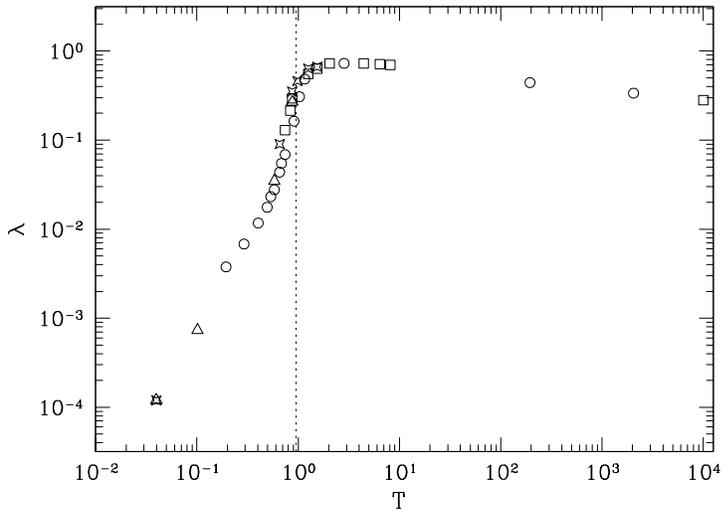,height=8cm,clip=true}}  
\caption{Lyapunov exponent $\lambda$ {\em vs.} the temperature $T$ for the  
two-dimensional $XY$ model: the circles refer to a $10 \times 10$,
the squares to a $40 \times 40$, the triangles to a $50 \times 50$, and
the stars to a $100 \times 100$ lattice, respectively. 
The critical temperature of the BKT transition is 
$T_C \simeq 0.95$ and is marked by a dotted vertical line. 
From Ref.\ \protect\cite{cccp}.} 
\label{fig_lyap_xy_2d} 
\end{figure}

\begin{figure} 
\centerline{\psfig{file=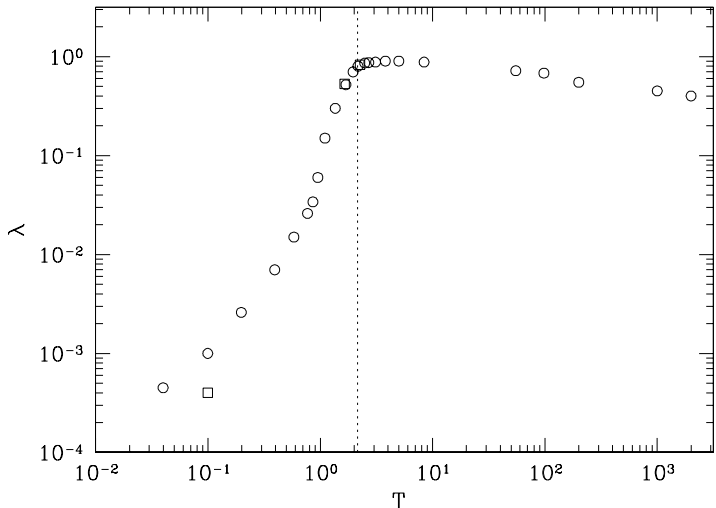,height=8cm,clip=true}} 
\caption{Lyapunov exponent $\lambda$ {\em vs.} the temperature $T$ for the  
three-dimensional $XY$ model, numerically computed on a 
$N = 10 \times 10 \times 10$ lattice (solid circles) and on a
$N = 15 \times 15 \times 15$ lattice (solid squares), respectively. 
The critical temperature of the phase 
transition is $T_C \simeq 2.15$ and is marked by a dotted vertical line. 
From Ref.\ \protect\cite{cccp}.} 
\label{fig_lyap_xy_3d} 
\end{figure}

BC's pioneering paper has been the only example of a study of this kind for a 
long period. However, very recently there has been a renewed interest in the 
study of the behaviour of Lyapunov exponents in systems undergoing phase 
transitions, and a number of papers appeared  
\cite{cccp,pre98,jpa98,Bonasera,Duke_pre,DellagoPosch_hd,DellagoPosch_lj,DellagoPosch_XY,DellagoPosch_hs,Mehra,Ruffo_prl,Firpo,Ruffo_talk,Antoni_rev}. 
The two-dimensional $XY$ model has been reconsidered, together with the  
three-dimensional case, in Ref.\ \cite{cccp}. We remark that in three spatial 
dimensions the $XY$ model undergoes a standard continuous (second-order) 
phase transition accompanied by the breaking of the $O(2)$ symmetry of the 
potential energy (\ref{V_XY}). The behaviour of the Lyapunov exponent 
$\lambda$ as a function of the temperature $T$ is shown in 
Figs.\ \ref{fig_lyap_xy_2d} and  
\ref{fig_lyap_xy_3d}.  
The behavior found for the two-dimensional case confirms 
the BC results. The three-dimensional case shows a similar behavior, 
but the change of the shape of the $\lambda(T)$ function near $T_C$ is somehow 
sharper than in the previous case. 
 
Dellago and Posch \cite{DellagoPosch_XY} considered an extended 
$XY$ model, whose potential energy is  
\beq 
V = 2 - 2\sum_{\langle i,j\rangle} 
 \cos\left(\frac{\varphi_i - \varphi_j}{2}\right)^{p^2}~~, 
\label{V_XY_ext} 
\eeq 
which includes the standard $XY$ model (\ref{V_XY}) for $p^2 = 2$.  
On a two-dimensional lattice the transition, which is a continuous BKT 
transition for $p^2 = 2$, becomes a discontinuous transition as 
$p^2 \simeq 100$. The results for the Lyapunov exponent $\lambda$   
show that for any considered value of $p^2$ 
there is a change in the shape of $\lambda(T)$ close to the 
critical temperature. 
 
\begin{figure} 
\centerline{\psfig{file=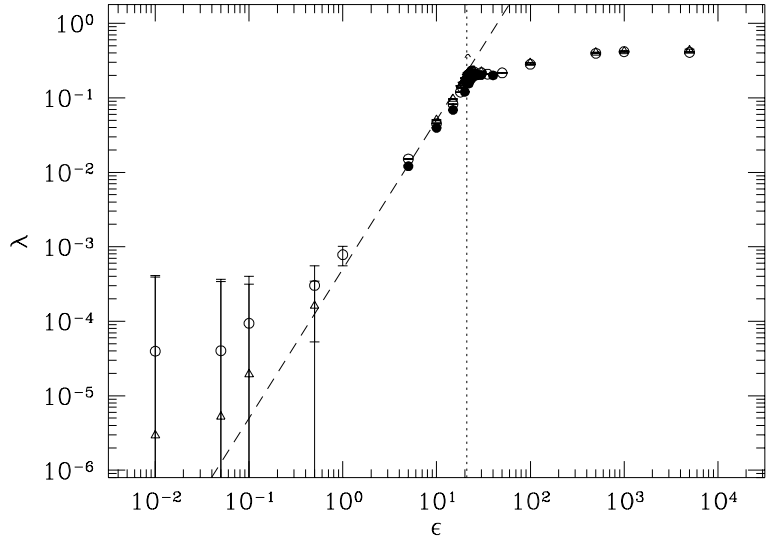,height=8cm,clip=true}} 
\caption{Lyapunov exponent $\lambda$ {\em vs.} the energy per particle 
$\varepsilon$, numerically computed for the  
two-dimensional $O(1)$ $\varphi^4$ model, with $N = 100$ (solid circles),  
$N=400$ (open circles),  
$N=900$ (solid triangles), and $N=2500$ (open triangles), respectively.  
The critical energy is marked by a vertical dotted line, and the 
dashed line is the power law $\varepsilon^2$.  
From Ref.\ \protect\cite{jpa98}.} 
\label{fig_lyap_phi4_2d} 
\end{figure}

\begin{figure} 
\centerline{\psfig{file=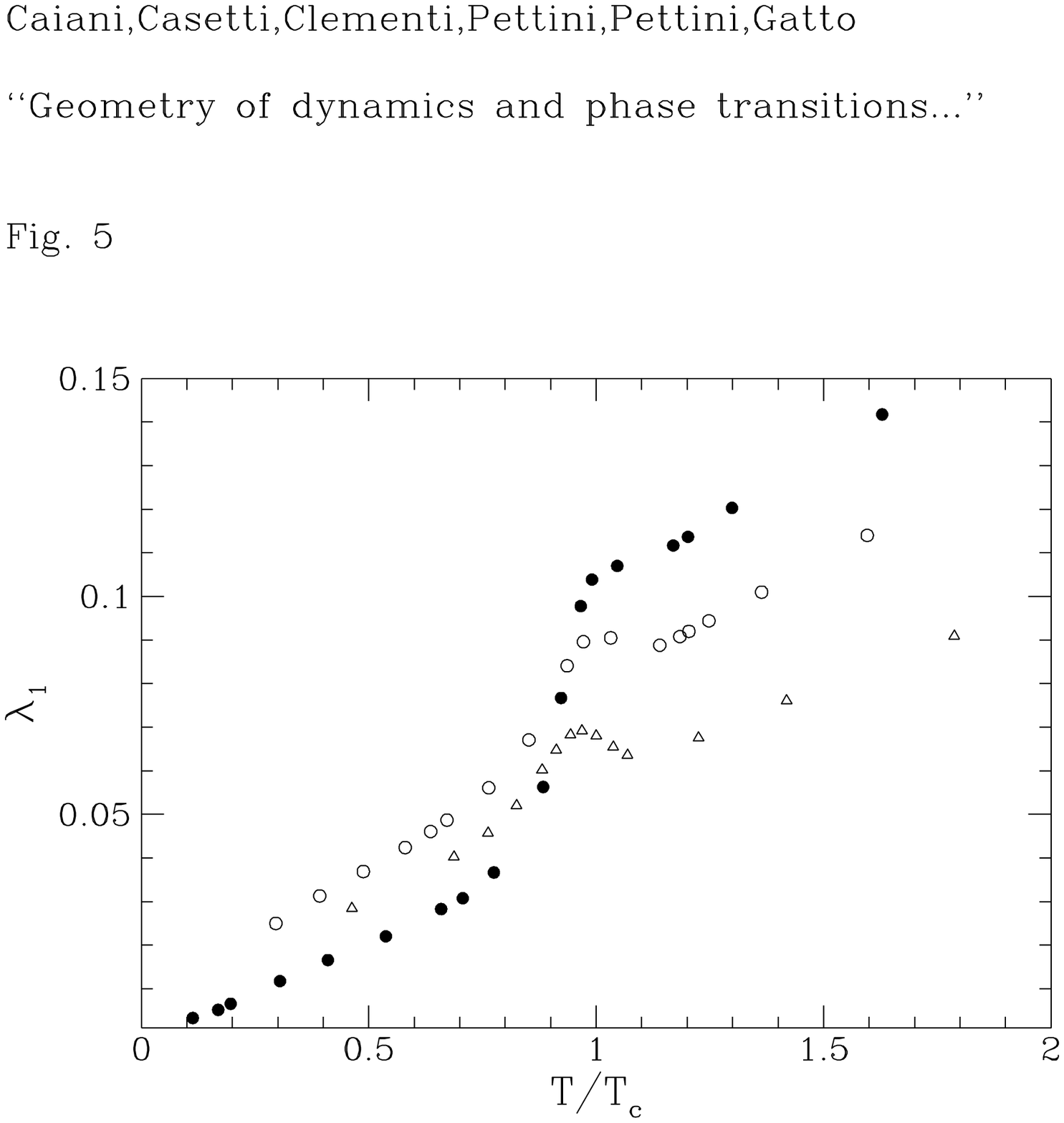,height=8cm,clip=true}} 
\caption{Lyapunov exponent $\lambda$ {\em vs.} the temperature $T$ for the  
three-dimensional $\varphi^4$ model. Full circles correspond to the $O(1)$ 
(scalar) case, open circles to the $O(2)$ case, and open triangles to the 
$O(4)$ case.  
From Ref.\ \protect\cite{pre98}.} 
\label{fig_lyap_phi4_3d} 
\end{figure}

One of the systems which have received considerable attention in this 
framework is 
the so-called lattice $\varphi^4$ model, i.e., 
a system with a Hamiltonian of the form (\ref{H}) and a potential energy 
given by 
\beq 
V = 
\frac{J}{2} \sum_{\langle i , j \rangle} 
(\varphi_{{i}}-\varphi_{j})^2 + \sum_i \left[  
-\frac{r^2}{2}  
\varphi_{i}^2 
+\frac{u}{4!}\varphi_{i}^4\right]~, 
\label{hfi4d} 
\eeq 
where the $\varphi$'s are scalar variables, $\varphi_i \in 
[-\infty,+\infty]$, defined on the sites of a 
$d$-dimensional lattice, and $r^2$ and $u$ are positive parameters. 
The lattice $\varphi^4$ model has a phase transition 
at a finite temperature provided that $d>1$.  
The existence of such a transition, which belongs to the universality class 
of the $d$-dimensional Ising model,  
can be proved by  
renormalization-group arguments (see e.g.\ \cite{KogutWilson,LiviMaritan}). 
The cases $d=2$ and $d=3$ have been considered in 
Refs.\ \cite{jpa98} and \cite{pre98}, respectively. Moreover,  
in Ref.\ \cite{pre98} also 
some vector versions of this model have been considered, 
namely, systems with potential energy given by 
\beq 
V = 
\frac{J}{2} \sum_{\langle i , j \rangle} 
\sum_{\alpha} (\varphi^\alpha_{{i}}-\varphi^\alpha_{j})^2  
+ \sum_i \left\{  
-\frac{r^2}{2}  
\sum_{\alpha} \left( \varphi^\alpha_{i} \right)^2 
+\frac{u}{4!} \left[ \sum_{\alpha} \left( \varphi^\alpha_{i}\right)^2 
\right]^2 \right\}~, 
\label{hfi4d_vector} 
\eeq 
where $\alpha$ runs from 1 to $n$, labelling the components of the vectors 
$\varphi_i = (\varphi_i^1,\ldots,\varphi^n_i)$. The potential energy 
(\ref{hfi4d_vector}) is $O(n)$-invariant; in  the case $n = 1$ we recover 
the scalar model  
(\ref{hfi4d}). 
Figures \ref{fig_lyap_phi4_2d} and \ref{fig_lyap_phi4_3d} 
 show the behaviour of $\lambda$ in the $\varphi^4$ model, in two and 
three dimensions, respectively. 
 

Again we see that the Lyapunov exponent is sensitive to the presence of 
the transition, and that the shape of $\lambda(T)$ close to the transition is 
highly model-dependent. Moreover, such a shape can be significantly different 
within the same model as its parameters are varied. For instance, in the 
$\varphi^4$ model, $\lambda$ can be either a monotonously increasing function 
of $T$ 
or can display a maximum close to $T_c$, depending on the values of $r^2$ 
and $u$ \cite{jpa98}. 
 
The Lyapunov exponents of systems undergoing phase transitions of 
the solid-liquid type have been recently determined numerically: 
Dellago and Posch (DP) considered, in two dimensions, a system of hard disks 
\cite{DellagoPosch_hd}, a Lorentz-gas-like
model  
and a Lennard-Jones fluid \cite{DellagoPosch_lj}, and, in 
three dimensions,  a system of hard spheres\cite{DellagoPosch_hs}. DP 
found that in all these systems  the Lyapunov exponent is  sensitive to 
the phase transition, 
and again the shape of $\lambda$ is different for different models,
the common feature being that $\lambda$ attains a maximum close, 
if not at, the transition. Similar results have been obtained by Mehra 
and Ramaswamy \cite{Mehra}. 
Bonasera {\em et al.} \cite{Bonasera} considered a classical model 
of an atomic cluster,  
whose particles interact via phenomenological pair potentials of the form 
\beq 
v(r) = a\, e^ {-\left(\frac{b r}{\sigma}\right)} -  
c \left(\frac{\sigma}{r} \right)^6~,  
\eeq 
and of a nuclear cluster, with nucleons interacting via 
Yukawa pair potentials. 
Such systems undergo a so-called ``multifragmentation'' transition at a  
critical (model-dependent) temperature $T_c$. Bonasera {\em et al.}  
computed the Lyapunov exponents of these systems by means 
of numerical simulations at different temperatures. The resulting 
$\lambda(T)$ of both systems develops a sharp maximum close to $T_c$. 
 
The numerical evidence that we have reviewed above clearly shows that 
the Lyapunov exponent of a Hamiltonian dynamical system is sensitive  
to the presence of  
a phase transition. However, the interpretation of the observed behavior 
as it now stands is very difficult, because each model behaves differently 
and the behavior of $\lambda$ close to the transition does not apparently 
exhibit any universal feature: on the contrary, the shape of $\lambda(T)$ can 
depend on the values of the parameters of the model. Moreover, the qualitative
behavior of $\lambda(T)$ appears to be only weakly dependent on whether the 
transition is accompanied by a symmetry breaking or not,
as in the case of the $XY$ model: the shape of $\lambda(T)$  in two 
dimensions, where there is {\em not} any breaking of the $O(2)$ symmetry of 
the potential energy below the BKT transition temperature, is not dramatically
different from that of the three-dimensional case where the phase transition 
is accompanied by a symmetry breaking. In the latter case the ``knee'' of the 
$\lambda(T)$ curve is sharper, but it would be difficult to discriminate 
between  
the two cases only by looking at the $\lambda(T)$ curve. 
Therefore, though clearly sensitive to the presence of a phase 
transition, the Lyapunov exponent does not seem a ``good'' probing observable
for the occurrence of a symmetry-breaking phase transition. 
 
\subsection{Curvature and phase transitions} 
 
In Section \ref{sec_geochaos} we have seen that the origin of 
chaos in Hamiltonian mechanics can be understood from a geometrical 
point of view,  
and that the Lyapunov exponents are closely related to a geometric quantity, 
i.e., to the fluctuations of the Ricci curvature of the configuration space.  
Thus, it is natural to investigate whether such a geometric observable 
also has some peculiar behaviour close to the phase transition. As we shall 
see in the following, the fluctuations of the curvature do indeed 
 have such  a peculiar behaviour which, 
in turn, suggests a topological intepretation of the phase transition itself. 
 
\begin{figure} 
\centerline{\psfig{file=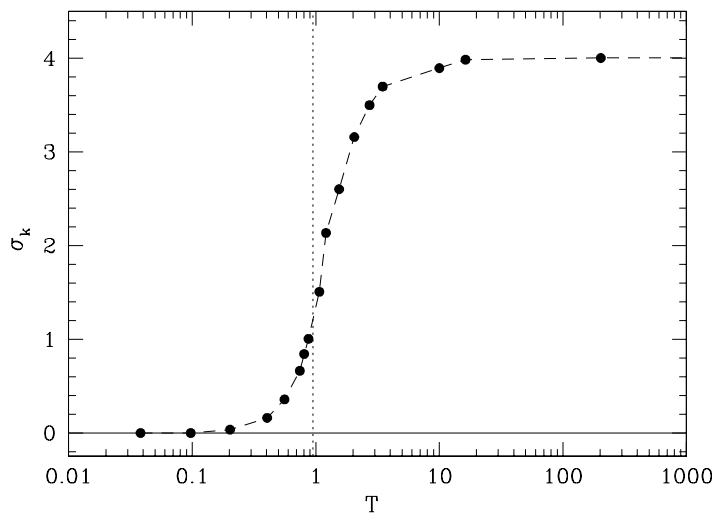,height=8cm,clip=true}} 
\caption{Fluctuations of the Ricci curvature (Eisenhart metric), 
$\sigma_k(T)$,  {\em vs.} the temperature $T$ for the  
two-dimensional $XY$ model. The solid circles are numerical values
obtained for a $40 \times 40$ lattice; the dashed line is only a guide 
to the eye.
The critical temperature of the BKT 
transition is $T_C \simeq 0.95$ and is marked by a dotted line. From Ref.\ \protect\cite{cccp}.} 
\label{fig_fluct_xy_2d} 
\end{figure}

\begin{figure} 
\centerline{\psfig{file=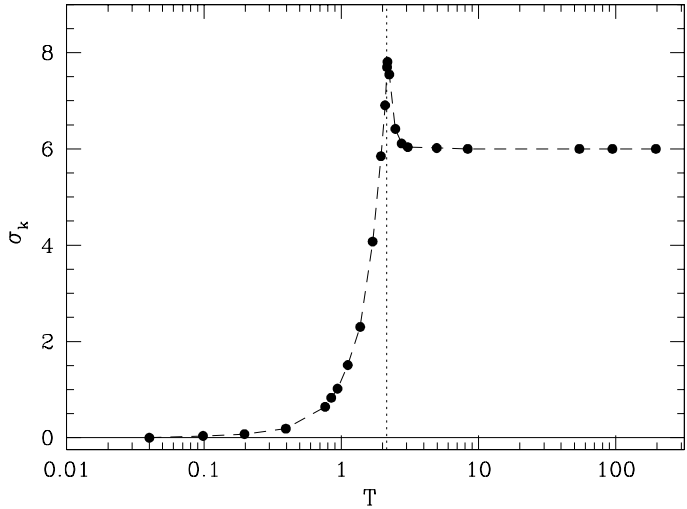,height=8cm,clip=true}} 
\caption{As Fig.\ \protect\ref{fig_fluct_xy_2d}, for the  
three-dimensional $XY$ model. Here $N = 10 \times
10 \times 10$, and the critical temperature of the phase transition 
is $T_c \simeq 2.15$. From Ref.\ \protect\cite{cccp}.} 
\label{fig_fluct_xy_3d} 
\end{figure}

The Ricci curvature along a geodesic of the enlarged configuration 
spacetime equipped with the Eisenhart metric, which we denoted by $K_R$ in 
the previous Sections, is given by the Laplacian of the potential energy ---  
see Eq.\ \ref{k_R_obs}. In the case of the $XY$ model we obtain, as already 
shown in  
\S \ref{sec_XY1d}, 
\beq 
K_R = 2N - 2V = 2 \sum_{\langle i,j\rangle} \cos(\varphi_i - \varphi_j)~. 
\eeq 
The root mean square fluctuation of $K_R$ divided by the number of 
degrees of freedom $N$, i.e.,  
\beq 
\sigma_k = \left(\frac{1}{N}\langle K_R^2 \rangle -  
\langle K_R \rangle^2\right)^{1/2}  ~, 
\eeq 
is plotted in Figs.\ \ref{fig_fluct_xy_2d} and \ref{fig_fluct_xy_3d} for the  
2-$d$ and 3-$d$ cases, respectively.  
 
\begin{figure} 
\centerline{\psfig{file=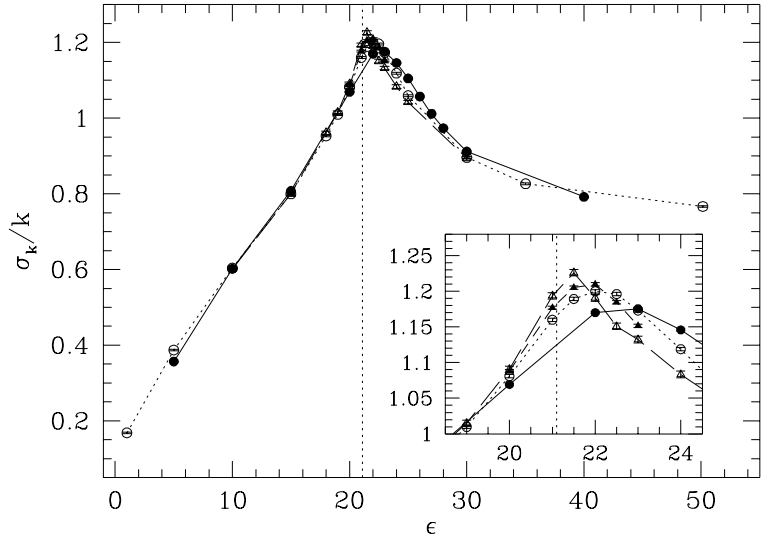,height=8cm,clip=true}} 
\caption{Root mean square fluctuation of the Ricci curvature  
(Eisenhart metric) $\sigma_k$, divided by the average curvature $k_0$,  
numerically computed for the  
two-dimensional $O(1)$ $\varphi^4$ model. The inset shows a magnification 
of the region close to the transition.
Symbols as in  
Fig.\ \protect\ref{fig_lyap_phi4_2d}. 
From Ref.\ \protect\cite{jpa98}.} 
\label{fig_fluct_phi4_2d} 
\end{figure}

\begin{figure} 
\centerline{\psfig{file=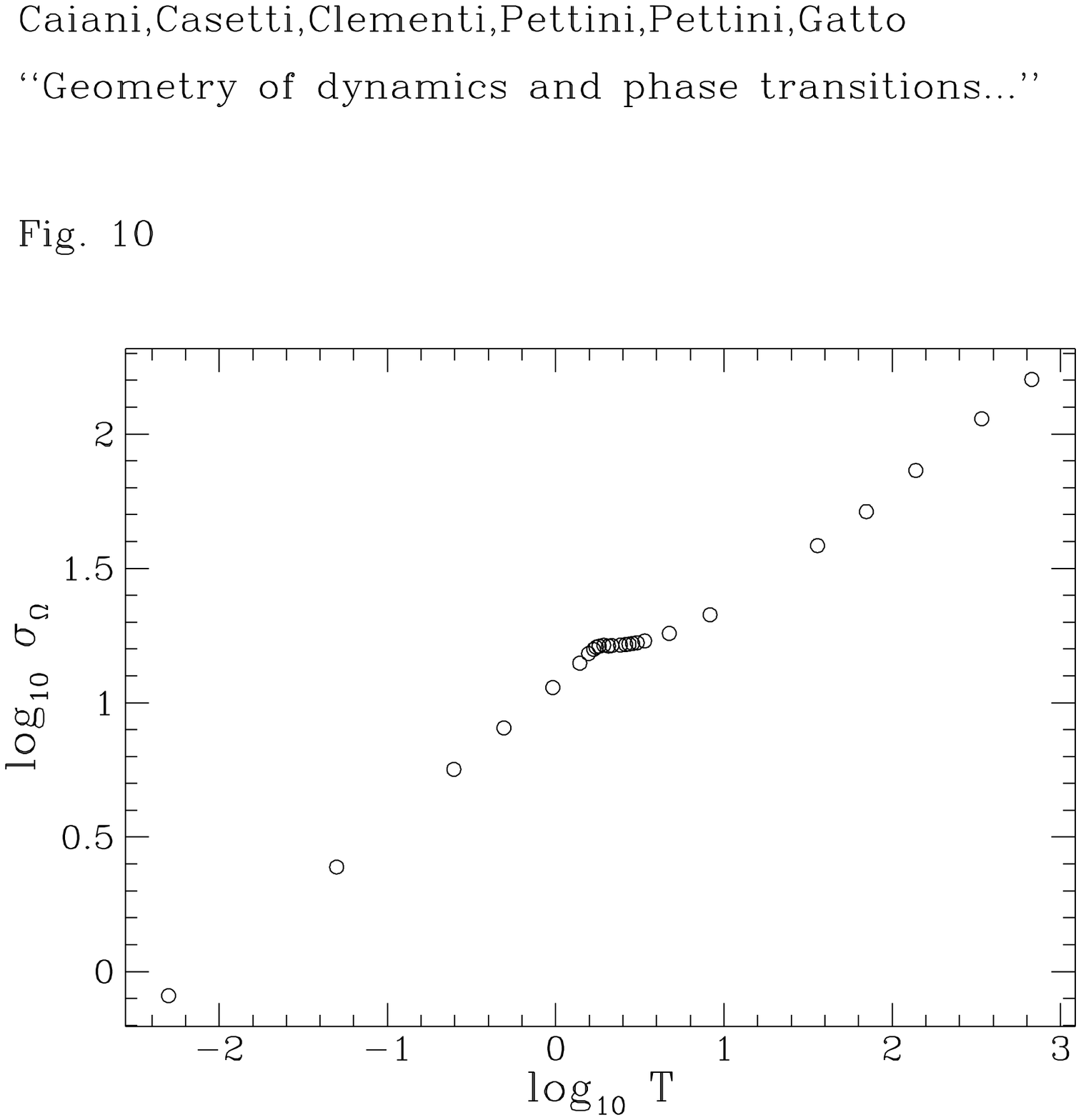,height=8cm,clip=true}}  
\caption{Curvature fluctuations $\sigma_k$ 
{\em vs.} the temperature $T$ for the  
two-dimensional $O(2)$ $\varphi^4$ model, numerically computed  
on a square lattice of $30\times 30$ sites. The critical 
temperature $T_c$ of the 
BKT transition is located at $T_c\simeq 1.5$. 
From Ref.\ \protect\cite{pre98}.} 
\label{fig_fluct_phi4_2_2d} 
\end{figure}

\begin{figure} 
\centerline{\psfig{file=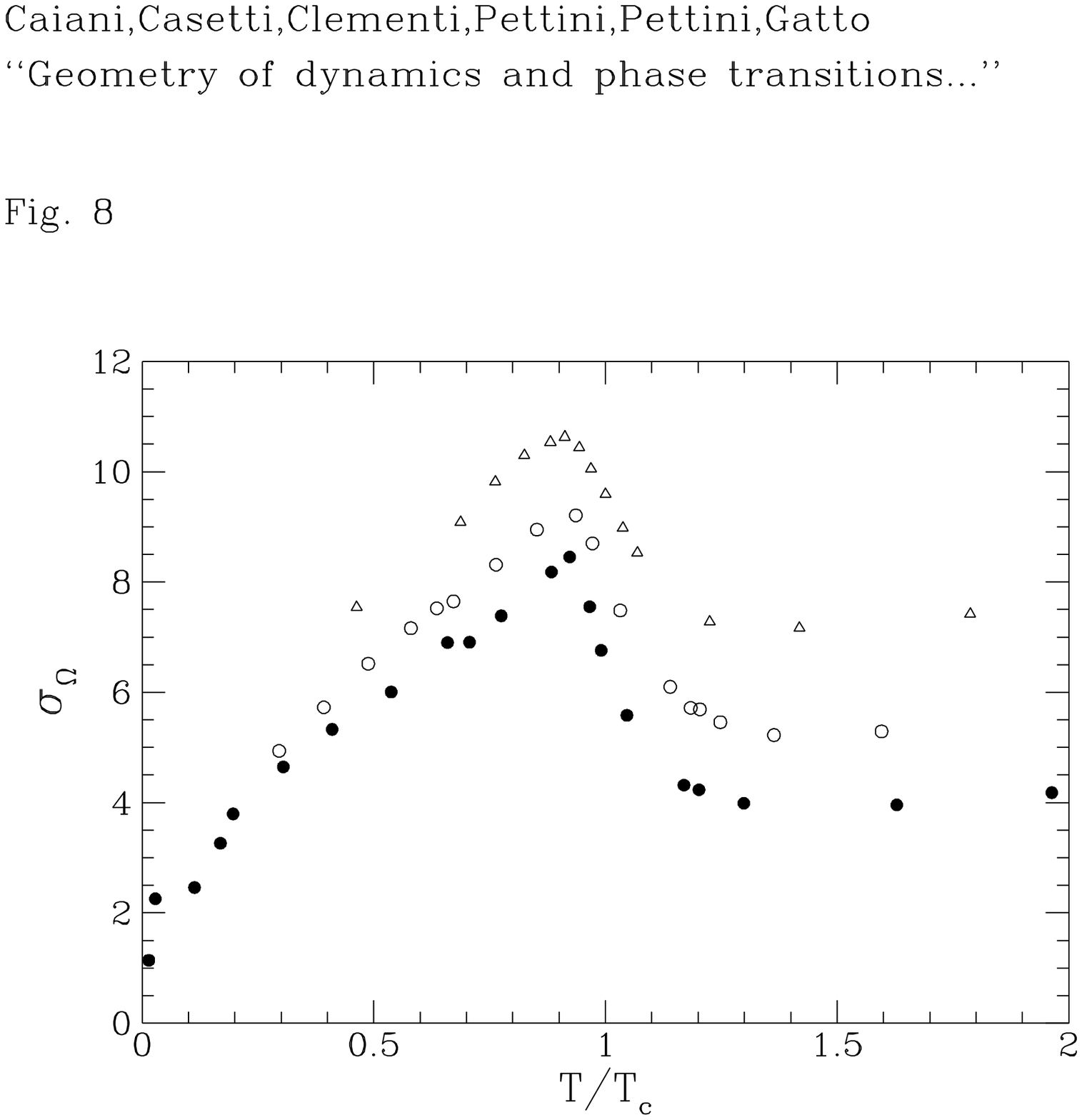,height=8cm,clip=true}}  
\caption{Curvature fluctuations $\sigma_k$   
{\em vs.} the temperature $T$ for the  
three-dimensional $\varphi^4$ model. Full circles correspond to the $O(1)$ 
(scalar) case, open circles to the $O(2)$ case, and open triangles to the 
$O(4)$ case.  
From numerical simulations performed on an 
$8\times 8\times 8$ cubic lattice, reported in Ref.\ \protect\cite{pre98}.} 
\label{fig_fluct_phi4_3d} 
\end{figure}

In the case of the $\varphi^4$ model with $O(n)$ symmetry,  
the Ricci curvature $K_R$ is given by \cite{pre98,jpa98} 
\begin{equation} 
K_R = \sum_{\alpha =1}^n\sum_{i = 1}^N\frac{\partial^2V}{\partial 
(\varphi^\alpha_{i})^2} = Nn(2Jd -r^2) + \lambda({n+2})\sum_{\alpha 
=1}^n 
\sum_{i=1}^N(\varphi^\alpha_{i})^2~. 
\label{riccifi4} 
\end{equation} 
The r.m.s.\ fluctuation of $K_R$, $\sigma_k$, is plotted against the energy 
per degree of freedom, $\varepsilon$, in the case of the two-dimensional 
$O(1)$ $\varphi^4$ model in Fig.\ \ref{fig_fluct_phi4_2d}, 
and against the temperature $T$ in the case of the two-dimensional 
$O(2)$ $\varphi^4$ model in Fig.\ \ref{fig_fluct_phi4_2_2d}, and for  
the three-dimensional 
$O(n)$ $\varphi^4$ models in Fig.\ \ref{fig_fluct_phi4_3d}. 
 
Looking at Figs.\ \ref{fig_fluct_xy_2d}-\ref{fig_fluct_phi4_3d}, one can  
clearly see 
that when a symmetry-breaking phase transition occurs, a
cusp-like (``singular'') 
behavior of the curvature fluctuations is found at the  
phase transition point (Figs.\ \ref{fig_fluct_xy_3d}, \ref{fig_fluct_phi4_2d} 
and \ref{fig_fluct_phi4_3d}), while, when only a BKT transition is present, 
no cusp-like pattern is observed\footnote{Although the cusp-like behaviour 
is lost, indeed some change of behavior is still visible in 
Figs.\ \protect\ref{fig_fluct_xy_2d}  
and \protect\ref{fig_fluct_phi4_2_2d} close to the critical temperature, 
so that a BKT transition appears as ``intermediate'' 
between the absence of a phase transition  and the presence
of a symmetry-breaking  phase transition.} 
(Figs.\ \ref{fig_fluct_xy_2d} 
and \ref{fig_fluct_phi4_2_2d}). We can summarize these results by saying that, 
in general, 
curvature fluctuations always show a cusp-like behavior 
 when a continuous symmetry-breaking phase transition 
is present, and, within numerical accuracy, the cusp occurs 
at the critical temperature. No counter-examples have yet been 
found to this general rule. 
 
The fact that the Lyapunov exponent is sensitive to the phase transition  
can now be understood, in the light of the fact that, as shown in the 
first part of the present Report, chaos can be described geometrically and, 
under rather general assumptions, the Lyapunov 
exponent is closely related to the fluctuations of the Ricci curvature  
(see Eq.\ \ref{lyap_fluct}). However, contrary to the Lyapunov exponent, 
which, although sensitive to the phase transition, is not a good probing 
observable of the presence of a symmetry-breaking phase transition, 
the fluctuation of the Ricci curvature, $\sigma_k$, {\em is} 
a good probing observable, for, as plotted as a function of the temperature,  
it exhibits a clearly peculiar (``cuspy'') 
pattern when a symmetry-breaking phase transition is present, and a  
rather smooth pattern otherwise. The difference among $\lambda(T)$ and 
$\sigma_k(T)$ as probing observables of the phase transition can be  
appreciated by comparing  
Figs.\ \ref{fig_lyap_phi4_3d} and \ref{fig_fluct_phi4_3d}, where $\lambda(T)$ 
and $\sigma_k(T)$ are reported, respectively, for the $O(n)$ $\varphi^4$ 
models. In Fig.\ \ref{fig_lyap_phi4_3d}, the $\lambda(T)$ curves for 
different values of $n$ are {\em qualitatively} different, while in  
Fig.\ \ref{fig_fluct_phi4_3d} the $\sigma_k(T)$ curves 
look strikingly similar, while being clearly different from the curve for 
the 2-$d$ $O(2)$ case (Fig.\ \ref{fig_fluct_phi4_2_2d}), where only a 
BKT transition is present. The same can be said in the case of the 2-$d$
and the 3-$d$ 
   
\subsubsection{Geometric estimate of the Lyapunov exponent}

At this point, it is worthwile to point out that we can apply 
the geometric formula (\ref{lambda_gauss}) for the Lyapunov exponent to
estimate $\lambda$ for all these models, since both $k_0$ and $\sigma_k$
have been numerically computed. As shown in Refs.\ \cite{cccp,pre98,jpa98},
one finds that in general, although the qualitative behavior of 
the Lyapunov exponent is well reproduced, the quantitative 
agreement between the values of $\lambda$
extracted from the numerical simulations and those obtained applying 
Eq.\ (\ref{lambda_gauss}) is {\em not} good, in a neighborhood of the phase 
transition.

However, this is to be expected, because among the assumptions under which 
the formula (\ref{lambda_gauss}) was derived there was the hypothesis that
the fluctuations of the curvature should be not too large, and this
is clearly not true close to a phase transition, as we have just
shown\footnote{The results of the formula (\protect\ref{lambda_gauss}) can be 
improved using procedures which are specific of the model under consideration
and which we are not going to describe here
(see Ref.\ \protect\cite{cccp} for the $XY$ case and 
Ref.\ \protect\cite{pre98} for the $\varphi^4$ case, respectively).}.

\subsection{The mean-field $XY$ model} 
 
The mean-field XY model \cite{Antoni} describes a system of  
$N$ equally coupled planar classical rotators. It is defined by a  
Hamiltonian of the class (\ref{H}) where the potential energy is  
\beq 
V(\varphi) = \frac{J}{2N}\sum_{i,j=1}^N  
\left[ 1 - \cos(\varphi_i - \varphi_j)\right] -h\sum_{i=1}^N \cos\varphi_i ~. 
\label{V_xymf} 
\eeq 
Here $\varphi_i \in [0,2\pi]$ is the rotation angle of the $i$-th rotator  
and $h$ is an external field. Defining at each site $i$ a  
classical spin vector ${\bf s}_i = (\cos\varphi_i,\sin\varphi_i)$ the model  
describes a planar (XY) Heisenberg system with interactions of equal strength  
among all the spins. We consider only the ferromagnetic case $J >0$;  
for the sake of simplicity, we set $J=1$. The equilibrium statistical 
mechanics  
of this system is exactly described, in the thermodynamic limit, by 
mean-field  
theory \cite{Antoni}. In the limit $h \to 0$,  
the system has a continuous phase transition, with classical  
critical exponents, at $T_c = 1/2$, or $\varepsilon_c = 3/4$,  
where $\varepsilon = E/N$ is the energy per particle.  
 
The Lyapunov  
exponent $\lambda$  
of this system is extremely sensitive to the phase transition. In fact, 
according to numerical simulations reported  
in Refs.\ \cite{Ruffo_prl,Ruffo_talk,Yamaguchi,AnteneodoTsallis}, 
$\lambda(\varepsilon)$ is positive for 
$0 < \varepsilon < \varepsilon_c$, shows 
a sharp maximum immediately below the critical energy, and drops to zero at 
$\varepsilon_c$ in the thermodynamic limit, where it remains zero 
in the whole region $\varepsilon > \varepsilon_c$, which corresponds to the 
thermodynamic disordered phase. 
In fact in this phase the system is integrable, 
reducing to an assembly of uncoupled rotators. These results are valid in the 
thermodynamic limit $N\to\infty$ 
in the sense that they have been obtained by estimating the 
infinite $N$ limit of finite $N$ numerical simulations 
\cite{Ruffo_prl,Ruffo_talk}: in the whole region $\varepsilon > \varepsilon_c$ 
the Lyapunov exponent, numerically computed for systems with different  
numbers of particles $N$, behaves as $\lambda \propto N^{-1/3}$, so that it 
extrapolates to zero at $N\to\infty$.  

\begin{figure} 
\centerline{\psfig{file=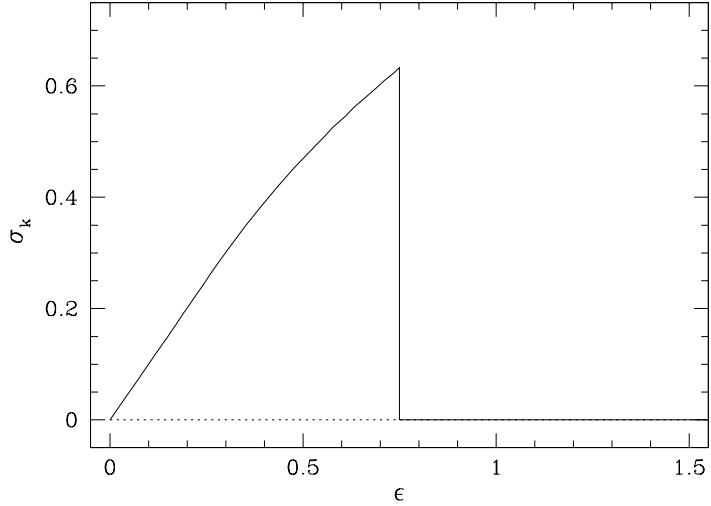,height=8cm,clip=true}} 
\caption{Mean field XY model: analytic expression for the microcanonical 
averages of the Ricci curvature (solid curve) and of its fluctuations 
(dot-dashed curve). 
From Ref.\ \protect\cite{Firpo}.} 
\label{fig_firpo_1} 
\end{figure}

\begin{figure} 
\centerline{\psfig{file=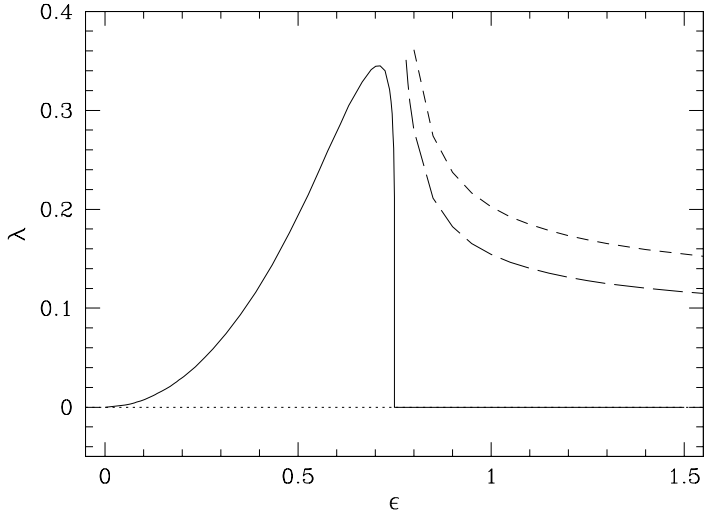,height=8cm,clip=true}} 
\caption{Mean field XY model: analytic expression for the Lyapunov exponent 
(solid curve). The curves above the transition are finite-$N$ results for 
$N=80$ and $N=200$: here $\lambda \propto N^{-1/3}$.  
From Ref.\ \protect\cite{Firpo}.} 
\label{fig_firpo_2} 
\end{figure}

These results have received a theoretical confirmation 
in recent work by M.-C.\ Firpo \cite{Firpo}  
based on the application of the geometric techniques described in the
first part of the present paper. 
Firpo has computed analytically $\langle k_R \rangle$ and  
$\langle \delta^2 k_R \rangle$ in the thermodynamic limit for
the mean-field $XY$ model, showing that such quantities indeed 
have a singular behavior at $\varepsilon_c$ 
(see Fig.\ \ref{fig_firpo_1}). Using these quantities and Eq.\ (\ref{lambda_gauss}),
Firpo has obtained the analytical estimate for 
$\lambda(\varepsilon)$ reported in Fig.\ \ref{fig_firpo_2}; it is remarkable
that also the behavior $\lambda \propto N^{-1/3}$ at 
$\varepsilon > \varepsilon_c$ has been extracted from this 
theoretical calculation. 
This result gives a theoretical confirmation to the qualitative behavior of 
the Lyapunov exponent extrapolated from the numarical simulations.
Moreover, Firpo's analytical results are in good quantitative agreement  
with numerical results reported in Refs.\ \cite{Ruffo_talk,Yamaguchi}, also
close to the phase transition, at variance with the cases of the 
nearest-neighbor $XY$ and $\varphi^4$ models considered earlier.
A tentative explanation of that the application of the geometric formula 
(\ref{lambda_gauss}) gives such good quantitative results in the present case
can be that the mean-field character of the model
prevents the curvature fluctuations from being too wild.

\section {Phase transitions and topology} 
\label{sec_topo} 
In the previous Section we have reported results of 
numerical simulations for the fluctuations of 
observables of a geometric nature (e.g., configuration-space curvature 
fluctuations) related to the Riemannian geometrization of the dynamics in 
configuration space\footnote{More precisely, we considered the enlarged 
configuration space-time, endowed with the Eisenhart metric.}. These
quantities  have been computed, using time averages, 
for many different models undergoing continuous phase transitions, 
namely $\varphi^4$ lattice models with discrete and continuous symmetries 
and  $XY$ models. In particylar, when plotted  
as a function of either the temperature or the energy, the fluctuations 
of the curvature have an apparently singular behavior at the transition 
point.  
Moreover, we have seen that the presence of a singularity in  
the statistical-mechanical fluctuations of the curvature at the  
transition point has  
been proved analytically for the mean-field $XY$ model. 
  
The aim of the present Section is to try to understand 
on a deeper level the origin of this 
peculiar behaviour. In \S \ref{sec_abs_models}, we will show, using  
abstract geometric models, that a singular behaviour in the fluctuations of 
the curvature of a Riemannian manifold can be associated with a change in  
the {\em topology} of the manifold itself. By ``topology change'' 
we mean the following. Let us consider 
a surface ${\cal S}_\varepsilon$ which depends on a parameter  
$\varepsilon$ in such a way that, upon varying the 
parameter, the surface is continuously deformed: as long as the different  
deformed surfaces can be mapped {\em smoothly} one onto  
another\footnote{The different surfaces 
are then said to be diffeomorphic to each other (see Appendix  
\protect\ref{app_morse}).}, the topology 
does not change; on the contrary, the topology changes if there is a 
critical value of the parameter, say $\varepsilon_c$, such that 
the surface ${\cal S}_{\varepsilon > \varepsilon_c}$ can {\em not} any more 
be mapped smoothly onto ${\cal S}_{\varepsilon < \varepsilon_c}$. 
 
The observation that a singularity in the curvature fluctuations of 
a Riemannian manifold, of the same type of those observed numerically 
at phase transitions, can be associated with a change 
in the topology of the manifold, leads us to conjecture that 
it is just this mechanism that
could be at the basis of thermodynamic phase transitions. 
Such a conjecture was 
originally put forward in Ref.\ \cite{cccp} as follows:  
a thermodynamic  
transition might be related to a change in the topology of the configuration  
space, and the observed singularities in the  
statistical-mechanical equilibrium measure and in the thermodynamic 
observables at the phase transition might be interpreted as a ``shadow'' of  
this major topological change that happens 
at a more basic level. We will refer to this conjecture as the {\em 
Topological Hypothesis (TH)}. 
 
The remaining part of the present Section is devoted to a discussion of the 
TH and of its validity. In \S \ref{sec_franz} we will report on a   
purely geometric, and thus still indirect, further indication  
that the topology of the configuration space might change at the 
phase transition, which has been obtained from numerical calculations 
for the $\varphi^4$ model on a two-dimensional lattice 
\cite{Franz,Franz_thesis}. Then, in \S \ref{sec_prl}, we will describe 
a {\em direct} confirmation of the TH, i.e., we will show that 
a topological change in configuration space, which can be related 
with a phase transition,  
indeed occurs in the particular case of the mean-field  
XY model \cite{prl99}. Finally, in \S \ref{sec_TH} we will reformulate 
the TH in a more precise way, taking advantage of the previously 
discussed examples, and \S \ref{sec_open} will be devoted to a general  
discussion of the many points that are still open and on the future 
perspectives of the geometrical and topological approach to 
statistical mechanics. 
 
\subsection{From geometry to topology: abstract geometric models} 
 
\label{sec_abs_models} 
Let us now describe how a singular behavior of the curvature fluctuations 
of a manifold can be put in correspondence with a change in the topology 
of the manifold itself. For the sake of clarity, we shall first discuss 
a simple example concerning two-dimensional surfaces \cite{cccp,pre98},  
and then we will generalize it to  
the case of  
$N$-dimensional hypersurfaces \cite{Franz,Franz_thesis}. 
 
The simple geometric model we are going to describe concerns surfaces 
of revolution. A surface of revolution  ${\cal S} \in {\bf R}^3$ is  
obtained by revolving the graph of a function $f$ around one of the axes 
of a Cartesian plane, and can be defined, in parametric form, 
as follows \cite{Spivak}: 
\beq 
{\cal S} (u,v) \equiv (x(u,v),y(u,v),z(u,v)) = 
(a(u)\cos v, a(u) \sin v, b(u))~, 
\label{def_surface} 
\eeq 
where either $a(u) = f(u)$ and $b(u) = u$, if the graph of $f$ is revolved 
around the vertical axis, or $a(u) = u$ and $b(u) = f(u)$, if the graph 
is revolved around the horizontal axis; in both cases,  
$u$ and $v$ are local coordinates on the surface ${\cal S}$:  
$v\in [0,2\pi]$ and $u$ belongs to the domain of definition of the 
function $f$. 
 
\begin{figure} 
\centerline{\psfig{file=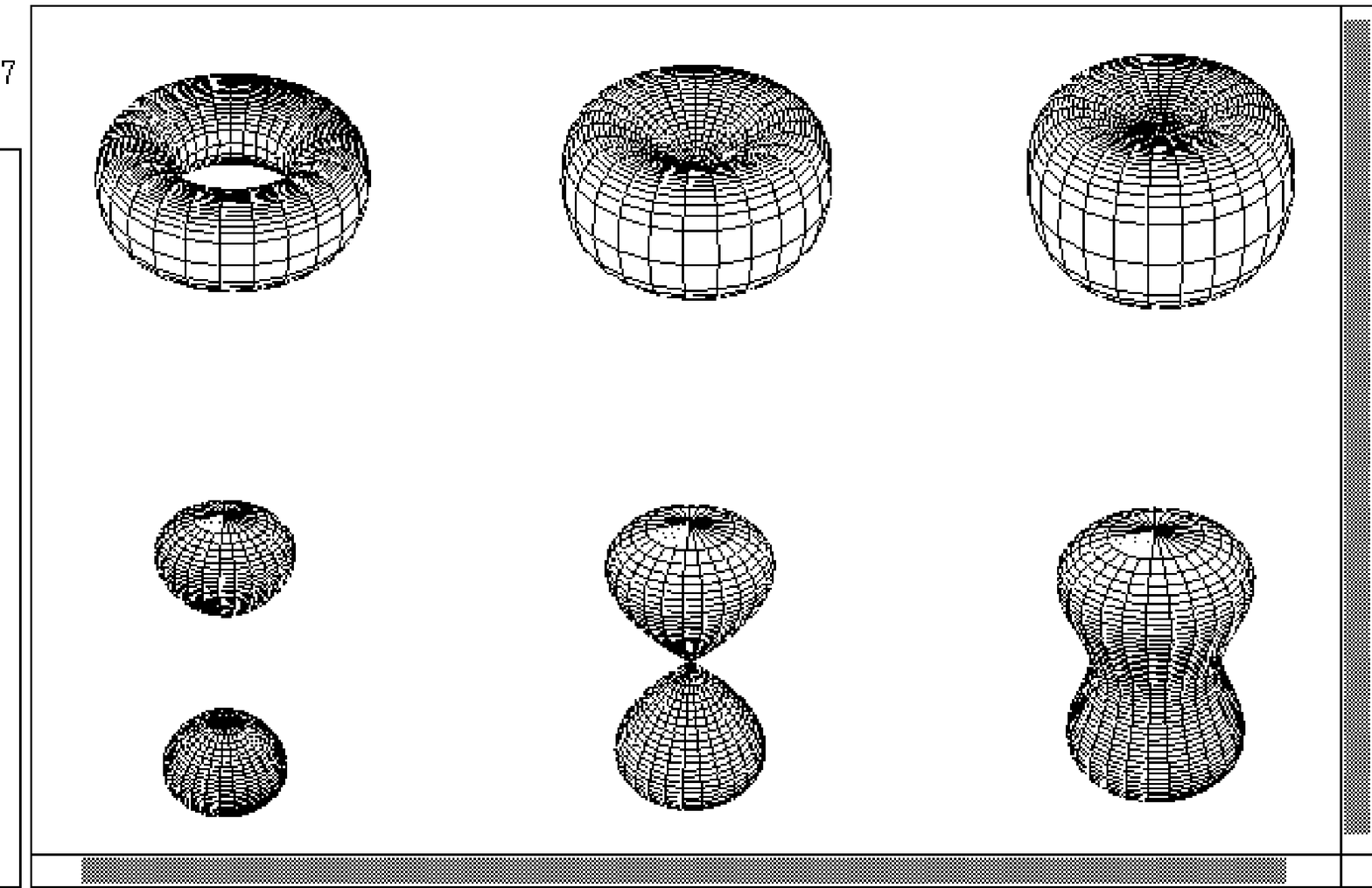,height=8cm,clip=true}} 
\caption{Some representatives of the two families of surfaces 
${\cal F}_\varepsilon$ and ${\cal G}_\varepsilon$ defined  
in Eqs.\ (\protect\ref{F}) and (\protect\ref{G}) respectively. 
Each family is divided into two 
subfamilies by the critical surface corresponding to $\varepsilon_c=0$ (middle 
members in the picture). Members of the same subfamily are diffeomorphic, 
whereas the two subfamilies are not diffeomorphic to each other.   
From Ref.\ \protect\cite{pre98}.} 
\label{figfamilies} 
\end{figure}

Let us consider now in particular
the two families of surfaces of revolution defined as: 
\begin{mathletters} 
\beq 
{\cal F}_\varepsilon   = 
(f_\varepsilon (u) \cos v, f_\varepsilon (u) \sin v, u) ~, 
\label{F} 
\end{equation} 
and 
\begin{equation} 
{\cal G}_\varepsilon  = 
(u \cos v, u \sin v, f_\varepsilon (u))~, 
\label{G} 
\end{equation} 
\end{mathletters} 
where 
\begin{equation} 
f_\varepsilon (u) = \pm \sqrt{\varepsilon + u^2 - u^4}~,\qquad 
\varepsilon \in [\varepsilon_{\rm min},+\infty)~, 
\end{equation} 
and $\varepsilon_{\rm min} = -\frac{1}{4}$. Some cases 
are shown in Fig.\ \ref{figfamilies}.  
 
There exists for  both families of surfaces a 
critical value of  $\varepsilon$, $\varepsilon_c = 0$, 
corresponding to a change in the {\em topology} of the surfaces: 
the manifolds ${\cal F}_\varepsilon$ 
are diffeomorphic to a torus ${\bf T}^2$ for 
$\varepsilon < 0$ and to a sphere ${\bf S}^2$ for 
$\varepsilon > 0$; the manifolds 
${\cal G}_\varepsilon$ are diffeomorphic to {\em two} spheres 
for $\varepsilon < 0$ 
and to one sphere for $\varepsilon > 0$. The Euler-Poincar\'e 
characteristic (see Appendix \ref{app_morse}, Eq.\ \ref{chi}) is 
$\chi({\cal F}_\varepsilon)  =  0$ if $\varepsilon < 0$, and 
$\chi({\cal F}_\varepsilon)  =  2$ otherwise, while 
$\chi({\cal G}_\varepsilon)$ is  $4$ or $2$ for $\varepsilon$ 
negative or positive, respectively.  

\begin{figure} 
\centerline{\psfig{file=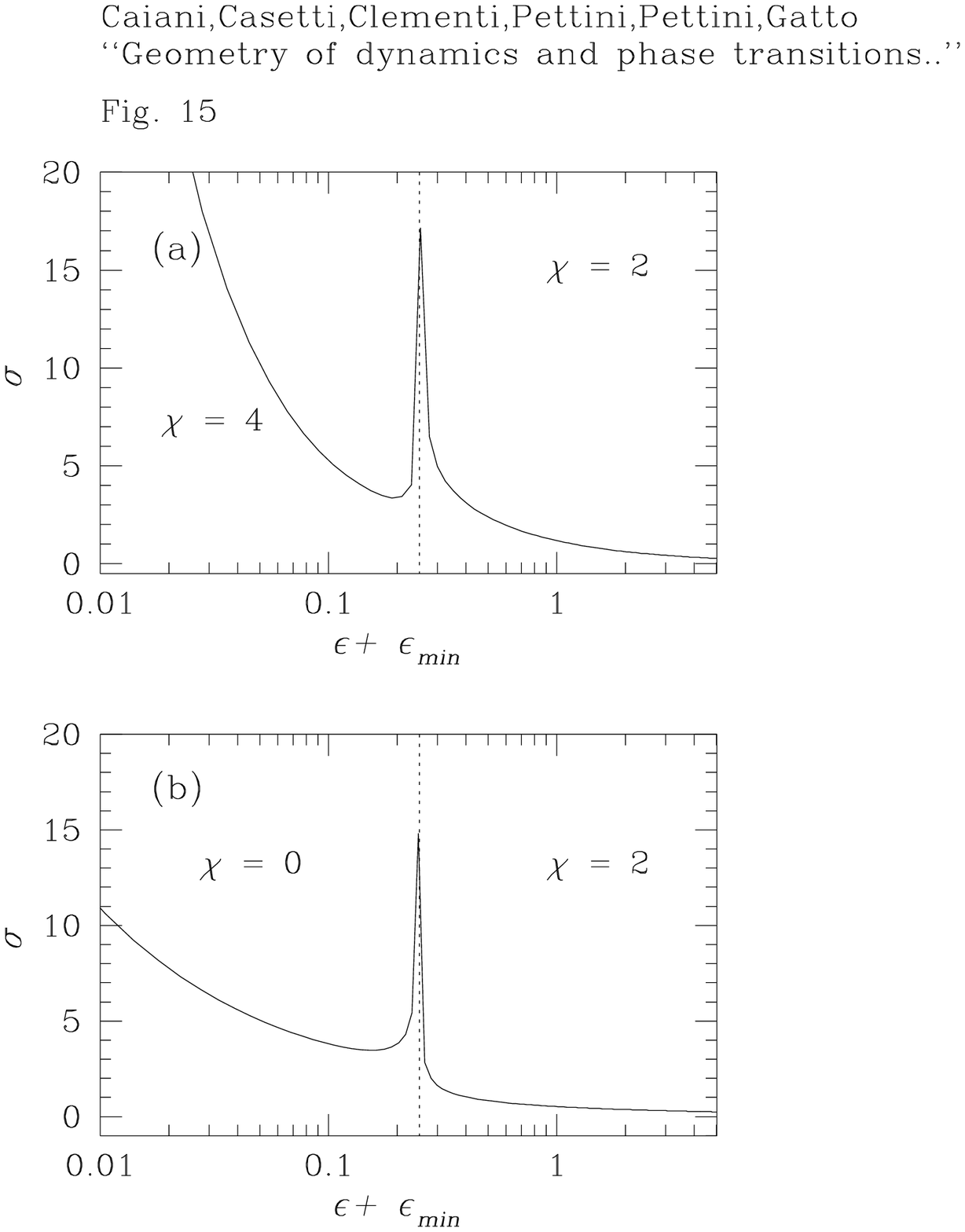,height=13cm,clip=true}} 
\caption{The fluctuation $\sigma_K$ of the gaussian curvature of the surfaces 
${\cal F}_\varepsilon$ and ${\cal G}_\varepsilon$ is plotted {\it vs} 
$\varepsilon$. 
$\sigma$ is defined in Eq.\ (\protect\ref{rms_modelli}), 
$\varepsilon$ is shifted 
by $\varepsilon_{\rm min}=0.25$ for reasons of clarity
of presentation.  
{\it (a)} refers to ${\cal G}_\varepsilon$ and  
{\it (b)} refers to ${\cal F}_\varepsilon$. 
The cusps 
appear at $\varepsilon =0$ where the topological transition takes place
for both ${\cal F}_\varepsilon$ and ${\cal G}_\varepsilon$. 
From Ref.\ \protect\cite{pre98}.} 
\label{figmodels} 
\end{figure}

We now turn to the definition and the calculation of the curvature 
fluctuations on these surfaces.
Let $M$ belong to one of the two families; 
its gaussian curvature $K$ is  \cite{Spivak} 
\begin{equation} 
K= {{a^\prime (a^{\prime\prime}b^\prime - b^\prime a^{\prime\prime})}\over 
{a (b^{\prime 2} + a^{\prime 2})^2}} 
\label{curvgauss} 
\end{equation} 
where  $a(u)$ and $b(u)$ are the coefficients of Eq.\ (\ref{def_surface}), and 
primes denotes differentiation, with respect to $u$. 
The fluctuations of $K$ can be then defined as
\begin{equation} 
\sigma_K^2 = \langle K^2 \rangle - \langle K \rangle^2 = 
A^{-1}\int_M K^2 \, dS - \left(A^{-1} \int_M K \, dS\right)^2~, 
\label{rms_modelli} 
\end{equation} 
where $A$ is the area of $M$ and $dS$ is the invariant surface element. 
Both families of surfaces exhibit a singular behavior in  $\sigma_K$ as 
$\varepsilon \to \varepsilon_c$, as shown in Fig.\ \ref{figmodels}, 
in spite of 
their different curvature 
properties on the average\footnote{For instance, 
$\langle K \rangle(\varepsilon) = 0$ for 
${\cal F}_\varepsilon$  as $\varepsilon < 0$, 
while for ${\cal G}_\varepsilon$ the same average curvature is positive and 
diverges as $\varepsilon \to 0$.}. 
 
We are now going to 
show that the result we have just obtained for two-dimensional 
surfaces has a much more general validity: a {\em generic} topology change 
in an $n$-dimensional manifold is accompanied by a 
singularity in its curvature fluctuations \cite{Franz}. 
In order to do that, we have to
make use of some concepts belonging to Morse theory,  
which will also be used in \S \ref{sec_prl} below; the main concepts 
of Morse theory are sketched in Appendix \ref{app_morse}, where also  
references to the literature are given. 
 
\begin{figure} 
\centerline{\psfig{file=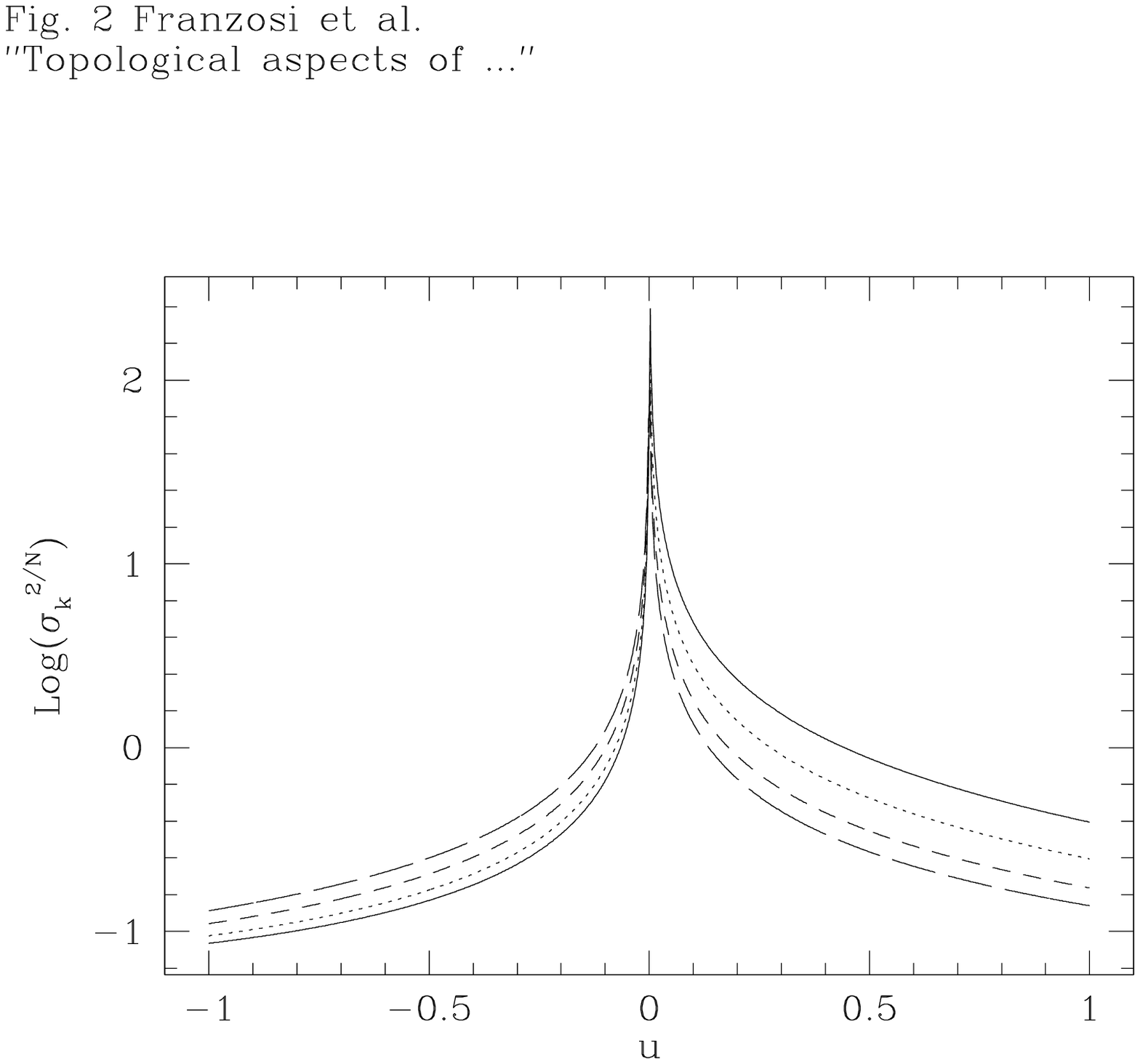,height=8cm,clip=true}}  
\caption{Fluctuations of the 
Gauss curvature of a hypersurface $f^{-1}(u)$ of ${\bf R}^N$ 
{\it vs}. $u$ close to a critical point. 
$\sigma^{2/N}_K$ is reported because it has the same dimensions of
the 
scalar curvature. Here $\dim (f^{-1}(u))=100$,  
and the 
Morse indexes are: $k=1,15,33,48$, represented by solid, dotted, dashed,   
long-dashed lines respectively. } 
\label{fig_franz}  
\end{figure}

We consider then a hypersurface of ${\bf R}^N$ which is the $u$-level set 
of a function $f$ defined in ${\bf R}^N$, i.e., 
a submanifold of ${\bf R}^N$ of dimension $n = N-1$ defined by the equation 
\beq 
f(x_1,\ldots,x_N) = u~; 
\eeq 
such a hypersurface can then be referred to as $f^{-1}(u)$. Let us now  
assume\footnote{This is not a strong assumption: in fact, it can be shown 
that Morse functions are generic (see Appendix \protect\ref{app_morse}).} 
that $f$ is a {\em Morse function}, i.e., such 
that its critical points (i.e., the points of ${\bf R}^N$  
where the differential $df$ 
vanishes) are isolated. One of the most important results of Morse theory 
is that the topology of the hypersurfaces $f^{-1}(u)$ can change {\em only} 
crossing a critical level $f^{-1}(u_c)$, i.e., a level set containing  
at least one critical point of $f$. This means that a generic change  
in the topology of the hypersurfaces can be associated with critical 
points of $f$. 
Now, the hypersurfaces $f^{-1}(u)$ can 
be given a Riemannian metric in a standard way \cite{Thorpe}, and  
it is possible to analyze the  
behavior of the curvature fluctuations in a neighborhood of 
a critical point. Let us assume, for the sake of simplicity, that  
this critical point is   
located at $x_0 = 0$ and belongs to the level $u_c = 0$.  
Any   
Morse function can be parametrized, in the neighborhood of a $x_0$,   
by means of the so-called Morse chart, 
i.e., a system of local coordinates $\{y_i\}$ such that  
$f(y)=f(x_0)-\sum_{i=1}^k y_i^2+\sum_{i=k+1}^N y_i^2$ ($k$ is the Morse index 
of the critical point). 
Then standard formul{\ae} for the Gauss curvature $K$ of  
hypersurfaces of ${\bf R}^N$ 
\cite{Thorpe} can be used to compute explicitly the fluctuations
of the curvature, $\sigma_K$, of the
level set $f^{-1}(u)$.
Numerical results for the curvature
fluctuations are reported in Fig.\ \ref{fig_franz} and  
show that also at high dimension 
$\sigma_K^2$ develops a sharp, singular peak as the critical surface 
is approached 
(for computational details, see Ref.\ \cite{Franz_thesis}). 
 
\subsection{Topology changes in configuration space and phase transitions} 
 
As we have discussed in Section \ref{sec_geopt}, the curvature  
fluctuations of the configuration space exhibit cusp-like 
patterns in presence of a second order phase 
transition. A  truly cuspy pattern, i.e., an analytic discontinuity, is 
mathematically proven in the case of mean-field $XY$ model. 
In \S \ref{sec_abs_models}, we have shown that singular patterns in the  
fluctuations of the curvature of a Riemannian manifold can be seen 
as consequences of the presence of a topology change.  
Hence, we are led to the Topological Hypothesis (TH), i.e., to conjecture 
that at least continuous, symmetry-breaking 
phase transitions are associated with topology changes in the
configuration space of the system. 
 
However, an important question arises, in that 
the fluctuations of the curvature considered in Sec.\ \ref{sec_geopt} 
have been  
obtained as time averages, computed along the dynamical trajectories of the  
Hamiltonian systems under investigation (or as statistical averages  
computed analytically, as in the 
case of the mean-field $XY$ model).  
Now, time averages of geometric observables  
are usually found to be in excellent agreement with ensemble averages  
\cite{pre93,pre96,cccp,pre98,jpa98}  
so that one could argue that the  
above mentioned singular-like patterns 
of the fluctuations of geometric observables are simply 
the precursors of truly 
singular patterns due to the fact that the measures of all the statistical  
ensembles tend to become singular in the limit $N\rightarrow\infty$ when a  
phase transition is present. In other words, geometric observables, like any  
other ``honest'' observable, already at finite $N$ would feel the eventually 
singular character of the statistical measures, i.e., of the
probability distribution functions of the statistical-mechanical 
ensembles. If this were the correct
explanation, we could not attribute  
the cusp-like patterns of the curvature fluctuations to 
any special geometric features  
of configuration space, and the cusp-like patterns observed in 
the numerical simulations could not be considered as (indirect) 
confirmations of the TH. 
 
In order to elucidate this important point, three different paths have been 
followed: $(i)$  
{\em purely geometric} information about certain  submanifolds of  
configuration space has been worked out {\it independently} of the  
statistical measures in the case of the two-dimensional $\varphi^4$ 
model, and the results lend  
{\em indirect} support to the TH \cite{Franz}; $(ii)$ a {\em direct} 
numerical confirmation of the TH has been given in \cite{fps1} by means
of the computation of a topologic invariant, the Euler characteristic,
in the case of a $2d$ lattice $\varphi^4$ model; 
 $(iii)$ a {\em direct} analytic 
confirmation of the TH has been found in the particular case of the  
mean-field $XY$ model \cite{prl99}.  We report on items 
$(i)$, $(ii)$ and $(iii)$ in \S \ref{sec_franz}, in \S \ref{sec_euler} 
and in \S \ref{sec_prl}, respectively. 
 
\subsubsection{Indirect numerical  
investigations of the topology of configuration space} 
 
\label{sec_franz} 
In order to separate the singular effects due to the singular character of
statistical measures at a phase transition from the singular effects due
to some topological transition in configuration space, the first natural
step is to consider again $\sigma^2_K$ as observable, and to compute its
integral value on suitable submanifolds of configuration space by means of
a geometric measure, i.e. by means of a measure which has nothing to do with
statistical ensemble measures.
   
Consider as ambient space the $N$-dimensional configuration space 
$M$ of a Hamiltonian system 
with $N$ degrees of freedom, which, when $N\to\infty$, 
undergoes a phase transition at a certain finite temperature $T_c$ (or  
critical energy 
per degree of freedom $\varepsilon_c$), and let ${\cal V} = 
V(\varphi)$ be its potential energy per degree of freedom. 
 
Then the relevant geometrical objects are the submanifolds of $M$ defined by 
\beq 
M_u = {\cal V}^{-1} (-\infty,u] =  
\{ \varphi \in M : {\cal V}(\varphi) \leq u\}~, 
\label{M_v} 
\eeq 
i.e., each $M_u$ is the set $\{\varphi_i\}_{i=1}^N$ such that the potential  
energy per particle does not exceed a given value $u$.  
As $u$ is increased from $-\infty$ to $+\infty$, this 
family covers successively the whole manifold $M$.  
All the submanifolds $M_u$ can be given a Riemannian  
metric $g$  
whose choice is largely arbitrary.  
On all these manifolds $(M_u,g)$ there is a standard 
invariant volume measure:  
\beq 
d \eta =\sqrt{{\rm det}(g)}\,dq^1\dots dq^N~, 
\eeq 
 which 
has nothing to do with statistical measures. Let us finally 
define the hypersurfaces $\Sigma_u$ as the $u$-level sets of $\cal V$, i.e., 
\beq 
\Sigma_u = {\cal V}^{-1} (u)~,
\label{hypsurf} 
\eeq 
which are nothing but the boundaries of the submanifolds $M_u$. 

According to the discussion reported in \S \ref{sec_abs_models},  
an indirect way to study the presence of topology changes in 
the family $\{(M_u,g)\}$ is to look at the behaviour of the  
fluctuations of the Gaussian curvature, $\sigma_K^2$, defined 
as 
\beq 
\sigma_K^2=\langle K_G^2\rangle_{\Sigma_u} - 
\langle K_G\rangle_{\Sigma_u}^2 
\eeq 
where $\langle\cdot\rangle$ stands for 
integration over the surface $\Sigma_u$, as a function of $u$.  
The presence of cusp-like singularities of $\sigma_K^2$ for some 
critical value of $u$, $u_c$, would eventually signal the presence of a 
topology change of the family $\{(M_u,g)\}$ at $u_c$. 
Such an indirect geometric probing of the presence of critical points seems 
an expedient way to probe the possible topology changes of the manifolds 
$(M_u, g)$. 
In fact, the properties of the manifolds $M_u$ are closely related to 
those of the hypersurfaces $\{\Sigma_u\}_{u\leq u_c}$, as can be inferred 
from the equation  
\beq 
\int_{M_u}f d\eta =\int_0^u\,dv\int_{\Sigma_v}f\vert_{\Sigma_v}  
d\omega/\Vert\nabla V\Vert 
\eeq 
where $d\omega$ is the induced measure\footnote{If a surface is parametrically
defined through the equations $x^i=x^i(z^1,\dots ,z^k),~i=1,\dots ,N$ then
the metric $g_{ij}$ induced on the surface is given by $g_{ij}(z^1,\dots ,z^k)
=\sum_{n=1}^N\frac{\partial x^n}{\partial z^i}\frac{\partial x^n}{\partial z^j}
$.} on $\Sigma_u$ and $f$ a generic  function \cite{Federer}.   
From Morse theory (see Appendix B) we know that the surface $\Sigma_{u_c}$ 
defined by ${\cal V}=u_c$ is a degenerate quadric, 
so that in its vicinity some of the principal curvatures \cite{Thorpe}  
of the surfaces $\Sigma_{u\simeq u_c}$ tend to 
diverge\footnote{The principal curvatures are the inverse of the curvature 
radii measured, at any given point of a surface, in suitable directions. At a
Morse critical point some of these curvature radii vanish.}. 
Such a divergence is generically detected by any function of the  
principal curvatures and thus, for 
practical computational reasons, instead of the Gauss curvature (which is the  
product of all the principal curvatures) 
we shall consider the total second variation of the {\it scalar} curvature  
${\cal R}$ (i.e., the sum of all the possible 
products of two principal curvatures) of the manifolds $(M_u, g)$, according 
to the definition  
\begin{equation} 
\sigma_{\cal R}^2(u) = [V\!ol(M_u)]^{-1} \int_{M_u}d\eta \left[ {\cal R} -  
[V\!ol(M_u)]^{-1} \int_{M_u}d\eta \, {\cal R} \right]^2  
\label{sigma2R}  
\end{equation} 
with ${\cal R}= g^{kj}R^l_{klj}$, where $R^l_{kij}$ are the components of the 
Riemann curvature tensor [see Appendix \ref{app_geo},  
Eq.(\ref{curv_components})] 
and $V\!ol(M_u)= \int_{M_u}d\eta$.  
The subsets $M_u$ of configuration space are given the structure of  
Riemannian manifolds $(M_u,g)$ by endowing all of them with the {\it same}  
metric tensor $g$. However, the choice of the metric $g$ is arbitrary in 
view of probing possible effects of the topology on the geometry of these 
manifolds. 
 
In Ref.\ \cite{Franz} the configuration spaces of a $\varphi^4$ model, 
defined on a one-dimensional lattice (linear chain) and on a  
two-dimensional square lattice have been considered. We 
recall that in the 2-$d$ case this system undergoes a phase transition at 
a finite temperature, while in the 1-$d$ case no phase transition is 
present. Three different types of metrics have been considered, i.e., 
\begin{itemize} 
\item[{$(i)$}] $g^{(1)}_{\mu\nu}=[A - V(\varphi)]\delta_{\mu\nu}$, i.e., 
a conformal 
deformation (see Appendix \ref{app_geo}, \S \ref{app_sec_curvature}) 
of the Euclidean flat metric $\delta_{\mu\nu}$, where  
$A>0$ is an arbitrary constant chosen large enough to be sure that in the
relevant interval of values of $u$ the determinant of the metric is always
positive;   
\item[{$(ii)$}]  $g^{(2)}_{\mu\nu}$ and  
$g^{(3)}_{\mu\nu}$ are generic metrics (no longer conformal deformations of
the flat metric) defined by 
\begin{equation} 
(g_{\mu\nu}^{(k)})=\left(
\begin{array}{ccc}
f^{(k)}&0&1 \\
0&{\bf I}&0 \\ 
1&0&1 
\end{array}
\right)~~,~~~~k=2,3 
\label{metriche} 
\end{equation} 
where ${\bf I}$ is the $N-2$ dimensional identity matrix,  
$g^{(2)}$ is obtained by setting  
$f^{(2)}=\frac{1}{N}\sum_{\alpha\in{\bf Z}^d}\varphi_\alpha^4 + A$,  
and $g^{(3)}$ by setting $f^{(3)}=\frac{1}{N}\sum_{\alpha\in{\bf Z}^d} 
\varphi_\alpha^6 + A$, with $A>0$, and $\alpha$ labels the $N$ lattice sites 
of a linear chain ($d=1$) or of a square lattice ($d=2$, $N=n\times n$). 
\end{itemize}
These choices are completely arbitrary, however, and only if metrics of very
simple form are chosen, both analytical and numerical computations are
feasible also for rather large values of $N$. Thus the first metric has been
chosen diagonal, and the other two metrics concentrate in only one matrix 
element all the non-trivial geometric information. Moreover, the first metric
still contains a reference to the physical potential, whereas the other two
define metric structures that are completely independent of the physical
potential and only contain monomials of powers sufficiently high that they do
not vanish after two successive derivatives have been taken 
(needed to compute curvatures).
The topology of the subsets of points $M_u$ and $\Sigma_u$ of ${\bf R}^N$
is already determined (though well concealed) by the definitions of 
Eqs.(\ref{M_v}) and 
(\ref{hypsurf}); the task is to ``capture'' some information about their 
topology through a mathematical object or structure, defined on these
sets of points, which is capable of mirroring the variations of topology 
through the
$u$-pattern of an analytic function. This idea follows the philosophy of
standard mathematical theories of differential topology, for example, within 
Morse theory, the information about topology is extracted through the 
critical points of any function -- defined on a given manifold -- fulfilling 
some conditions (necessary to be a good Morse 
function), or, whithin cohomology theory \cite{Nakahara}, 
topology is probed through vector 
spaces of differential forms (the de Rham's cohomology vector spaces)
``attached'' to a given manifold.
Provided that good mathematical quantities are chosen as 
topology-variation detectors, arbitrary 
Riemannian metric structures could work as well.

For the above defined metrics $g^{(k)},~k=1,2,3$, simple algebra leads from
the definition of the scalar curvature (see Appendix \ref{app_geo}) to the
following explicit expressions:
\begin{eqnarray} 
{\cal R}^{(1)}&=&(N-1)\left[\frac{\triangle V}{(A-V)^2} - \frac{\Vert\nabla  
V\Vert^2}{(A-V)^3}\left(\frac{N}{4}-\frac{3}{2}\right)\right]  
\label{R1} \\ 
{\cal R}^{(k)}&=&\frac{1}{(f^{(k)}-1)}\left[ \frac{\Vert{\tilde\nabla}  
f^{(k)}\Vert^2}{2(f^{(k)}-1)} -{\tilde\triangle} f^{(k)} \right]~,~~~k=2,3  
\label{R23} 
\end{eqnarray} 
where $\nabla$ and $\triangle$  
are the euclidean gradient and laplacian respectively, and $\tilde\nabla$ 
and $\tilde\triangle$ lack the derivative $\partial /\partial  
\varphi_\alpha$ with $\alpha =1$ in the $d=1$ case, and lack the derivative 
$\partial /\partial\varphi_\alpha$ with $\alpha =(1,1)$ in the $d=2$ case.
 
\begin{figure} 
\centerline{\psfig{file=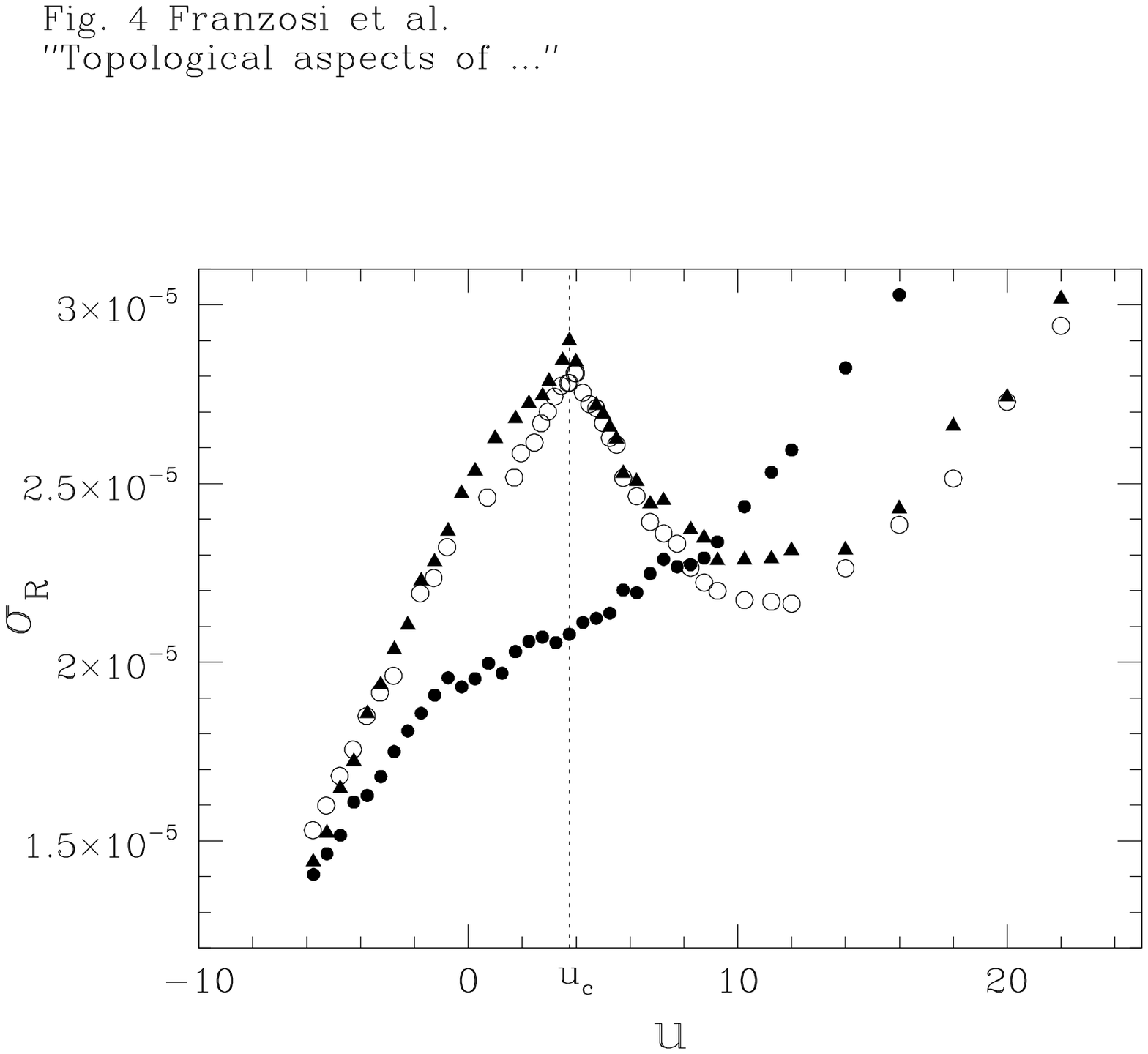,height=8cm,clip=true}}  
\caption{Variance of the scalar curvature of $M_u$ {\it vs} $u$ computed with 
the metric $g^{(1)}$. Full circles 
correspond to the $1d$-$\varphi^4$ model with $N=400$. Open circles refer to 
the $2d$-$\varphi^4$ model with $N=20\times 20$ lattice sites, and full  
triangles refer to $40\times 40$ lattice sites (whose values are rescaled for 
graphic reasons). } 
\label{fig_franz_1}  
\end{figure}

\begin{figure} 
\centerline{\psfig{file=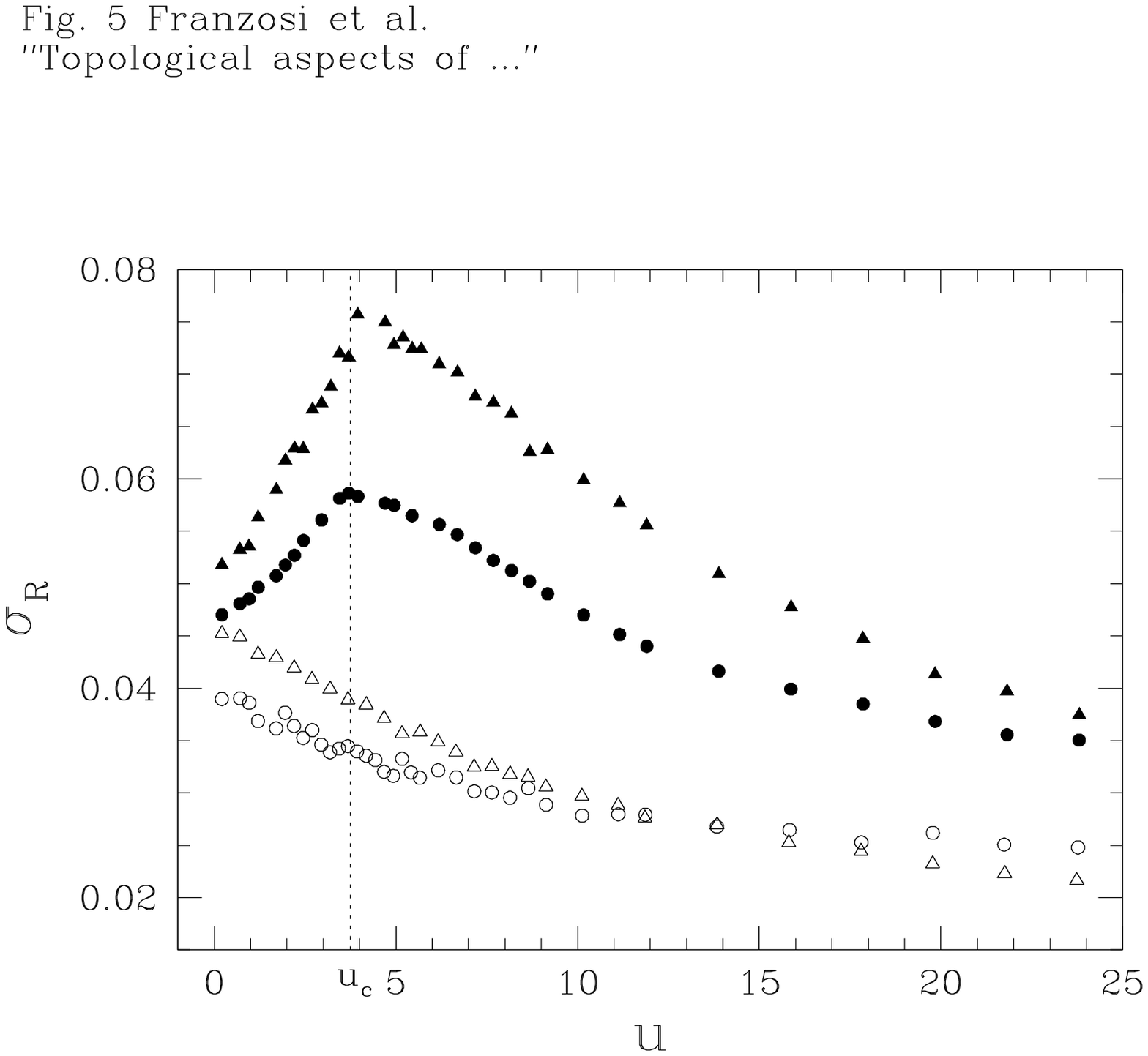,height=8cm,clip=true}}  
\caption{$\sigma_{\cal R}^2(u)$ of $M_u$ {\it vs} $u$ computed for 
the $\varphi^4$ model with: metric $g^{(2)}$ in $1d$, $N=400$ 
(open triangles); 
metric $g^{(2)}$ in $2d$, $N=20\times 20$ (full triangles); 
metric $g^{(3)}$ in $1d$, $N=400$ (open circles); metric $g^{(3)}$ in $2d$,  
$N=20\times 20$ (full circles). } 
\label{fig_franz_2}  
\end{figure}

The numerical computation of the geometric integrals in Eq.(\ref{sigma2R})
is worked out by means of a  MonteCarlo algorithm 
\cite{lapo_thesis,Franz_thesis} to sample the geometric measure $d\eta$ 
by means of an ``importance sampling'' algorithm \cite{Binder} suitably 
modified. 
In Figs.\ \ref{fig_franz_1} and \ref{fig_franz_2}  
$\sigma_{\cal R}^2(u)$ are given 
for the one and two dimensional cases obtained for two different lattice sizes 
with $g^{(1)}$ (Fig.\ \ref{fig_franz_1}), and at given lattice size with 
$g^{(2,3)}$ (Fig.\ \ref{fig_franz_1}).  
Peaks of $\sigma_{\cal R}^2(u)$ appear at a certain value $v_c$, of $v_c$ in 
the two-dimensional case, whereas only smooth patterns are found in the 
one-dimensional case, where no phase transition is present.  
 
\begin{figure} 
\centerline{\psfig{file=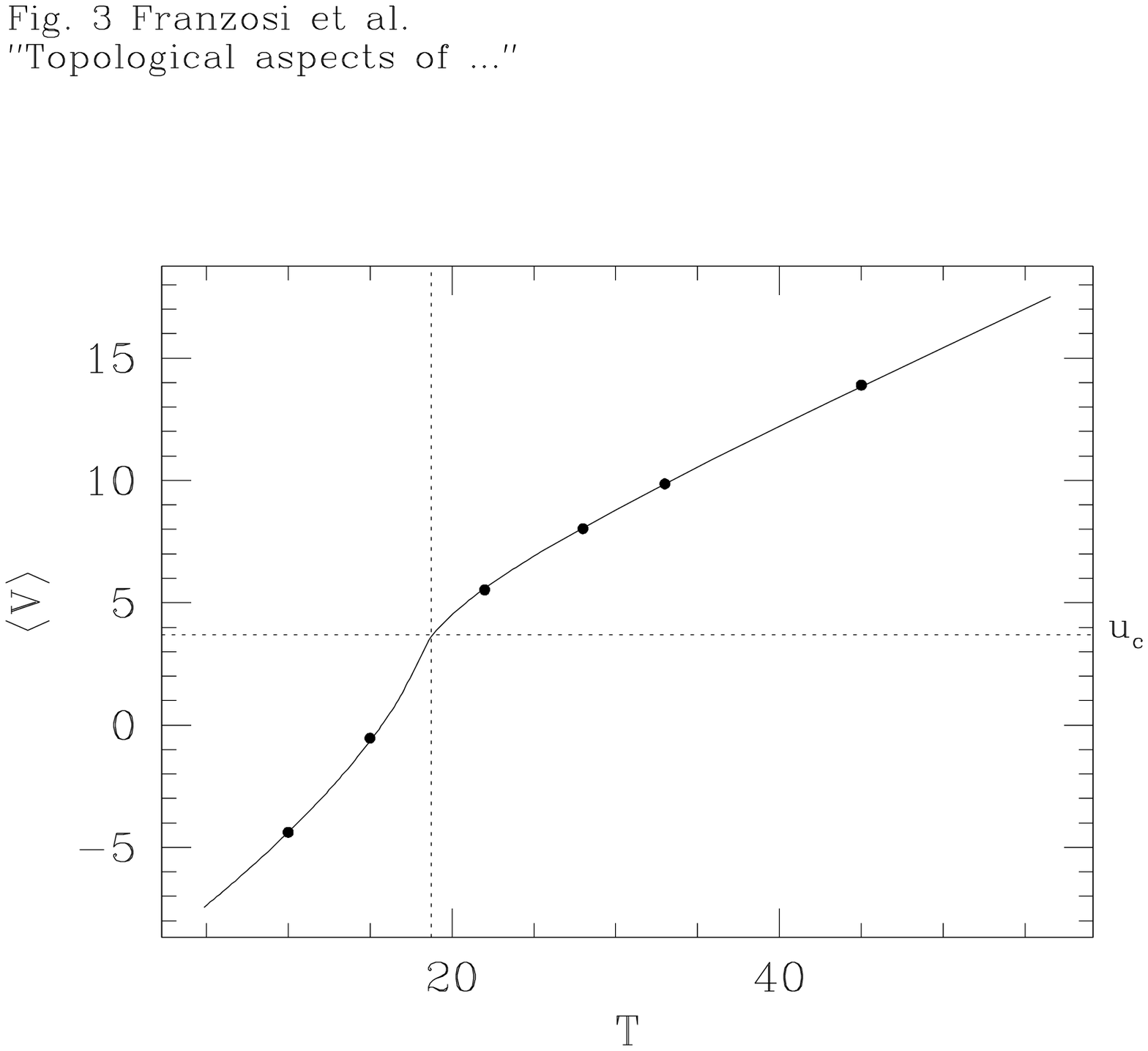,height=8cm,clip=true}}  
\caption{Average potential energy {\it vs} temperature for the 2-$d$ lattice 
$\varphi^4$ model with $O(1)$ symmetry.
Lattice size $N=20\times 20$. The solid line is made out of 
200 points obtained as time averages. Full circles represent MonteCarlo 
estimates of canonical ensemble averages. The dotted lines locate the phase 
transition.} 
\label{fig_franz_3}  
\end{figure}

According to the discussion above, these peaks can be considered as 
indirect evidence of the presence of a topology transition  
in the manifolds $M_u$ at $u = u_c$ in the case 
of the two-dimensional $\varphi^4$ model. It is in particular the persistence
of cusp-like patterns of $\sigma_{\cal R}^2(u)$ independently of the metric
chosen that lends credit to the idea that this actually  reflects a
topological transition. Now we want to argue 
that the topological transition occurring at  
$u_c$ is related to a {\it thermodynamic phase transition} which occurs
in the $\varphi^4$ model.  
In order to do that, in Ref.\ \cite{Franz} the  
average potential energy per particle 
\beq 
u(T) = \langle {\cal V} \rangle  
\eeq  
has been numerically computed, as a function of $T$, by means of both 
MonteCarlo 
averaging with the canonical configurational measure, and Hamiltonian 
dynamics. 
In the latter case the temperature $T$ is given by the average kinetic 
energy per degree of freedom, and $u$ is obtained as 
time average. 
Fig.\ \ref{fig_franz_3} shows a perfect 
agreement between time and ensemble averages. The phase transition 
point is well visible at $u = u_c \simeq 3.75$. Looking at  
Figs.\ \ref{fig_franz_1} and \ref{fig_franz_2}, we realize that, within 
the numerical accuracy, 
the critical value of the potential energy  
per particle $u_c$ where the  
topological change occurs equals the statistical-mechanical  
average value of the potential energy at the phase transition. 

At this point the doubt, formulated at the beginning of this Section, about
the possible non-geometrical origin of the ``singular'', cusp-like patterns 
of $\sigma_{\cal R}^2(u)$ has been dissipated. These results have been found
{\it independently} of statistical mechanical measures and of their singular 
character in presence of a phase transition. These results are also 
{\it independent} -- at least to the limited extent of the three metric
tensors reported above --  of the geometric structure given to the family 
$\{M_u\}$.
Thus they seem most likely to have their origin at a deeper 
level than the geometric one, i.e. at the topologic level. 
Hence the observed phenomenology strongly hints that some 
{\it major} change in the topology of the configuration-space-submanifolds 
$\{M_u\}$ occurs when a second-order phase transition takes place.  

\subsubsection{Direct numerical investigation of the topology of configuration
space}
\label{sec_euler}
Though still based on numerical computations for a special model, a 
{\it direct} evidence of the tight relation between topology and phase 
transitions has been obtained by computing the $u$-dependence of a topologic 
invariant of the
leaves $\Sigma_u$ in the foliation of configuration space into a family of
equipotential surfaces.

In order to directly probe if and how the topology change -- in the sense of
a breaking of {\it diffeomorphicity} of the surfaces $\Sigma_u$ -- is actually
the counterpart of a phase transition,  a {\it diffeomorphism invariant} 
has to be computed. This is a very challenging task because of the high 
dimensionality of the manifolds involved. Moreover, any algorithm of a
combinatorial type (like those implied by simplicial decompositions, i.e.
high dimensional analogs of tesselations with triangles that are used, for
example, in numerical quantum gravity for low dimensional manifolds) is here
hopeless. Only through a link between analytic and topologic mathematical
objects can one hope to work out some direct information about topology.
One such a link is provided by the Gauss-Bonnet-Hopf theorem that relates 
the Euler characteristic (see Appendix B) $\chi (\Sigma_u)$
with the total Gauss-Kronecker curvature of the manifold, i.e. \cite{Spivak}
\begin{equation}
\chi (\Sigma_u)  = \gamma \int_{\Sigma_u} K_G \,d\sigma
\label{gaussbonnet}
\end{equation}
which is valid in general for even dimensional hypersurfaces of euclidean 
spaces
${\bf R}^N$ [here ${\rm dim}(\Sigma_u)=n\equiv N-1$],  and where:
$\gamma =2/Vol({\bf S}^n_1)$ is twice the inverse of
the volume of an $n$-dimensional sphere of unit radius; $K_G$ is the
Gauss-Kronecker curvature of the manifold;
$d\sigma =\sqrt{det(g)}dx^1dx^2\cdots dx^n$  is the invariant volume
measure of $\Sigma_u$ and $g$ is the Riemannian metric induced from
${\bf R}^N$.
The Gauss-Kronecker curvature at a given point of a hypersurface is the product
of the eigenvalues of its so-called shape operator; these eigenvalues are the
principal curvatures of the hypersurface at the given point. The shape operator
is constructed through the directional derivatives of the unit normal vector
to the hypersurface at the given point computed in the $n$ directions of the 
basis vectors of the plane tangent to the surface at the same point 
\cite{Thorpe}.

\begin{figure} 
\centerline{\psfig{file=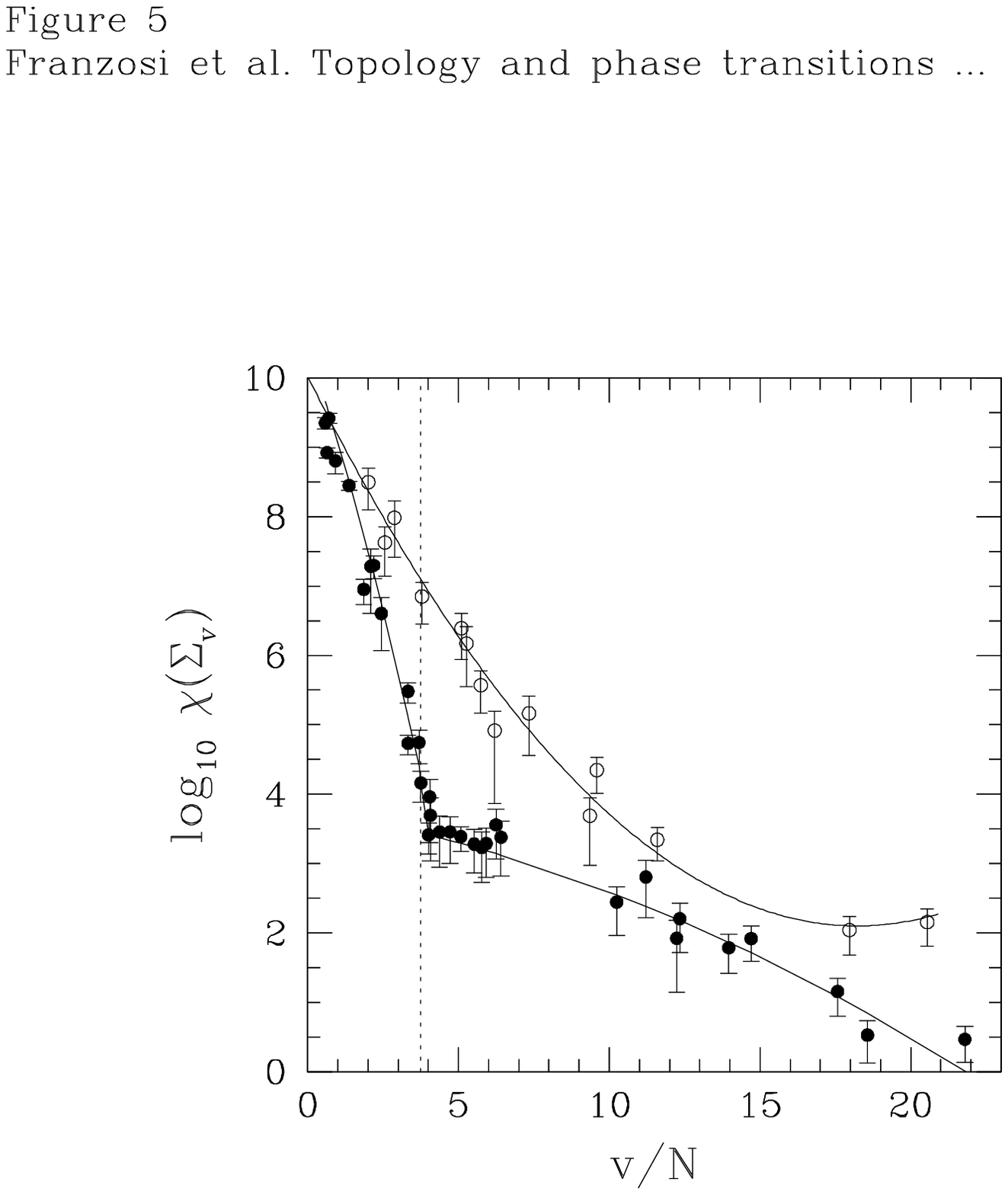,height=8cm,clip=true}}  
\caption{Euler characteristic $\chi (\Sigma_v)$ for 1-$d$ and 2-$d$ 
$\varphi^4$ lattice models.
Open circles:  1-$d$ case, $N = 49$; full circles: 2-$d$ case, $N =
7 \times 7$. The
vertical dotted line, computed separately for larger $N$, accurately
locates the phase transition and the parameters are the same as in 
Figs.\ \protect\ref{fig_franz_1}, \protect\ref{fig_franz_2} and 
\protect\ref{fig_franz_3}.} 
\label{fig_chi(v)}  
\end{figure}

The numerical application of the Gauss-Bonnet-Hopf theorem has been worked out
for a lattice $\varphi^4$ model in one and two spatial dimensions 
\cite{Franz_thesis,Lionel_thesis,fps1}.
The main result is shown in Fig.\ \ref{fig_chi(v)} where $\chi (\Sigma_v)$ is 
reported {\it vs} $v$. In the $1d$ case (open circles) a ``smooth'' pattern
of $\chi (v)$ is found, whereas in the $2d$ case a cusp-like shaped $\chi (v)$
shows up with the singular point corresponding to the phase transition point
marked by the vertical dotted line. The parameters are those of the preceding
Section, therefore the phase transition point is at $v_c/N\simeq 3.75$.

These results have two important consequences: {\it i)} the non constant value
of $\chi (v)$ in the $1d$ case clearly shows that topology changes are there
even in the absence of phase transitions; {\it ii)} an {\it abrupt change} 
in the rate of variation of topology with $v$  seems the hallmark of a phase 
transition. Thus we have direct numerical evidence about the actual implication
of topology in the appearance of a phase transition. At the same time we have
evidence of a non simple one-to-one correspondence between topology changes 
and phase transitions. This is in full agreement with what is discussed in the
next Section about a different kind of potential which can be analytically 
investigated.
 
\subsubsection{Topological origin of the phase transition 
in the mean-field $XY$ model} 
\label{sec_prl} 
 
Until now we have not yet given any {\em direct} analytic evidence of 
the validity of the TH. 
Let us now consider again the mean-field $XY$ model (\ref{V_xymf}). 
In the case of this particular model it is possible to show analytically  
that a topological change in the configuration space exists and that it can be 
related to the thermodynamic phase transition \cite{prl99}. 
 
Let us consider again, as was 
already done in \S \ref{sec_franz}, the family $M_v$ 
of submanifolds of the configuration space defined in Eq.\ (\ref{M_v});  
now the potential energy per degree 
of freedom is that of the mean-field $XY$ model, i.e.,  
\beq 
{\cal V}(\varphi) = \frac{V(\varphi)}{N} =  
\frac{J}{2N^2}\sum_{i,j=1}^N  
\left[ 1 - \cos(\varphi_i - \varphi_j)\right] -h\sum_{i=1}^N \cos\varphi_i ~, 
\label{calV} 
\eeq 
where $\varphi_i \in [0,2\pi]$.  
Such a function can be considered a Morse function on $M$, so that, according 
to Morse theory (see Appendix \ref{app_morse}), 
all these manifolds have  
the same topology  
until a critical level ${\cal V}^{-1}(v_c)$  
is crossed, where the topology of $M_v$ changes.  
 
A change in the topology of $M_v$ can only occur when $v$ passes through a  
critical value of ${\cal V}$. Thus in order to detect topological  
changes in $M_v$ we have to find the  
critical values of ${\cal V}$, which means solving the equations 
\beq 
\frac{\partial {\cal V}(\varphi)}{\partial \varphi_i} = 0~,  
\qquad i = 1,\ldots,N~. 
\label{crit_eqs} 
\eeq 
For a general potential energy function $\cal V$, the solution of the 
Eqs.\ (\ref{crit_eqs}) would be a formidable task \cite{method_note}, but 
in the case of the mean-field XY model, 
the mean-field character of the interaction greatly simplifies the analysis, 
allowing an analytical treatment of the Eqs.\ (\ref{crit_eqs}); moreover, a 
projection of the configuration space onto a two-dimensional plane is 
possible.  
 
We recall that in the limit $h \to 0$, 
the system has a continuous phase transition, with classical critical 
exponents, at $T_c = 1/2$, or $\varepsilon_c = 3/4$,  
where $\varepsilon = E/N$ is the energy per particle. We aim at showing that 
this phase transition has its foundation in a 
basic topological change that occurs in the configuration space $M$  
of the system. Let us remark that since ${\cal V}(\varphi)$ is bounded, 
$-h \leq {\cal V}(\varphi) \leq 1/2 + h^2/2$, the manifold is empty as long 
as $v< -h$, and when $v$  
increases beyond  $1/2 + h^2/2$ no changes in its topology  
can occur so that the manifold $M_v$ remains the same for any 
$v > 1/2 + h^2/2$, and is then 
an $N$-torus. To detect topological changes we have to solve 
Eqs.\ (\ref{crit_eqs}). To this end it is useful to define the  
magnetization vector, i.e., the collective spin  
vector ${\bf m} = \frac{1}{N} \sum_{i=1}^N {\bf s}_i$, which as a  
function of the angles is given by  
\beq 
{\bf m} = (m_x,m_y) = \left(\frac{1}{N}\sum_{i=1}^N \cos\varphi_i, 
\frac{1}{N}\sum_{i=1}^N \sin\varphi_i  \right)~. 
\label{m} 
\eeq 
Due to the mean-field character of the model, the potential energy 
(\ref{V_xymf}) can be expressed as a function of ${\bf m}$ alone  
(remember that $J=1$), so that the potential energy per particle reads 
\beq 
{\cal V}(\varphi) = {\cal V}(m_x,m_y) = \frac{1}{2} 
(1 - m_x^2 - m_y^2) - h\, m_x~. 
\label{V(m)} 
\eeq 
This allows us to write the Eqs. (\ref{crit_eqs}) in the form 
($i = 1,\ldots,N$) 
\beq 
(m_x + h) \sin\varphi_i - m_y \cos\varphi_i = 0 ~ . 
\label{crit_eqs_i} 
\eeq 
Now we can solve these equations and find 
all the critical values of ${\cal V}$. 
The solutions of Eqs.\ (\ref{crit_eqs_i}) can be grouped in three classes: 
 
$(i)$ The minimal energy configuration $\varphi_i = 0 ~  
\forall i$, with a critical value $v=v_0=-h$, which tends to 0 as $h \to 0$. 
In this case, $m_x^2 + m_y^2 = 1$. 
 
$(ii)$ Configurations such that $m_y = 0, \sin\varphi_i = 0 ~ 
\forall i$. These are the configurations in which $\varphi_i$ equals either 
$0$ or $\pi$; i.e., we have again $\varphi_i = 0 ~ \forall i$, but also the 
$N$ configurations with $\varphi_k = \pi$ and $\varphi_i = 0 ~ \forall i 
\not = k$, as well as the $N(N-1)$ configurations with 2 angles equal to 
$\pi$ and all the others equal to 0, and so on, up to the configuration with 
$\varphi_i = \pi ~ \forall i$. The critical values corresponding to these 
critical points depend only on the number of $\pi$'s, $n_\pi$, so that 
$v(n_\pi) = \frac{1}{2}[1 - \frac{1}{N^2}(N - 2n_\pi)^2] - 
\frac{h}{N}(N - 2n_\pi)$. We see that the largest critical value is, for $N$ 
even, $v(n_\pi = N/2) = 1/2$ and that the number of critical points 
corresponding to it is ${\cal O}(2^N)$.  
 
$(iii)$ Configurations such that $m_x = -h$ and $m_y = 0$, which correspond 
to the critical value $v_c = 1/2 + h^2/2$, which tends to $1/2$ as $h \to 0$. 
The number of these configurations grows with $N$ not slower than $N!$ 
\cite{prl99}. 
 
Configurations $(i)$ are the absolute minima of ${\cal V}$, $(iii)$ are 
the absolute maxima, and $(ii)$ are all the other stationary configurations 
of $\cal V$.  
 
Since for $v < v_0$ the manifold is empty, the topological change that 
occurs at $v_0$ is the one corresponding to the ``birth'' of the manifold 
from the empty set; subsequently there are many topological changes at values 
$v(n_\pi) \in (v_0,1/2]$ till at $v_c$ 
there is a final topological change which corresponds to the ``completion'' 
of the manifold. We remark that the number of critical values in the 
interval $[v_0, 1/2]$ grows with $N$ and that eventually the set of these 
critical values becomes dense in the limit $N \to \infty$. However, the 
critical value $v_c$ remains isolated from other critical values 
also in that limit. 
We observe that it is necessary to consider a nonzero external field $h$ 
 in order that $\cal V$ is a Morse function, because if 
$h = 0$ all the critical points of classes $(i)$ and $(ii)$ are 
degenerate, in which case topological changes do not necessarily 
occur\footnote{It would also be possible to avoid this problem by considering 
an improved version of Morse theory, referred to as {\em equivariant 
Morse theory} \protect\cite{equivariant_Morse_theory}.}. 
This degeneracy is due to the  
$O(2)$-invariance of the potential energy in the absence of an external 
field. To be sure, for $h \not = 0$, $\cal V$ may not be a Morse function 
on the whole of $M$ either, but only on $M_v$ with $v < v_c$, 
because the critical points of class $(iii)$ may also be degenerate, 
so that $v_c$ does not necessarily correspond to a topological change.  
However, this difficulty could be dealt with by using that the 
potential energy can be written in terms of the collective 
variables $m_x$ and $m_y$ --- as in Eq. (\ref{V(m)}). This implies 
that we consider the system of $N$ spins projected onto the two-dimensional 
configuration space of the collective spin variables. According to 
the definition (\ref{m}) of ${\bf m}$, the accessible configuration 
space is now not the whole plane, but only the disk  
\beq 
D = \{(m_x,m_y) : m_x^2 + m_y^2 \leq 1\}~. 
\eeq 
Thus we want to study the topology of the submanifolds  
\beq 
D_v = \{ (m_x,m_y) \in D : {\cal V}(m_x,m_y) \leq v\}~. 
\eeq 
\begin{figure} 
\vspace{0.25cm} 
\centerline{\psfig{file=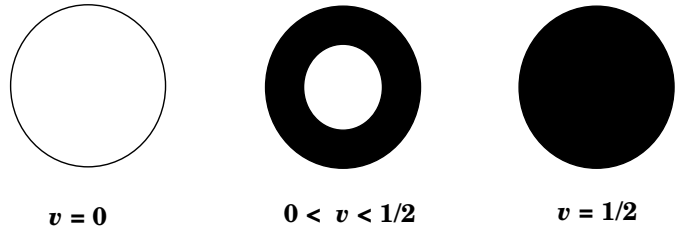,height=3cm}} 
\vspace{0.25cm} 
\caption{The sequence of topological changes undergone by the manifolds 
$D_v$ with increasing $v$ in the limit $h \to 0$. 
\label{fig_effective}} 
\end{figure} 
The sequence of topological transformations undergone by $D_v$ can now be 
very simply determined in the limit $h \to 0$  
(see Fig.\ \ref{fig_effective}), as follows. 
As long as $v < 0$, $D_v$ is the empty set. The first topological 
change occurs at $v = v_0 = 0$ where the manifold appears 
as the circle $m_x^2 + m_y^2 = 1$, i.e., the boundary $\partial D$ of $D$. 
Then as $v$ grows $D_v$ is given by the conditions 
\beq 
1 - 2v \leq m_x^2 + m_y^2 \leq 1~,  
\eeq  
i.e., it is the ring with a hole centered at $(0,0)$ (punctuated disk)  
comprised between two circles 
of radii $1$ and $\sqrt{2v}$, respectively. As $v$ continues to grow the 
hole shrinks and is eventually completely filled as 
$v = v_c = 1/2$, where the second topological change occurs.   
In this coarse-grained two-dimensional description in $D$, all the 
topological changes that occur in $M$ between $v=0$ and $v=1/2$ disappear, 
and only the two topological changes corresponding to the extrema of 
$\cal V$, occurring at $v = v_0$ and $v = v_c$, survive. This means 
that the topological change at $v_c$ should be present also in 
the full $N$-dimensional configuration space, so that the degeneracies 
mentioned above for the critical points of class $(iii)$ should not 
prevent a topological change. 
 
Now we want to argue that the topological change occurring at  
$v_c$ is related to the thermodynamic phase transition of the mean-field 
XY model. Since the  
Hamiltonian is of the standard form (\ref{H}), the temperature $T$, the  
energy per particle $\varepsilon$ and the average potential energy per  
particle $u = \langle {\cal V} \rangle$ obey,  
in the thermodynamic limit, the following equation: 
\beq 
\varepsilon = \frac{T}{2} + u(T)~, 
\eeq 
where we have set Boltzmann's constant equal to 1. Substituting the 
values of the critical energy per particle  
$\varepsilon_c = 3/4$ and of the critical temperature $T_c = 1/2$ we get  
$u_c = u(T_c) = 1/2$, so that the critical value of the potential energy  
per particle $v_c$ where the last 
topological change occurs equals the statistical-mechanical  
average value of the potential energy at the phase transition, 
\beq 
v_c = u_c~. 
\label{vc=uc} 
\eeq 
Thus although a topological change in $M$ occurs at any $N$, and 
$v_c$ is {\em independent} of $N$,  
there is a connection of such a topological change and a thermodynamic 
phase  
transition {\em only} in the limit $N\to\infty$, $h \to 0^+$, when indeed 
a thermodynamic phase transition can be defined.  
 
A similar kind of difference, as here between topological changes in  
mathematics (for all $N$) and phase transitions in physics (for $N\to\infty$  
only), also occurs in other contexts in 
statistical mechanics, e.g.\ in nonequilibrium stationary 
states \cite{CohenRondoni}. 

Since {\em not all} topological changes  
correspond to phase transitions, those that  
do correspond, remain to be determined to make the  
conjecture of Ref.\ \cite{cccp} more precise.  
In this context, we consider one example where there are topological changes 
very similar to the ones of our model but no phase  
transitions, i.e., the one-dimensional XY model with nearest-neighbor 
interactions, whose Hamiltonian is of the class (\ref{H}) with interaction 
potential 
\beq 
V (\varphi) = \frac{1}{4}\sum_{i=1}^N \left[ 1 -  
\cos(\varphi_{i+1} - \varphi_i) \right] -h\sum_{i=1}^N \cos\varphi_i ~. 
\label{V_nn} 
\eeq 
In this case the configuration space $M$ is still an $N$-torus, and using 
again the potential energy per degree of freedom ${\cal V} = V/N$ as a Morse 
function, we can see  
that also here there are many topological changes in the submanifolds  
$M_v$ as $v$ is varied in the interval $[0,1/2]$ (after taking $h \to 0^+$). 
However there are critical points of the type 
$\varphi_j = \varphi_k = \varphi_l = 
\ldots = \pi$, $\varphi_i = 0 ~ \forall i \not = j,k,l,\ldots$;   
at variance with the mean-field XY model, it is now no longer the number 
of $\pi$'s that determines the value of $\cal V$ at the critical point, but 
rather the number of domain walls, $n_d$, i.e., the number of boundaries 
between ``islands'' of $\pi$'s and ``islands'' of $0$'s: $v(n_d) = n_d/2N$. 
Since $n_d \in [0,N]$, the critical values lie in the same interval as in the 
case of the mean-field XY model; but now the maximum critical value $v = 1/2$,
 instead of corresponding to a huge number of critical points, which rapidly 
grows with $N$, corresponds to {\em only two} configurations with $N$ domain 
walls, which are $\varphi_{2k} = 0$, $\varphi_{2k + 1} = \pi$, with 
$k = 1,\ldots,N/2$, and the reversed one. 
There are also ``spin-wave-like'' critical points, i.e., such that
$e^{i\theta_k} = {\rm const}~ e^{2\pi i k n/N}$ with $n = 1,\ldots,N$ 
\cite{Dhar}; their critical energies are comprised in the interval  above but 
again there is
not a critical value associated to a huge number of critical points.
 
Thus this example suggests the conjecture that a topological change in the 
configuration space submanifolds $M_v$ occurring at a critical value $v_c$, 
is associated with a phase transition in the thermodynamic limit only 
if the number 
of critical points corresponding to the critical value $v_c$ is sufficiently 
rapidly growing with $N$. On the basis of the behavior of the mean-field XY 
model we expect then that such a growth should be at least exponential. 
Furthermore,
a relevant feature appears to be that $v_c$ remains an isolated critical 
value also in the limit $N \to \infty$: in the mean-field XY model this holds 
only if the thermodynamic limit is taken {\em before} the $h \to 0^+$ limit: 
this appears as a topological counterpart of the  
non-commutativity of the limits $h \to 0^+$ and $N \to \infty$ in order to get
a phase transition in statistical mechanics. 
 
The sequence of topological changes occurring with growing $\cal V$ 
makes the configuration space larger and larger, till at $v_c$ the whole 
configuration space becomes fully accessible to the system through the last 
topological change. From a physical point of view, this corresponds to the 
appearance of more and more disordered configurations as $T$ grows, which 
ultimately lead to the phase transition at $T_c$. We remark that the 
connection between the topology of the configuration space and the physics of 
continuous phase transitions made here via the potential energy, in particular
 Eq.\ (\ref{vc=uc}), only makes sense in the thermodynamic limit, where the 
potential energy per particle $u(T)$ is well-defined since its fluctuations 
vanish then at least as $1/\sqrt{N}$. This holds for our mean-field model, 
since for such a model, all fluctuations are absent. 
In the case of a real continuous (critical) phase transition the non-trivial 
role of fluctuations may complicate the present picture.

\subsection{The Topological Hypothesis} 
 
\label{sec_TH} 

The statistical behaviour of physical systems described by standard 
Hamiltonians is encompassed, in the canonical ensemble, by the  
partition function in phase space  
\begin{eqnarray} 
Z_N(\beta ) 
  &=&\int\,  dp\, dq\, e^{-\beta H(p,q)} 
 =\left(\frac{\pi}{\beta}\right)^\frac{N}{2}
\int  dq\,
e^{-\beta V(q)}   \nonumber \\ 
&=&\left(\frac{\pi}{\beta}\right)^\frac{N}{2} \int_0^\infty du\, e^{-\beta u} 
\int_{\Sigma_{u}}\frac{d\sigma}{\Vert \nabla V\Vert}      
\label{zeta} 
\end{eqnarray}
where $p=(p_1,\dots ,p_N)$, $dp=\prod_{i=1}^N dp_i$, $q=(q_1,\dots ,q_N)$,
$dq=\prod_{i=1}^N dq_i$; $\Sigma_u= 
\{(q_1,\dots,q_N)\in{\bf R}^N\vert V(q_1,\dots,q_N)=u\}$ are equipotential 
hypersurfaces of configuration space and $d\sigma$ is the measure on 
$\Sigma_u$ induced from ${\bf R}^N$. 
The last term of the equation above shows that for Hamiltonians (\ref{H}) 
the relevant statistical information is contained in the canonical 
configurational partition function $Z_N^C=\int dq\exp [-\beta V(q)]$.  
Moreover, the last term of Eq.\ (\ref{zeta}), written using a co-area 
formula \cite{Federer}, decomposes $Z_N^C$ into an infinite summation of  
geometric integrals $\int_{\Sigma_u} d\sigma\,/\Vert\nabla V\Vert$.
This decomposition provides the point of attack for the formulation of a
general hypothesis about a deep link between geometry, topology and 
thermodynamics and, obviously, phase transitions. In fact,    
once the potential energy $V(q)$ is given, the 
configuration space of the system is automatically foliated into the family  
$\{ \Sigma_u\}_{u\in{\bf R}}$ of these equipotential hypersurfaces. 
Now, from standard statistical mechanical arguments concerning the equivalence
of canonical and microcanonical ensembles we know that, at any  
given value of the inverse temperature $\beta$, the larger the number $N$  
of particles, the closer to some $\Sigma_{\overline u}$ are the  
microstates that significantly contribute to the statistical averages of 
thermodynamic observables. The hypersurface  
$\Sigma_{\overline u}$ is the one associated with  ${\overline u}\equiv  
u_\beta =(Z_N^C)^{-1}\int\prod dq_i V(q) e^{-\beta V(q)}$, the average  
potential energy computed at a given $\beta$. 
Thus, at any $\beta$, if $N$ is very large the effective support of the  
canonical measure shrinks very close to a single  
$\Sigma_v=\Sigma_{u_\beta}$.   
Hence and on the basis of what was found in \cite{cccp,prl99,Franz}, it was
formulated the following 

\smallskip
\noindent{\it {\underline{Topological Hypothesis}}: The basic origin of a 
phase transition lies in a topological change of the support of the 
measure describing a system. 
This change of topology induces a change of the measure itself at the 
transition point. }
\smallskip
 
In other words, this hypothesis stipulates that some  
change of the topology  of the $\{ \Sigma_u\}$, occurring at some 
$u_c=u_c(\beta_c)$, could be the origin of the singular behavior of 
thermodynamic observables at a phase transition rather than measure
singularities which in this view are induced from a deeper level where the
topology changes take place.
  
In other words, the claim is that the canonical measure  
should ``feel'' a big and sudden change -- if any -- of the topology   
of the equipotential hypersurfaces of its underlying support, with as a
consequence, the appearance of the typical signals of a phase transition, i.e. 
almost singular (at finite $N$) energy or temperature dependences of  
the averages of appropriate observables. 
The larger $N$, the narrower the effective support is of the measure 
-- as discussed above -- and  
hence the sharper the mentioned signals can be, until true singularities 
appear in the $N\rightarrow\infty$ limit. 
 
We emphasize though that not all topological transitions lead to physical phase
transitions. At present the precise connection between topological transitions
and phase transitions still has to be clarified in many respects. 
Certain is that not every topological
transition corresponds to a phase transition, as has been discussed in 
Sections \ref{sec_euler} and \ref{sec_prl}. Rather it seems that, on the basis
of present evidence, a phase transition corresponds to a super-combination
of many simultaneous topological transitions taking place, where many might 
mean at least exponentially growing with the number of particles. It seems 
therefore more like a super-topologically constructed  transition. This is 
illustrated analytically by the above discussed XY mean-field model, where an
exponential crowding of topological transitions occurs on one side of the 
phase transition. Though such an analysis has not been possible for a 
numerically treated lattice $\varphi^4$ model, on the other hand, like the 
Euler characteristic $\chi $ clearly shows, a phase transition corresponds to
an abrupt transition between different rates of change in the topology above 
and below of the phase transition point; no such information is available
for the analytic XY mean-field model where the calculation of $\chi$ is for
the moment very difficult.

\subsection{Open questions and future developments} 
 
\label{sec_open}

{\bf 1.} The phase space trajectories of dynamical systems described by Hamiltonian
functions of the form (\ref{H}), i.e., 
\[ 
{\cal H} = \frac{1}{2}\sum_{i=1}^N \pi_i^2 + V(\varphi)~,  
\] 
where the $\varphi$'s and the $\pi$'s are, respectively, the coordinates and 
the conjugate momenta, are bound to the constant energy hypersurfaces 
$\Sigma_E$ of the $2N$-dimensional phase space spanned by the $\varphi$'s and 
the $\pi$'s. Therefore it would be natural to investigate the relationship
between the topology variations of these hypersurfaces in phase space and 
phase transitions. The Hamiltonian ${\cal H}$ would be the Morse function
in this case. Moreover, having already considered the role played by the
$\Sigma_u$ of configuration space, we can wonder what is the relationship 
between topology changes in configuration space and phase space respectively.  
Such a relationship is somehow subtle. At first sight all the 
critical points of configuration space are embodied also in critical points
of phase space, in fact 
all the critical points of ${\cal H}$ are such that, $\forall i$, $\pi_i = 0$
and $\nabla_i V(\varphi )=0$. However, the critical points of ${\cal H}$ 
physically correspond to vanishing kinetic energy, so that, if those topology 
changes of the $\Sigma_u$ that are associated to a phase transition
were to correspond to topology changes of the $\Sigma_E$ -- because the 
critical points of ${\cal H}$ incorporate those of $V(\varphi )$ -- then the 
critical potential energy density and the critical total energy density at a 
phase transition should coincide, which is not the case.
An argument to clarify this point can be given \cite{CSCP} by observing that 
the dynamics does not equally sample the whole $\Sigma_E$: 
the larger $N$, the smaller the relative fluctuations of the potential
energy $\langle\delta^2V\rangle^{1/2}/\langle V\rangle$ and kinetic energy
$\langle\delta^2K\rangle^{1/2}/\langle K\rangle$, respectively, are. Thus, by
putting $v\equiv\langle V\rangle$ and $t\equiv\langle K\rangle$, 
we can assume that, at large $N$ and given energy $E$, the momenta mainly 
live close to the hypersphere ${\bf S}_t^{N-1}=\{(p_1,\dots 
,p_N)\in{\bf R}^N\vert\sum_{i=1}^N\frac{1}{2}p_i^2=t\}$ and the lagrangian
coordinates mainly live close to the equipotential hypersurface 
$\Sigma_v^{N-1}$ with $v+t=E$.
Therefore, though the microcanonical measure 
mathematically extends over a whole energy surface,  
as far as physics is concerned, at very large $N$ a non-negligible 
contribution to the microcanonical measure is
in practice given only by a small subset of an energy surface. This subset can
be approximately modeled by the product manifold $\Sigma^{N-1}_v\times{\bf S}
_t^{N-1}$. Since the kinetic energy submanifold ${\bf S}_t^{N-1}$ is a
hypersphere at any $t$, a change in the topology of $\Sigma^{N-1}_v$ directly 
entails a change of the topology of $\Sigma^{N-1}_v\times {\bf S}_t^{N-1}$, 
that is of the effective model-manifold for the subset of $\Sigma_E$ where 
the dynamics mainly ``lives'' at a given energy $E$.

This confirms that, as long as we are interested only in classical Hamiltonian 
systems of the standard form (\ref{H}), we can restrict our geometrical and 
topological investigation to the submanifolds of the $N$-dimensional 
configuration space $M$.

{\bf 2.}
A recent and important advance has been achieved at a more mathematical 
level. In Refs.\ \cite{Franz_thesis,Lionel_thesis,fps2} a theorem has been 
proved that establishes the {\it necessity} of topological changes of the 
equipotential hypersurfaces $\Sigma_u$ for the appearance of first or second
order thermodynamic phase transitions. The theorem applies to a wide class of
finite-range potentials, bound below, describing systems confined in finite 
regions of space with continuously varying coordinates. The proof proceeds by
showing that, under the crucial
assumption of diffeomorphicity of the $\Sigma_u$ in an arbitrary interval of
values for $u$, the Helmoltz free energy is uniformly convergent in $N$ to
its thermodynamic limit, at least within the class of twice 
differentiable functions, in a corresponding interval of temperature. 

This theorem confirms the general validity of the TH and ensures that for a 
wide class of physical potentials the mathematical framework of differential 
topology is adequate to describe, at least, first and second order phase
transitions. There is no proof of {\it sufficiency}. On the basis of the
discussions in Sections \ref{sec_euler}, \ref{sec_prl} and \ref{sec_TH} 
we already know that
a simple loss of diffeomorphicity of the $\Sigma_u$ is not sufficient to
lead to a phase transition.

A thorough investigation of those classes of topology changes that are 
responsible for the appearance of phase transitions is at present the main 
challenge of this new point of view about phase transitions and certainly
represents a topic that will remunerate the efforts addressed to it. 

{\bf 3.}
Let us finally highlight some interesting related topics.

\begin{itemize}
\item[{$(i)$}] We might speculate about the possibility of relating universal
quantities of the theory of critical phenomena, like critical exponents,
to some topological counterpart; in fact a notion of universality arises 
quite naturally in a topological framework.

\item[{$(ii)$}] Topology provides a 
common ground for the roots of both dynamics
and thermodynamics:
insofar as the dynamics of the system, i.e. the motion of the trajectory 
in phase space, takes place in what was called before the support of the
statistical measure, it is clear that the nature of the trajectory will 
crucially depend on the topology of the manifold to which it belongs.
This therefore strenghtens the interest of a dynamical treatment of phase 
transitions giving new emphasis to the microcanonical ensemble and thereby 
joining other recent developments in the field \cite{Gross}.
Since the dynamical approach does not depend on whether a system is in
statistical equilibrium, non-equilibrium, or in a metastable state (like a
glass and, more generally, amorphous materials) a dynamical approach to
phase transitions might also be important for systems whose thermodynamical 
state is not well defined.

\item[{$(iii)$}] The fact that the topological changes appear at any $N$ 
opens a new possibility to study transitional phenomena in {\em finite} 
systems, like nuclear and atomic clusters, polymers and proteins, or other  
biological systems, as well as for nano and mesoscopic structures.  

\end{itemize}

\acknowledgments 
We want to dedicate this paper 
to the memory of our friend and  
brilliant collaborator Lando Caiani, prematurely deceased, who gave  
seminal contributions to many aspects of the approach described here, 
and especially to the early development of the connection between
topology and phase transitions.  
 
During their development, the main ideas described in the present Report 
have been discussed with many colleagues, among which we would like to thank, 
for their particular interest and for many useful suggestions and criticism, 
and sometimes for an active collaboration, 
S.\ Caracciolo, M.\ Cerruti-Sola, C.\ Clementi,   
Y.\ Elskens, M.-C.\ Firpo,  
R.\ Franzosi, R.\ Gatto, R.\ Livi, G.\ Mussardo,  
H.\ A.\ Posch, V.\ Penna, G.\ Pettini,  
M.\ Rasetti, S.\ Ruffo, L.\ Spinelli, G.\ Vezzosi, 
A.\ M.\ Vinogradov, 
and the DOCS research group in Firenze (http://docs.de.unifi.it/$\sim$docs/). 
 
Part of this work has been done at the Erwin Schr\"odinger Institut f\"ur 
Mathematische Physik (ESI) in Vienna, whose kind hospitality is gratefully 
acknowledged.  
LC thanks the Rockefeller University in New York  
for its kind hospitality and for partial financial support,  
the 1997-1999 SINTESI program of the Italian Ministry of University and of 
Scientific and Technologic Research (MURST) 
for partial financial support, and Prof.\ H.\ A.\ Posch for hospitality at the 
Institut f\"ur Experimental Physik of the University of Vienna.  
EGDC is indebted to the Basic Engineering Section of 
the US Department of Energy for support under  
grant DE-FG02-88-ER13847, as well as to the Politecnico di Torino and to the  
INFM, UdR Torino Politecnico, for hospitality and partial financial support.

\appendix 
 
\section{Summary of Riemannian geometry} 
\label{app_geo} 
In the following we briefly recall some essential concepts and notations 
of Riemannian differential geometry which are used 
in the main text. The present section is only meant to facilitate
the reader 
to follow the main text of the Report, so that our discussion will not be
a rigorous treatment of the subject.
For a more elaborate discussion, we refer the reader to a textbook of 
general relativity
(e.g., the classic Landau and Lifshitz's book \cite{Landau} or the more 
recent and complete, but still very clear and readable, textbook by Wald 
\cite{Wald}).
A more mathematically oriented, but still readable by physicists, 
introduction 
to the subject is given by do Carmo \cite{doCarmo}; a comprehensive and 
rigorous 
treatment, which, however, goes far beyond what is needed 
to follow the exposition
in the main text, can be found in Kobayashi and Nomizu \cite{Kobayashi}. 

The Einstein summation convention over repeated indices
is always understood unless explicitly stated to the contrary. Moreover, 
we follow throughout the paper the usual convention to suppress the
dependence of the components of vector and tensor quantities on the 
(proper) time and, in general, only indicate it explicitly when 
this dependence is absolutely relevant.

\subsection {Riemannian manifolds}

A set $M$ is called a {\em differentiable manifold} if it can be 
covered with a 
collection, either finite or denumerable, of {\em charts}, 
such that each point
of $M$ is represented at least on one chart,
and the different charts are differentiably connected to each other.
A chart is a set of coordinates on the manifold, i.e., it is a
set of $n$ real numbers $(x^1,\ldots,x^n)$ 
which denote the ``position'' of a point 
on the manifold. The number $n$ of coordinates of a chart is
the same for each connected part of the manifold (and for the whole
manifold if the latter is connected, i.e., it cannot be split in 
two disjoint parts which are still manifolds); such a number is 
called the {\em dimension} of the manifold $M$. The union of the charts
on $M$ is called an {\em atlas} of $M$. 

\subsubsection{Vectors and tensors}

A vector (more precisely, a {\em tangent} vector), can be defined using
curves on the manifold $M$. Given
a curve $\gamma$ in $M$, represented in local coordinates
by the parametric equations 
$x = \varphi(t)$, we define a tangent vector at $P\in M$ as the velocity 
vector of the curve in $P$, i.e.,
\beq
v = \dot \gamma = \lim_{t \to 0} \frac{\varphi(t) - \varphi(0)}{t}~,  
\qquad \varphi(0) = P,
\eeq
so that the $n$ components of the tangent vector $v$ are given by
\beq
v^i = \frac{d\varphi^i}{dt}~.
\eeq
The set of all the tangent vectors of $M$ in $P$ is a linear space, 
referred to as the {\em tangent space} of $M$ in $P$, and denoted by $T_P M$. 
Each tangent space is isomorphic to an $n$-dimensional Euclidean space. 
Given a chart $(x^1,\ldots,x^n)$ in a neighborhood of $P$, 
a basis $(X_1,\ldots,X_n)$ 
of $T_P M$ can be defined,  
so that a generic vector $v$ is expressed as a sum of the $X_i$'s 
weighted by its components,
\beq
v = v^i X_i~.
\eeq
The basis $\{ X_i \}$ is called a {\em coordinate basis} of $T_P M$, 
and its components $X_i$ are often denoted\footnote{The origin of this
notation is in the fact that vectors can be defined as directional
derivatives on $M$ (see e.g.\ Ref.\ \protect\cite{Wald}).} 
by $\partial / \partial x^i$. 
The basis depends on the chart: choosing another chart, $(x'^1,\ldots,x'^n)$, 
we get another basis $\{ X'_i \}$. The components of $v$ 
in the two different bases are connected by the following rule,
\beq
v'^i = v^j \frac{\partial x'^i}{\partial x^j}~,
\label{vector_rule}
\eeq
referred to as the {\em vector transformation rule}. 
Indeed, one can define a 
vector as a quantity whose components transform according 
to Eq.\ (\ref{vector_rule}).
The union of all the tangent spaces of the manifold $M$,
\beq
TM = \bigcup_{P\in M} T_P M~,
\eeq
is a $2n$-dimensional manifold and is referred to as the 
{\em tangent bundle} of $M$. 

A {\em vector field} $V$ on $M$ is an assignment of a vector $v_P$ at 
each point  $P \in M$. If $f$ is a smooth function, 
\beq
V(f)|_P = v_P(f)
\eeq
is a real number for each $P \in M$, i.e., $v(f)$ is a function on $M$. 
If such a function is smooth, $V$ is called a {\em smooth vector field} 
on $M$. The curves $\varphi(t)$ which satify the differential equations 
\beq
\dot\varphi = V(\varphi(t))
\eeq
are called the {\em trajectories} of the field $V$, 
and the mapping $\varphi_t:
M \mapsto M$ which maps any point $P$ of $M$ along the trajectory of $V$ 
emanating from $P$ is called the $\em flow$ of $V$.
Given two vector fields $V,W$, one can define the {\em commutator} 
as the vector field $[V,W]$ such that
\beq
[V,W](f) = V(W(f)) - W(V(f))~,
\eeq
i.e., in terms of the local components,
\beq
[V,W]^j = V^i \frac{\partial W^j}{ \partial x^i} - 
W^i \frac{ \partial V^j}{\partial x^i}~.
\label{def_commutator}
\eeq
We note that, if $\{ X_i \}$ is a coordinate basis,
\beq
[X_i,X_j] = 0 \quad~~ \forall i,j~,
\eeq
and that, conversely, given $n$ nonvanishing and commuting vector 
fields which are linearly independent, there always exists a chart for which 
these vector fields are a coordinate basis.

Tangent vectors are not the only vector-like quantities that can be
defined on a manifold $M$: there are also {\em cotangent} vectors, which
can be defined as follows.
Let us recall that the {\em dual space} $V^*$ of a vector space $V$ 
is the space of {\em linear} maps from $V$ to the real numbers. 
Given a basis of $V$, $\{ u_i \}$, a basis of $V^*$, $\{ u^{i *} \}$,
 called the {\em dual basis}, is defined by
\beq
u^{i*} (u_j) = \delta^i_j~.
\eeq  
The dual space of $TM$, $T^*M$, is called the {\em cotangent bundle} of $M$. 
Its elements are called {\em cotangent vectors}, or sometimes 
{\em covariant vectors} (while the tangent vectors are sometimes denoted as
{\em contravariant} vectors). The dual basis elements are usually denoted as 
$dx^1,\ldots,dx^n$, i.e., $dx^i$ is such that $dx^i(\partial/\partial x^j) = 
\delta^i_j$. The components $\omega_i$ of cotangent vectors transform 
according to the rule 
\beq
\omega'_i = \omega_j \frac{\partial x^j}{\partial x'^i}~,
\label{covector_rule}
\eeq
to be compared with Eq.\ (\ref{vector_rule}). The common rule is to use 
subscripts to denote the components of dual vectors and superscripts for 
those of vectors.

A $(k,l)$-{\em tensor} $T$ over a vector space $V$ is a 
multilinear map
\beq
T : \underbrace{V^* \times \cdots \times V^*}_{k ~ \text{times}} 
\times \underbrace{V \times \cdots \times V}_{l ~ \text{times}}
 \mapsto {\bf R}~,
\eeq
i.e., acting on $k$ dual vectors and $l$ vectors, $T$ yields a number, 
and it does so in such a manner that if we fix all but one of the vectors or 
dual vectors, it is a linear map in the remaining variable.
A $(0,0)$ tensor is a scalar, a $(0,1)$ tensor is a vector, and a 
$(1,0)$ tensor is a dual vector. 
The space ${\cal T}(k,l)$ of the tensors of type $(k,l)$ is a linear space; 
a $(k,l)$-tensor is defined once its action on $k$ vectors of the dual 
basis and on $l$ vectors of the basis is known, and since there are 
$n^{k}n^{l}$ independent ways of choosing these basis vectors, 
${\cal T}(k,l)$ is a
$n^{k+l}$-dimensional linear space. 
Two natural operations can be defined on tensors. The first one is called 
{\em contraction} with respect to the $i$-th (dual vector) and the 
$j$-th (vector) arguments and is a map 
\beq
C : T \in {\cal T}(k,l) \mapsto CT \in {\cal T}(k - 1,l -1)
\eeq
defined by
\beq
CT = \sum_{\sigma = 1}^n T(\ldots,\underbrace{v^{\sigma 
\ast}}_{i},\ldots;\ldots,\underbrace{v_{\sigma}}_{j},\ldots)~.
\eeq
The contracted tensor $CT$ is independent of the choice of the basis, 
so that the contraction is a well-defined, invariant, operation. 
The second operation is the {\em tensor product}, which maps an element 
${\cal T}(k,l) \times {\cal T}(k',l')$ into an element of 
${\cal T}(k+k',l+l')$, i.e., two tensors $T$ and $T'$ into a new tensor, 
denoted by $T \otimes T'$, defined as follows: given $k+k'$ dual vectors 
$v^{1*},\ldots,v^{k+k'*}$ and $l+l'$ vectors $w_1, \ldots, w_{l+l'}$, then
\beq
T \otimes T' (v^{1*},\ldots,v^{k+k'*};w_1, \ldots, w_{l+l'}) = 
T (v^{1*},\ldots,v^{k*};w_1, \ldots, w_{l})\, 
T' (v^{k+1*},\ldots,v^{k+k'*};w_{l+1}, \ldots, w_{l+l'})~. 
\eeq
The tensor product allows one to construct a basis for ${\cal T}(k,l)$ 
starting from a basis $\{ v_\mu \}$ of $V$ and its dual basis 
$\{ v^{\nu *} \}$: such a
basis is given by the $n^{k+l}$ tensors $\{v_{\mu_1} \otimes \cdots
\otimes v_{\mu_k} \otimes v^{\nu_1 *}  \otimes \cdots  
\otimes v^{\nu_l *} \}$. 
Thus, every tensor $T \in {\cal T}(k,l)$ allows a decomposition
\beq
T = \sum_{\mu_1,\ldots,\nu_l = 1}^n
T^{\mu_1\cdots\mu_k}_{~~~~~~~~\nu_1\cdots\nu_l}
v_{\mu_1} \otimes \cdots \otimes v^{\nu_l *}~;
\label{tensor_components}
\eeq
the numbers $T^{\mu_1\cdots\mu_k}_{~~~~~~~~\nu_1\cdots\nu_l}$ 
are called the {\em components} of $T$ in the basis $\{ v_\mu \}$.
The components of the contracted tensor $CT$ are
\beq
(CT)^{\mu_1\cdots\mu_{k-1}}_{~~~~~~~~~~~\nu_1\cdots\nu_{l-1}} = 
T^{\mu_1\cdots\sigma\cdots\mu_k}_{~~~~~~~~~~~~\nu_1\cdots\sigma\cdots\nu_l}
\eeq
(remember the summation convention), and the components of the 
tensor product $T \otimes T'$ are
\beq
(T \otimes T')^{\mu_1\cdots\mu_{k+k'}}_{~~~~~~~~~~\nu_1\cdots\nu_{l+l'}} =
T^{\mu_1\cdots\mu_k}_{~~~~~~~~\nu_1\cdots\nu_l} \, 
T'^{\mu_{k+1}\cdots\mu_{k+k'}}_{~~~~~~~~~~~~~~~\nu_{l+1}\cdots\nu_{l+l'}}~.
\eeq
All these results are valid for a generic vector space, so that they hold in 
particular for the vector spaces of the tangent bundle $TM$ of $M$, over 
which tensors (and, analogously to vector fields, {\em tensor fields}) 
can be defined exactly as above.

\subsubsection{Riemannian metrics}

\label{sec_app_metrics}

The infinitesimal square distance on $M$, i.e., the length element $ds^2$ 
(also referred to as the {\em metric}) can be defined at each point $P 
\in M$ by means of a  $(0,2)$-tensor $g$, provided it is {\em symmetric}, 
i.e., $g(v,w) = g(w,v)$, and {\em nondegenerate}, i.e., $g(v,w) = 0 
~\forall v \in T_P M$ if and only if $w = 0$\footnote{Or, equivalently, 
$g(v,w) = 0 ~\forall w \in T_P M$ if and only if $v = 0$; 
the two statements are 
equivalent because $g$ is a symmetric tensor.}. In fact, a $g$ with these 
properties 
induces on the tangent bundle $TM$ a nondegenerate quadratic form 
(called the {\em scalar product}),
\beq
g(v,w) = \langle v , w \rangle : TM \times TM \mapsto {\bf R}~.
\label{scalprod}
\eeq
Then it is possible to measure lengths on the manifold. 
A manifold $M$, equipped with a scalar product, is called a 
(pseudo){\em Riemannian manifold}, and the scalar product is referred to 
as a (pseudo){\em Riemannian structure} on $M$.
If the quadratic form (\ref{scalprod}) is positive-definite, then one speaks 
of a (proper) {\em Riemannian metric}. In the latter case the squared length 
element is always positive. For instance, one can define the length of a 
curve  as
\beq
\ell (\gamma) = \int_\gamma \sqrt{\langle \dot\gamma,\dot\gamma \rangle} 
\, dt~.
\label{elle}
\eeq
The curves $\gamma$ which are extremals of the length functional are 
called the {\em geodesics} of $M$.

In a coordinate basis, we can expand the metric $g$ as
\beq
g = g_{ij}\, dx^i \otimes dx^j~,
\eeq
so that  one defines the invariant (squared) length element on the 
manifold, in local coordinates, as
\beq
ds^2 = g_{ij} dx^i dx^j~.
\eeq
The scalar product of two vectors $v$ and $w$ is given, in terms of $g$, by
\beq
\langle v,w \rangle = g_{ij} v^i w^j = v_j w^j = v^i w_i~.
\eeq
In the above equation we have made use of the fact that $g$ estabilishes a 
one-to-one correspondence between vectors and dual vectors, i.e., in 
components,
\beq
g_{ij}v^j = v_i ~.
\eeq
For this reason, the components of the inverse metric $g^{-1}$ are simply 
denoted by $g^{ij}$, instead of $(g^{-1})^{ij}$, and allow to pass from dual 
vector (covariant) components to vector (contravariant) components:
\beq
g^{ij}v_j = v^i ~.
\eeq
This operation of raising and lowering the indices can be applied not only 
to vector, but also to tensor components. This allows us to pass from $(k,l)$ 
tensor components to the corrseponding $(k+1,l-1)$ tensor components and vice 
versa. What does not change in the operation is the sum $k+l$ which is called 
the {\em rank} (or the {\em order}) of the tensor.

\subsection {Covariant differentiation}

The introduction of a differential calculus on a manifold which is not 
Euclidean
is complicated by the fact that ordinary derivatives do not map vectors into 
vectors, i.e., the ordinary derivatives of the components of a vector $w$, 
$dw^i/dt$, taken for instance at a point $P$ along a given curve $\gamma(t)$, 
are {\em not} 
the components of a vector in $T_P M$, because they do not transform according
to the rule (\ref{vector_rule}). The geometric origin of this fact is that the 
{\em parallel transport} of a vector from a point $P$ to a point $Q$ on a 
non-Euclidean manifold {\em does} depends on the path chosen to join $P$ and 
$Q$. 
Since in order 
to define the derivative of a vector at $P$, we have to move that vector 
from $P$ to a neighboring point  along a curve
and then  parallel-transport it back to the original point in order to 
measure the difference, we need a definition of parallel transport to 
define a derivative; conversely, given a (consistent) derivative, i.e., a 
derivative which maps vectors into vectors, one could define the parallel 
transport by imposing that a vector is parallel transported along a curve if 
its derivative along the curve is zero. The two ways are conceptually 
equivalent: we follow the first way, by introducing the notion of a 
{\em connection} and then using it to define the derivative operator. 
Such a derivative will be
referred to as the {\em covariant derivative}.

A (linear) {\em  connection} on the manifold $M$ is a map $\nabla$ such that, 
given two vector fields\footnote{One could also consider tensor fields, but 
for the sake of simplicity we define connections using vectors.} $A$ and $B$, 
it yields a third field $\nabla_A B$ with the following properties:
\begin{itemize}
\item[{\bf 1.}] $\nabla_A B$ is  bilinear in $A$ and $B$, i.e., 
$\nabla_A (\alpha B + \beta C) = \alpha \nabla_A B + \beta \nabla_A C$
and $\nabla_{\alpha A + \beta B} C = \alpha \nabla_A C + \beta \nabla_B C$;
\item[{\bf 2.}] $\nabla_{f(A)} B = f (\nabla_A B)$;
\item[{\bf 3.}] (Leibnitz rule) $\nabla_A f(B) = (\partial_A f) B  + f 
( \nabla_A B)$, where $\partial_A$ is the ordinary directional derivative in 
the direction of $A$.
\end{itemize}
The {\em parallel transport} of a vector $V$ along a curve $\gamma$, 
whose tangent vector field is $\dot\gamma$, is then defined as the (unique) 
vector field $W(t) = W(\gamma(t))$ along $\gamma(t)$ such that 
\begin{itemize}
\item[{\bf 1.}] $W(0) = V$;
\item[{\bf 2.}] $\nabla_{\dot\gamma} W = 0$ along $\gamma$.
\end{itemize}
The notion of covariant derivative now immediately follows: 
the {\em covariant derivative} $DV/dt$ of $V$ along $\gamma$ is given by 
the vector field
\beq
\frac{DV}{dt} = \nabla_{\dot\gamma} V~.
\label{def_D}
\eeq
On the basis of  Eq.\ (\ref{def_D}), with a certain abuse of language, 
one often refers to $\nabla_X Y$ as the covariant derivative of $Y$ along $X$,
where $X$ and $Y$ are generic vector fields.
Among all the possible linear connections, and given a metric $g$, there is 
one and only one which $(i)$ is {\em symmetric}, i.e., 
\beq
\nabla_X Y - \nabla_Y X = [X,Y] ~~~~~ \forall X,Y~,
\label{symm_connection}
\eeq 
and $(ii)$ conserves the scalar product, i.e., the scalar product of two 
{\em parallel} vector fields $P$ and $P'$ is constant along $\gamma$,
\beq
\frac{d}{dt} \langle P,P' \rangle \equiv 0~.
\eeq
Such a linear connection is obviously the natural one on a Riemannian 
manifold, and is referred to as the {\em Levi-Civita} (or {\em Riemannian}) 
connection. Whenever we refer to a {\em covariant derivative} without any 
specification, we mean the covariant derivative induced by the Riemannian 
connection.

The components
of the Riemannian connection $\nabla$ with respect to a coordinate basis 
$\{ X_i \}$  are the  {\em Christoffel symbols}, given by
\beq
\Gamma^i_{jk} = \langle dq^i, \nabla_{X_j} X_k \rangle~.
\eeq
and are given, in terms of the derivatives of the components of the metric,
by the following formula
\beq
\Gamma^i_{jk} = 
\frac{1}{2} g^{im} 
\left( \partial_j g_{km} + \partial_k g_{mj} - \partial_m 
g_{jk} \right)~,
\label{Gamma}
\eeq
where $\partial_i = \partial/\partial x^i$.
The expression in local coordinates of the covariant derivative (\ref{def_D}) 
of a vector field $V$ is then
\beq
\frac{DV^i}{dt} = \frac{dV^i}{dt} + \Gamma^i_{jk}\frac{dx^j}{dt} V^k~.
\label{D_local}
\eeq

\subsubsection{Geodesics}

The {\em geodesics}, which were already defined as the curves of extremal 
length on the manifold, 
can also be defined as {\em self-parallel curves}, i.e., curves 
such that the tangent vector $\dot\gamma$ is always parallel transported. 
Thus geodesics are the curves $\gamma(t)$ which 
satisfy the equation (referred to as the {\em geodesic equation})
\beq
\frac{D\dot \gamma}{dt} = 0
\eeq
whose expression in local coordinates follws from Eq.\ (\ref{D_local}), and is
\beq
\frac{d^2 x^i}{dt^2} + \Gamma^i_{jk} \frac{dx^j}{dt} \frac{dx^k}{dt} = 0~.
\label{eqs_geodet}
\eeq
Since the norm of the tangent vector $\dot\gamma$ of a geodesic is constant, 
$|d\gamma/dt| = c$, the arc lenght of a geodesic is proportional to the 
parameter:
\beq
s(t) = \int_{t_1}^{t_2} \left| \frac{d\gamma}{dt} \right| dt = c(t_2 - t_1)~.
\eeq
When the parameter is actually the arc length, i.e., $c = 1$, we say that the 
geodesic is {\em normalized}. Whenever we consider a geodesic, we assume it 
is normalized, if not explicitly stated otherwise. This means that the 
Eqs.\ (\ref{eqs_geodet}) are nothing but the Euler-Lagrange equations for the 
length functional along a curve $\gamma(s)$ parametrized by the arc length,
\beq
\ell(\gamma) = \int_\gamma ds~.
\eeq
Given a geodesic $\gamma(s)$ on $M$, there exists a unique vector 
field $G$ on $TM$ such that its trajectories are $(\gamma(s),\dot\gamma(s))$. 
Such a vector field is called the {\em geodesic field} and its flow the {\em 
geodesic flow} on $M$.

\subsection {Curvature}

\label{app_sec_curvature}
The curvature of a Riemannian manifold $(M,g)$ is --- 
intuitively --- a way of measuring how much
the manifold deviates from being Euclidean.  
The {\em curvature tensor}, also known as the 
{\em Riemann-Christoffel tensor},
is a tensor of order 4 defined as
\beq
R(X,Y)=\nabla_X\nabla_Y-\nabla_Y\nabla_X-\nabla_{[X,Y]}~,
\label{tens_curv}
\eeq
where $\nabla$ is the Levi-Civita connection of $M$. Observe 
that 
if $M = {\bf R}^N$, then $R(X,Y) = 0$ for all the pairs of 
tangent vectors $X,Y$, because of the commutativity of the ordinary
derivatives. In addition, $R$ measures the non-commutativity of the 
covariant
derivative: in fact, if we choose a coordinate system 
$\{x_1,\ldots,x_n\}$,  we have,
since $\left[\frac{\partial}{\partial 
x_i},\frac{\partial}{\partial x_j}\right] = 0$,
\beq
R \left( \frac{\partial}{\partial 
x_i},\frac{\partial}{\partial x_j}\right) = 
\nabla_{\partial/\partial x_i} \nabla_{\partial/\partial 
x_j} -
\nabla_{\partial/\partial x_j} \nabla_{\partial/\partial 
x_i}~.
\eeq
In local coordinates, the components of the Riemann 
curvature tensor (considered here as a (1,3)-tensor) are given by 
\beq
R^i_{~jkl} = \frac{\partial \Gamma^i_{jl}}{\partial x^k} - 
\frac{\partial \Gamma^i_{kl}}{\partial x^j} +
\Gamma^r_{jl} \Gamma^i_{kr} - \Gamma^r_{kl} \Gamma^i_{jr}~.
\label{curv_components}
\eeq
Thus, given a metric $g$, the curvature $R$ is uniquely defined.
A manifold $(M,g)$ is called {\em flat} when the curvature tensor vanishes. 

Given a {\em positive} function $f^2$, a {\em conformal transformation} is 
the transformation
\beq
(M,g) \mapsto (M,\tilde g)~; ~~~~~~ \tilde g = f^2 g~.
\eeq
Two Riemannian manifolds are said {\em conformally related} if they are 
linked by a conformal transformation. In particular, a manifold is $(M,g)$ 
{\em conformally flat} 
if it is possible to find a conformal transformation which sends $g$ into a 
flat metric. Conformally flat manifolds exhibit some remarkable 
simplifications for the calculation of the curvature tensor components 
(see e.g.\ \cite{Goldberg}; an application is given in \S \ref{curv_mech}).

Closely related to the curvature tensor is the sectional --- 
or Riemannian --- curvature, which we define now.
Let us consider two vectors $u,v \in T_P M$, and let us put 
\beq
|u \wedge v| = \left(|u|^2 |v|^2 - \langle u,v \rangle 
\right)^{1/2}~,
\eeq
which is the area of the two-dimensional parallelogram 
determined by
$u$ and $v$. If $|u \wedge v| \not = 0$ the vectors $u,v$ 
span a two-dimensional
subspace $\pi \subset T_P M$. We define the {\em sectional 
curvature} 
at the point $P$ relative to $\pi$, as the quantity:
\beq
K(P;u,v) = K(P,\pi) = \frac{\langle R(v,u)u,v \rangle}{|u 
\wedge v|^2}
\label{K}
\eeq
which can be shown to be independent of the choice of the two vectors
$u,v \in \pi$.
In local coordinates, Eq.\ (\ref{K}) becomes
\beq
K(P;u,v) = R_{ijkl} \frac{u^i v^j u^k v^l}{|u \wedge v|^2}~.
\label{sec_curv_comp}
\eeq
The knowledge of $K$ for the $N(N-1)$ planes 
$\pi$ spanned by a maximal set of linearly independent vectors
completely 
determines
$R$ at $P$. 

If $\dim (M) = 2$ then $K$ coincides with the 
Gaussian curvature of the surface, i.e., with the product of 
the reciprocals of two curvature radii.

A manifold is called {\em isotropic} if $K(P,\pi)$ does not 
depend on the
choice of the plane $\pi$. 
\label{Schur} The remarkable result --- Schur's 
theorem \cite{doCarmo} --- is that in this case $K$ is also 
constant, i.e. it does not depend on the point $P$ either.

Some ``averages'' of the sectional curvatures are very important. 
The {\em Ricci curvature} $K_R$ at $P$ in the direction $v$ 
is defined as the
sum of the sectional curvatures at $P$ relative to the 
planes determined by $v$ and the $N - 1$ directions orthogonal to 
$v$, i.e., if $\{e_1,\ldots,e_{N-1},v = e_N \}$ is an orthonormal basis of 
$T_P M$ and $\pi_i$ is the plane spanned by $v$ and $e_i$,
\beq
K_R(P,v) = \sum_{i = 1}^{N-1} K(P,\pi_i)~.
\label{ricci_curvature}
\eeq
The {\em scalar curvature} ${\cal R}$ at $P$ is the sum 
of the $N$ Ricci curvatures at $P$,
\beq
{\cal R}(P) = \sum_{i = 1}^{N} K_R(P,e_i)~.
\label{scalar_curvature}
\eeq
In terms of the components of the curvature tensor, such curvatures can be 
defined as follows (in the following formul{\ae}, we drop the dependence on
$P$, because it is understood that the components are local quantities). We 
first define the {\em Ricci tensor}  as the two-tensor whose
components, $R_{ij}$, are obtained by contracting the first and the 
third indices of the Riemann tensor,
\beq
R_{ij} = R^k_{~ikj}~; 
\label{ricci_tensor}
\eeq
then 
\beq
K_R(v) = R_{ij} v^i v^j~. 
\label{ricci_curv_comp} 
\eeq
The right hand side 
of Eq.\ (\ref{ricci_curv_comp}) is called ``saturation'' of
$R_{ij}$ with $v$.
The scalar curvature can be obtained as the trace of the 
Ricci tensor,
\beq
{\cal R} = R^i_{~i}~. 
\label{scalar_trace}
\eeq
In the case of a {\em constant curvature} 
--- or isotropic --- manifold, the components of the Riemann curvature 
tensor have the remarkably simple form 
\beq
R_{ijkl} = K \, (g_{ik}g_{jl} - g_{il}g_{jk})~,
\label{R_comp_const}
\eeq
where $K$ is the constant sectional curvature, so that
the components of the Ricci tensor are
\beq
R_{ij} = K \, g_{ij}~,
\label{Ricci_comp_const}
\eeq
and all the above defined
curvatures are constants, and are related by
\beq
K = \frac{1}{N-1} K_R = \frac{1}{N(N-1)} {\cal R}~.
\label{curv_relations}
\eeq

\subsection{The Jacobi equation}

\label{sec_app_jacobi}

In this subsection we give a derivation of the Jacobi equation, already 
introduced in the main text as Eq.\ (\ref{eq_jacobi_geo}).
We will proceed as follows: first, we will define the geodesic
separation vector field $J$, then we will show that the field $J$ is
actually a Jacobi field, i.e., obeys the Jacobi equation.

\begin{figure}
\centerline{\psfig{file=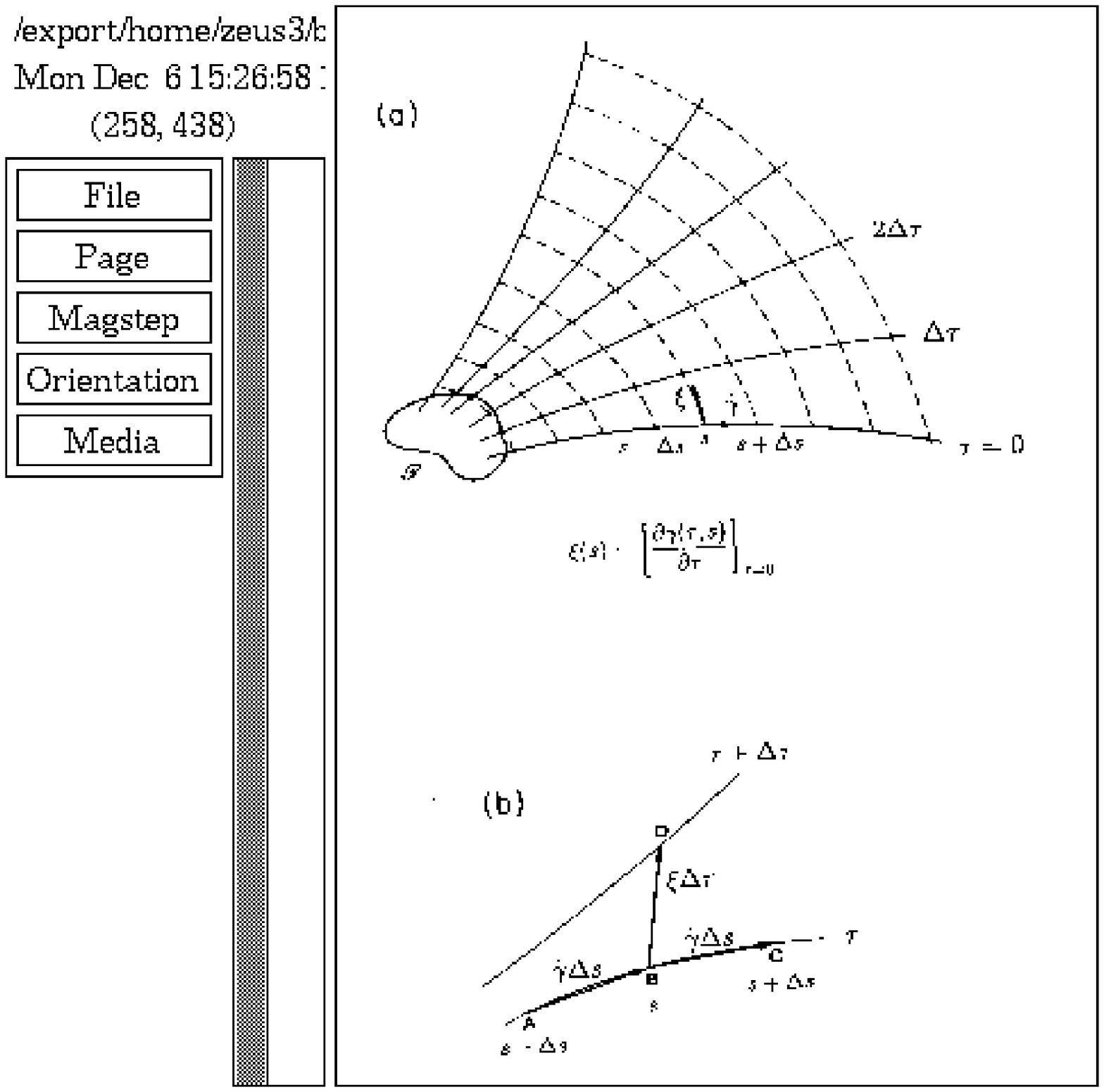,height=10cm,clip=true}} 
\caption{Pictorial description of the definition of the geodesic separation 
vector $J$. From Ref.\ \protect\cite{Pettini}.}
\label{fig_def_J}
\end{figure}

Let us define a 
{\em geodesic congruence} as a family of geodesics 
$\{ \gamma_\tau(s) = \gamma(s,\tau) \, | \, \tau \in {\bf R} \}$ 
issuing from a neighborhood ${\cal I}$ of a point of a
manifold, smoothly
dependent on the parameter $\tau$, and let us fix a reference 
geodesic $\bar\gamma (s,\tau_0)$. Denote then by 
$\dot\gamma(s)$ the vector field tangent to $\bar\gamma$ in $s$, i.e., 
the velocity vector field whose components are
\beq
\dot\gamma^i = \frac{dx^i}{ds}~,
\label{gamma_def_loc}
\eeq
and by $J(s)$  the vector field tangent in $\tau_0$ to the curves 
$\gamma_s(\tau)$ for a fixed $s$, i.e., the vector field of components
\beq
J^i = \frac{dx^i}{d\tau}~.
\label{J_def_loc}
\eeq
The field $J$ will be referred to as the {\em geodesic separation field},
and measures the distance between nearby geodesics, as is 
shown in Fig.\ \ref{fig_def_J}.

Let us now show that $J$ is a Jacobi field. 
First of all, we notice that the field $J$ commutes with $\dot\gamma$, i.e., 
$[\dot\gamma, J] = 0$.  
In fact, from the definition of the commutator (Eq.\ \ref{def_commutator})
and from the definitions of $J$, Eq.\ (\ref{J_def_loc}), and of $\dot\gamma$,
Eq.\ (\ref{gamma_def_loc}), we have
\bea
[\dot\gamma,J]^i & = & \dot\gamma^j \frac{\partial J^i}{\partial x^j} - 
J^j \frac{\partial \dot\gamma^i}{\partial x^j} \nonumber \\
& = & \frac{\partial x^j}{\partial s} \frac{\partial J^i}{\partial x^j} -
\frac{\partial x^j}{\partial \tau}\frac{\partial \dot\gamma^i}{\partial x^j} 
\nonumber \\
& = & \frac{\partial J^i}{\partial s} - 
\frac{\partial \dot\gamma^i}{\partial \tau}~,
\eea
and using again Eqs.\ (\ref{J_def_loc}) and (\ref{gamma_def_loc}), we find 
that
\beq
\frac{\partial J^i}{\partial s} = \frac{\partial}{\partial s} 
\frac{\partial x^i}{\partial \tau} = \frac{\partial}{\partial \tau} 
\frac{\partial x^i}{\partial s} = \frac{\partial \dot\gamma^i}{d\tau}~,
\eeq
so that $[\dot\gamma, J]=0$.
Now, let us compute the second covariant derivative of the field $J$,
$\nabla_{\dot\gamma}^2 J$.
First of all, let us recall that our covariant derivative
comes from a Levi-Civita connection, which is symmetric 
(see Eq.\ (\ref{symm_connection})), so that
\beq
\nabla_{\dot \gamma} J - \nabla_J {\dot\gamma} = [\dot\gamma,J]~,
\eeq
and having just shown that $[\dot\gamma,J] = 0$, we can write
\beq
\nabla_{\dot\gamma} J = \nabla_{J} \dot\gamma~.
\eeq
Now, using this result, and the fact that $\nabla_{\dot\gamma}
\dot\gamma = 0$ because $\bar\gamma$ is a geodesic, we can write
\beq
\nabla_{\dot\gamma}^2 J = \nabla_{\dot\gamma}\nabla_{\dot\gamma} J =
\nabla_{\dot\gamma}\nabla_J {\dot\gamma} = [\nabla_{\dot\gamma},\nabla_J]
\dot\gamma~,
\eeq
from which, using the definition of the curvature tensor 
(Eq.\ (\ref{tens_curv})) and, again, the vanishing of the commutator
$[\dot\gamma,J]$, we get
\beq
\nabla_{\dot\gamma}^2 J = R(\dot\gamma,J)\dot\gamma~,
\eeq
which is nothing but the Jacobi equation (\ref{eq_jacobi_geo}), written in
compact notation.

It is worth noticing that the normal component $J_{\perp}$ 
of $J$, i.e., the component of $J$ orthogonal to $\dot\gamma$ along 
the geodesic $\gamma$, is again a Jacobi field, since we can 
always write $J=J_{\perp}+\lambda\dot{\gamma}$: one immediately finds 
then that the velocity $\dot{\gamma}$ satisfies the Jacobi equation, so that
$J_{\perp}$ must obey the same equation. This allows us 
to restrict ourselves to the study of the normal Jacobi fields, as
we have already done in the main text.

\section{Summary of elementary Morse theory} 
\label{app_morse} 
The purpose of this Appendix is to recall the main ideas and concepts
of Morse theory which are relevant for the main text of the paper.
For a more elaborate discussion we refer the reader to 
Refs.\ \cite{Morse,Milnor,Palais}. 
 
Morse theory, also referred to as critical point theory, links the {\em  
topology} of a given manifold $M$ with the properties of the {\em critical  
points} of smooth (i.e., with infinitely many derivatives) 
functions defined on it. Morse theory links {\em local}  
properties (what happens at a particular point of a manifold) with 
{\em global} properties (the topology, i.e., the shape, of the manifold
as a whole). Two manifolds $M$ and $M'$ are topologically equivalent if
they can be smoothly deformed one into the other: a tea cup is topologically
equivalent to a doughnut, but it is {\em not} topologically equivalent 
to a ball. In fact a ball has no holes, while both a tea cup and a doughnut 
have one hole. To define precisely what a ``smooth deformation'' is, one has
to resort to the notion of a {\em diffeomorphism}. A diffeomorphism
is a smooth one-to-one map, whose inverse is smooth. Then $M$ can be
smoothly deformed into $M'$ if there exists a diffeomorphism $\psi$ which
maps $M$ into $M' = \psi(M)$. If such a diffeomorphosm exists, we say 
that $M$ and $M'$ are then {\em diffeomorphic}. Thus the notion of
``topological equivalence'' we referred to has now a precise meaning.
 
For the sake of simplicity, we shall consider only compact,  
finite-dimensional manifolds: most of the results can be extended not only to  
noncompact manifolds, but also to infinite-dimensional manifolds modeled on  
Hilbert spaces (see \cite{Palais}). 
 
The key ingredient of Morse theory is to look at the manifold $M$ 
as decomposed  
in the {\em level sets} of a function  $f$. 
Let us recall that the $a$-level set of a  
function $f: M \mapsto {\bf R}$ is the set 
\beq 
f^{-1}(a) = \{ x \in M : f(x) = a \}~,
\eeq 
i.e., the set of all the points $x \in M$ such that $f(x) = a$. 
Now, $M$ being compact, any function $f$ has a minimum, $f_{\rm min}$, and a  
maximum, $f_{\rm max}$, so that  
\beq 
f_{\rm min} \leq f(x) \leq f_{\rm max} \qquad \forall x \in M~.  
\eeq 
This means that the whole manifold $M$ can be decomposed in the level sets of  
$f$: in fact, one can build $M$ starting from $f^{-1}(f_{\rm min})$ and then  
adding continuously to it all the other 
level surfaces up to $f^{-1}(f_{\rm max})$. To be  
more precise, one defines the ``part of $M$ below $a$'' as 
\beq 
M_a = {f}^{-1} (-\infty,a] = \{ x \in M : f(x) \leq a\}~, 
\eeq 
i.e., each $M_a$ is the set of the points 
$x \in M$ such that the function $f(x)$ does  
not exceed a given value $a$; 
as $a$ is varied between $f_{\rm min}$ and $f_{\rm  
max}$, $M_a$ describes the whole manifold $M$. 
 
For our purposes, we need to restrict the class of functions we are  
interested in to the class of {\em Morse functions}, which are defined as  
follows. Given a manifold $M$ of dimension $n$
and a smooth function $f: M \mapsto {\bf R}$, a 
point $x_c \in M$ is called a {\em critical point} 
of $f$ if $df(x_c) = 0$, while 
the value $f(x_c)$ is called a 
{\em critical value}. The function $f$ is called  
a {\em Morse function} on $M$ if 
its critical points are nondegenerate, i.e., if  
the Hessian matrix of $f$ at $x_c$, whose elements in local coordinates are 
\beq 
H_{ij} = \frac{\partial^2 f}{\partial x^i \partial x^j}~, 
\eeq 
has rank $n$, i.e., has only nonzero eigenvalues. This means that 
there are no directions along which one could move the critical point,
so that there are no lines (or surfaces, or hypersurfaces) made of
critical points.  
As a consequence, one can  
prove that the critical points $x_c$ of a Morse function, and also 
its critical values, are isolated.
It can be proved also that  
Morse functions are generic: the space of the Morse functions is a  
{\em dense} subset of the space of the smooth functions from $M$ to $\bf R$. 
A level set $f^{-1}(a)$ of $f$ is called a {\em critical level} 
if $a$ is a critical value of $f$, i.e., if there is at least one  
critical point $x_c \in f^{-1}(a)$.  
 
The main results of Morse theory are the following: 
\begin{itemize} 
\item[{\bf 1.}] If the interval $[a,b]$ 
contains no critical values of $f$, then  
the topology of $f^{-1}[a,v]$ does not change for any $v \in (a,b]$. 
This result\footnote{We note that this result is valid even if $f$  
is not a Morse function; it is sufficient that it is a smooth function.}
is sometimes called the {\em  
non-critical neck theorem}. 
The reason for this terminology will be made clear  
in the following. 
\item[{\bf 2.}] If the interval $[a,b]$ contains critical values, 
the topology  
of $f^{-1}[a,v]$ changes in correspondence with the 
critical values themselves, in  
a way which is completely determined by the properties of the Hessian
of $f$ at the critical points.  
\item[{\bf 3.}] Some 
topological invariants of $M$, i.e., quantities that are the
same for all the manifolds which have the same topology as $M$, so that
they characterize unambigously 
the topology itself, can be estimated and sometimes  
computed exactly once all the critical points of $f$ are known. 
\end{itemize} 
 
Without giving explicit proofs, which can be found in Ref.\ \cite{Palais},
let now us discuss in more detail items {\bf 1}-{\bf 3}
above. 
 
\subsection{The non-critical neck theorem} 

\label{app_sec_neck} 
\begin{figure} 
\centerline{\psfig{file=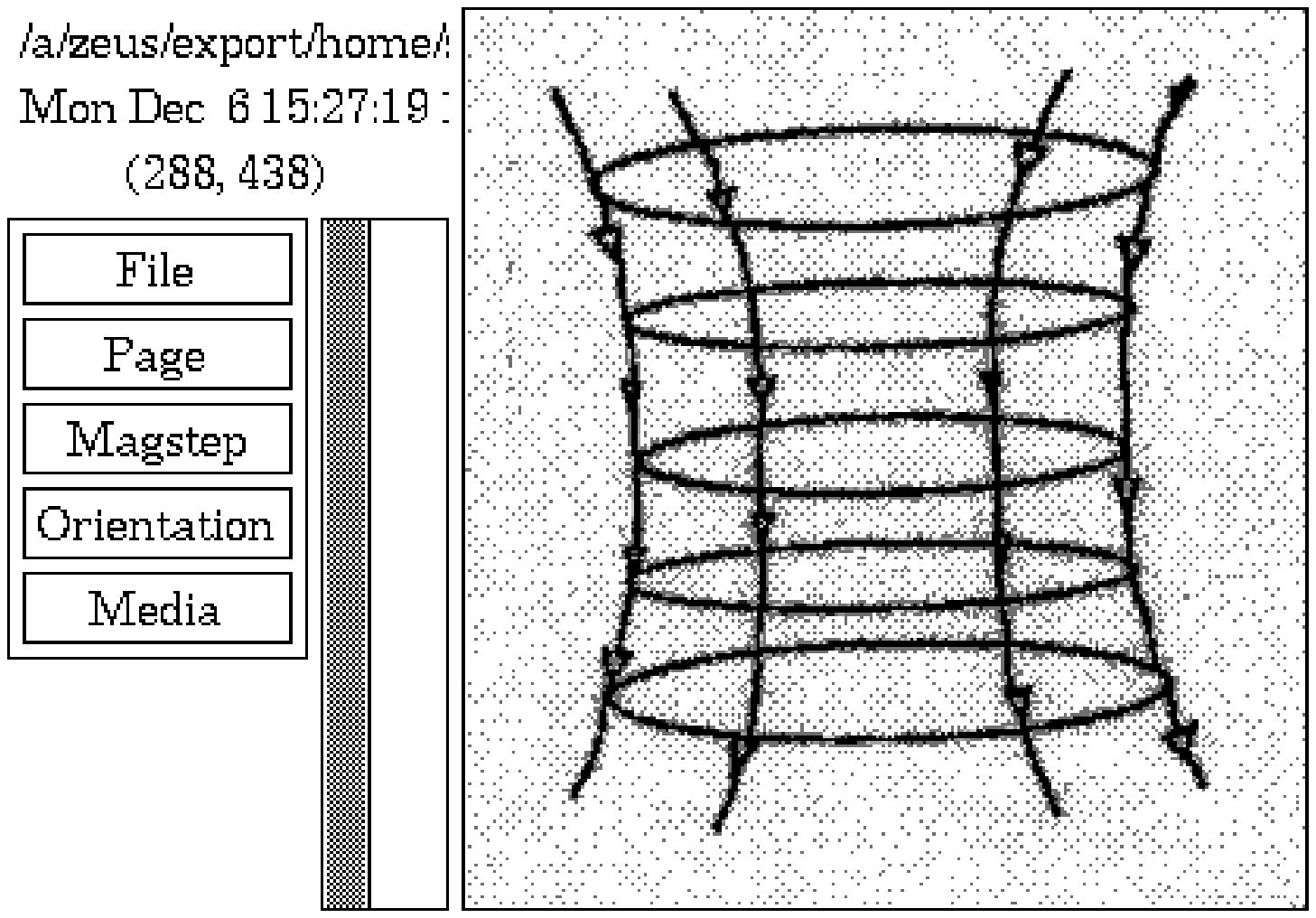,height=7cm,clip=true}} 
\caption{A non-critical neck. The lines with the arrows are the flow lines
of $\nabla f$, and the ellipses are the level sets of $f$.} 
\label{fig_neck} 
\end{figure}

If there are no critical values in the interval $[a,b]$,  
there exists a diffeomorphism 
which sends $f^{-1}[a,b]$ into the Cartesian product 
$f^{-1}(a) \times [a,b]$. This means that the  
shape of $f^{-1}[a,b]$ is that of a multi-dimensional 
cylinder, or a neck (from which the name ``non-critical neck''), 
if $f^{-1}(a)$ is simply connected, because the Cartesian product of
a circle and an interval is a cylinder. This  
might be better understood with the aid of a two-dimensional example. Suppose  
that $M$ is two-dimensional, and that the level set $f^{-1}(a)$ is 
topologically equivalent to a circle (see Fig.\ \ref{fig_neck}).  
Then one can construct a diffeomorphism explicitly as 
the flow of the gradient  
vector field of $f$, $\nabla f$, whose flow lines are orthogonal to the level  
surfaces of $f$ and are depicted as the lines with the arrows in 
Fig.\ \ref{fig_neck}. This flow has no singularities if there are no critical 
values of $f$, so that the level set $f^{-1}(a)$ is transported up to 
$f^{-1}(b)$ along the flow lines of $\nabla f$ without  
changing its topology. As a consequence, 
\beq 
f^{-1}[a,b] \approx f^{-1}(a) \times [a,b] \approx f^{-1}(b) \times [a,b]~,  
\eeq    
where ``$x \approx y$'' must be read as ``$x$ is diffeomorphic to $y$''. 
 
\subsection {Critical points and topology changes} 
 
In  
the neighborhood of a regular point $P$, $N(P)$,  there always exists a  
coordinate system such that $f$ can be written
as its first-order Taylor expansion\footnote{This follows 
from the implicit function theorem.},  
setting the origin of such coordinates in $P$, in the form 
\beq 
f(x) = f(0) + \frac{\partial f}{\partial x^i} x^i + \cdots \qquad \forall 
x\in N(P)~. 
\eeq 
Geometrically, this means that in the neighborhood of a regular 
point the level  
sets of $f$ look like hyperplanes in ${\bf R}^n$, because they are the
level sets of a {\em linear} function. 
 
\begin{figure} 
\centerline{\psfig{file=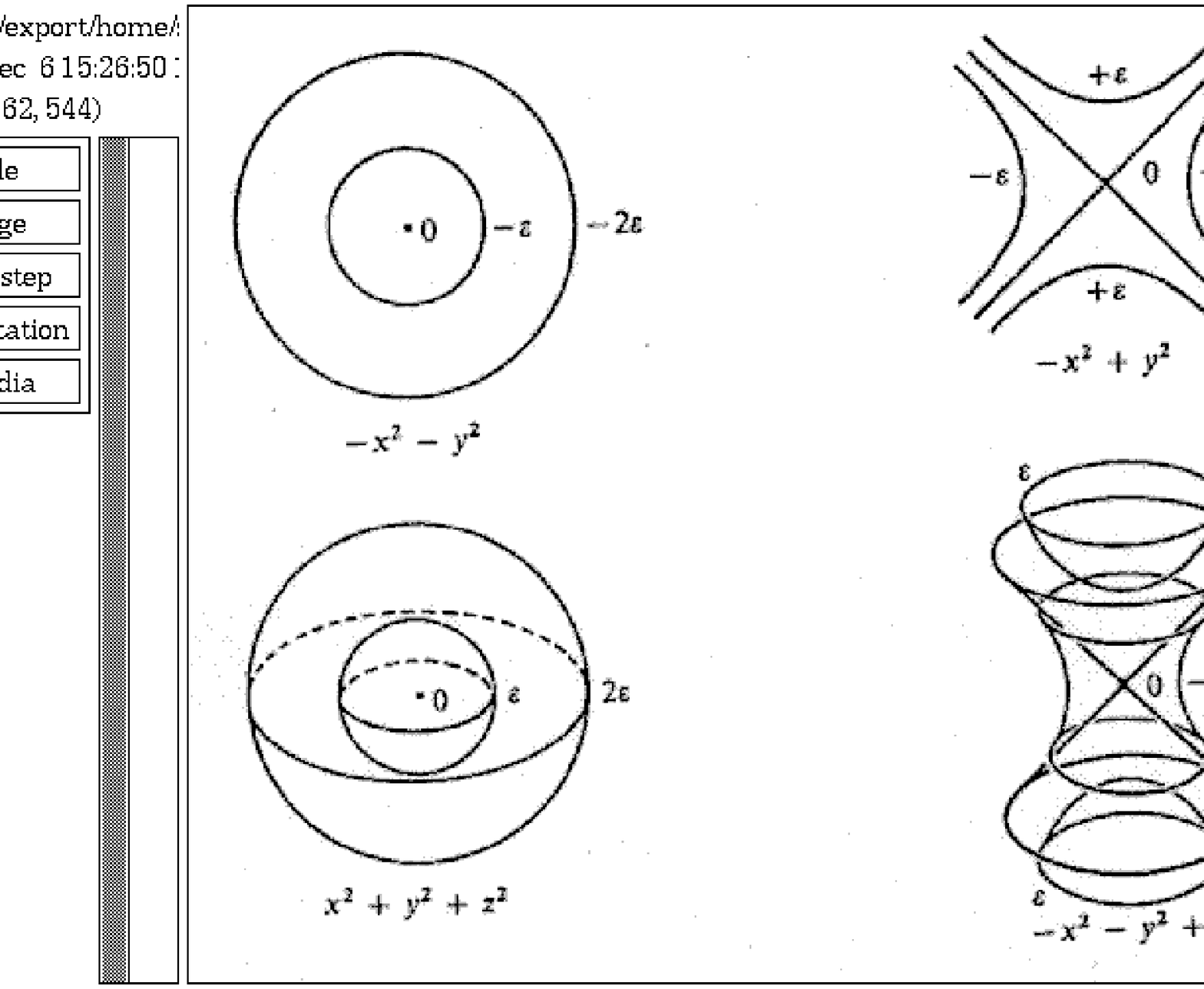,height=11cm,clip=true}} 
\caption{Some examples of $\varepsilon$-level sets near a 
critical point (the critical value
of the function is set to $0$). 
Upper left: $n = 2$, critical point of index $k = 2$; upper right: 
$n = 2$, critical point of index $ k = 1$; lower left:
$n=3$, critical point of index $k=0$; lower right: $n=3$, 
critical point of index $k=2$.} 
\label{fig_crit_points} 
\end{figure}

But what if $P$ is a critical point of $f$? A fundamental result by M. Morse,  
called the {\em Morse lemma}, is that if $f$ is a Morse function then there 
always exists in $N(P)$ a coordinate system (called a {\em Morse chart}) such  
that $f$ is given by its {\em second-order} Taylor polynomial: 
\beq 
f(x) = f(0) + \frac{\partial^2 f}{\partial x^i \partial x^j} x^i x^j + 
\cdots \qquad \forall x\in N(P)~. 
\label{morse_exp} 
\eeq 
With a suitable rotation  of the coordinate frame, $\{x^i\} \mapsto \{y^i \}$,
the expansion (\ref{morse_exp}) can always be reduced to the canonical 
diagonal form 
\beq 
f(y) = f(0) - \sum_{i = 1}^k (y^i)^2 + \sum_{i = k + 1}^n (y^i)^2  + 
\cdots \qquad \forall u\in N(P)~. 
\label{morse_can} 
\eeq 
Close to $P$, the level sets of $f$ are the
level sets of a {\em quadratic} function, so that, 
geometrically, they are  
non-degenerate quadrics, like hyperboloids or ellipsoids, 
which become degenerate at $P$. 
The number of negative  
eigenvalues of the Hessian matrix, $k$, is called the {\em index} of the  
critical point. Passing through the critical level, 
the shape of the level sets  
of $f$ changes dramatically, in a way that is completely 
determined by the index  
$k$. Some examples in two and three 
dimensions are given in Fig.\ \ref{fig_crit_points}. 
 
\begin{figure} 
\centerline{\psfig{file=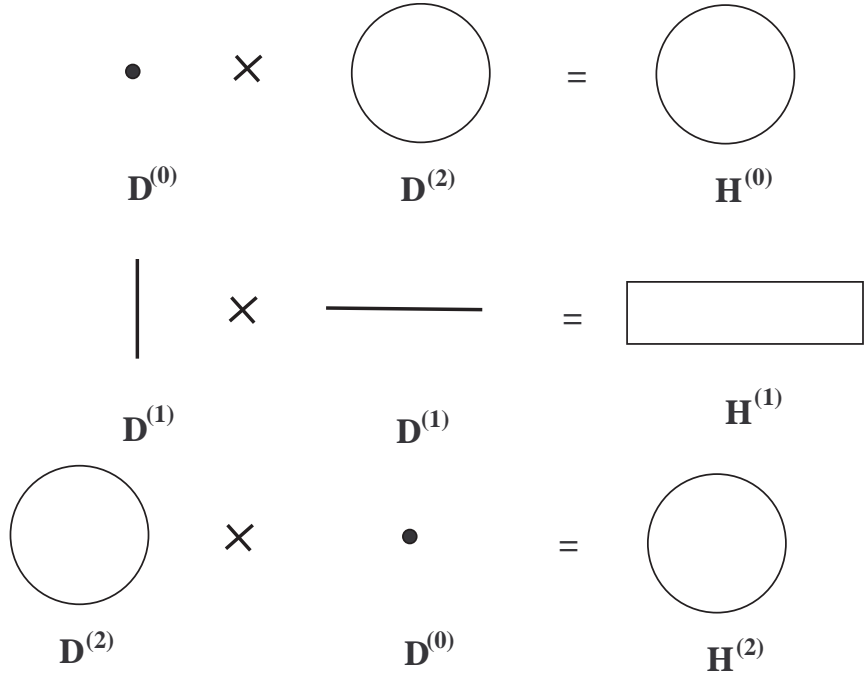,height=9cm,clip=true}} 
\caption{Two-dimensional handles: $H^{(0)}$ is the product of a 0-disk (a
point) and a 2-disk, so that it is a 2-disk; $H^{(1)}$ 
is the product of two 1-disks, i.e., of two intervals, so that it is
a strip; $H^{(2)}$ is again a 2-disk as $H^{(0)}$ is.} 
\label{fig_h_2} 
\end{figure}

The change undergone by the submanifolds $M_a$ as a critical level is passed
is  described using the concept of  
``attaching handles''. A $k$-handle $H^{(k)}$ in $n$ dimensions 
($0 \leq k \leq  
n$) is a product of two disks, one $k$-dimensional ($D^k$) and the other  
$(n-k)$-dimensional ($D^{n-k}$):  
\beq 
H^{(k)} = D^k \times D^{n - k}~. 
\eeq 
In two dimensions, we can have either 0-handles, which are 2-dimensional disks
or 1-handles, which are the product of two 1-dimensional disks, i.e., of two  
intervals, so that they are stripes, or 2-handles, which are 
again 2-dimensional
disks (Fig.\ \ref{fig_h_2}). In three dimensions, we have 0-handles which are  
solid spheres, 1-handles which are the product of a disk and an interval, so 
that they are solid cylinders, 2-handles which are the same as 1-handles,  
and 3-handles which are the same as 0-handles  (Fig.\ \ref{fig_h_2}). 
\begin{figure} 
\centerline{\psfig{file=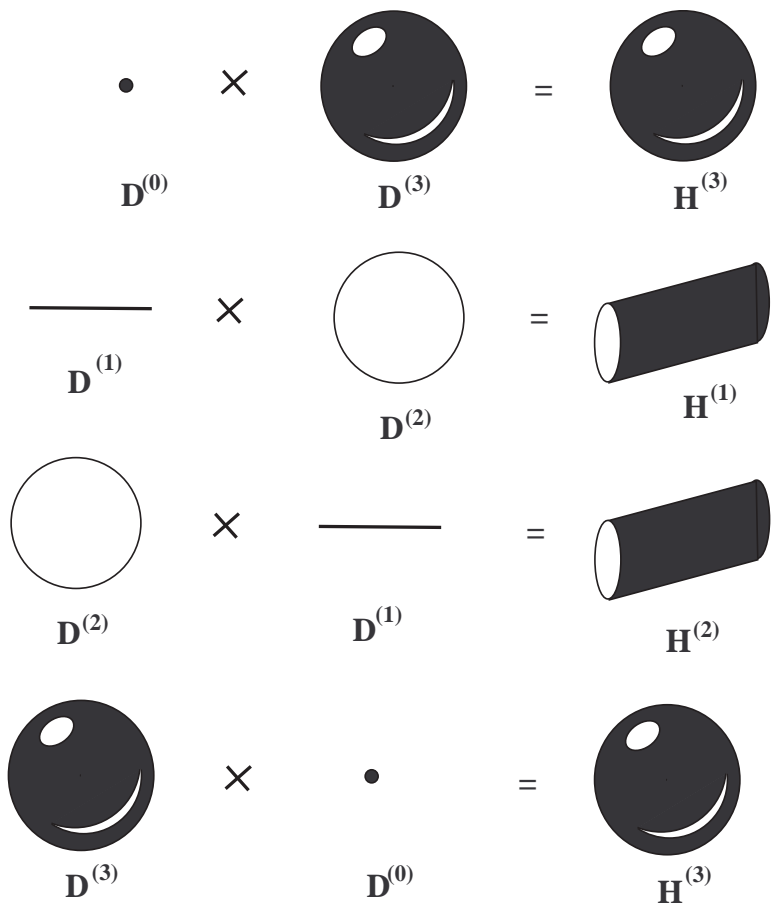,height=12cm,clip=true}} 
\caption{Three-dimensional handles: $H^{(0)}$ is the product of a 0-disk
(a point) and a 3-disk (a ball), so that it is a ball; $H^{(1)}$ is the
product of a 1-disk (an interval) and a 2-disk, so that it is a tube;
$H^{(2)}$ is as $H^{(1)}$, and $H^{(3)}$ is as $H^{(0)}$.} 
\label{fig_h_3} 
\end{figure} 
In more than three dimensions it is difficult to visualize handles: however,  
there is still the duality of the $n = 2$ and $n = 3$ cases, 
i.e., $k$ and $n-k$ handles are topologically equivalent. 
 
Having defined handles, 
we can state the main result of Morse theory as follows.  
 
\vspace{0.2cm} 
\noindent {\bf Theorem.} {\em Suppose that there is one (and only one) 
critical value $c$ in the interval $[a,b]$, and that it 
corresponds to only one critical point of index $k$. Then the manifold $M_b$ 
arises from $M_a$ by attaching a  
$k$-handle, and the transition occurs precisely at the critical level $c$. 
Everything goes in the same way if there are $m > 1$ critical points, with 
indices $k_1,\ldots,k_m$ on the critical level $f^{-1}(c)$; in this case $M_b$
arises from $M_a$ by attaching $m$ disjoint handles of types $k_1,\ldots,k_m$. 
} 
\vspace{0.2cm} 
 
\begin{figure} 
\centerline{\psfig{file=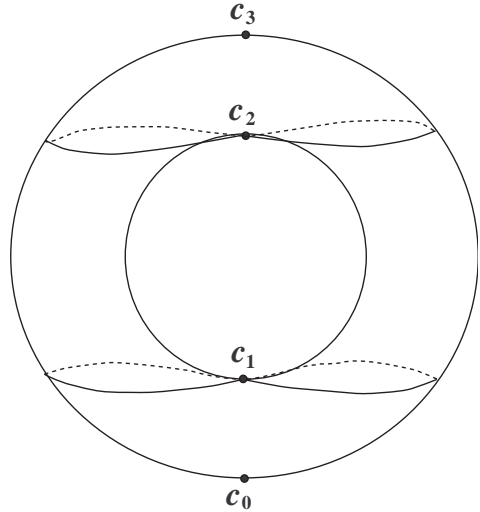,height=7cm,clip=true}} 
\caption{The critical points and critical levels of the height function on a   
two-dimensional torus.} 
\label{fig_torus} 
\end{figure}

\begin{figure} 
\centerline{\psfig{file=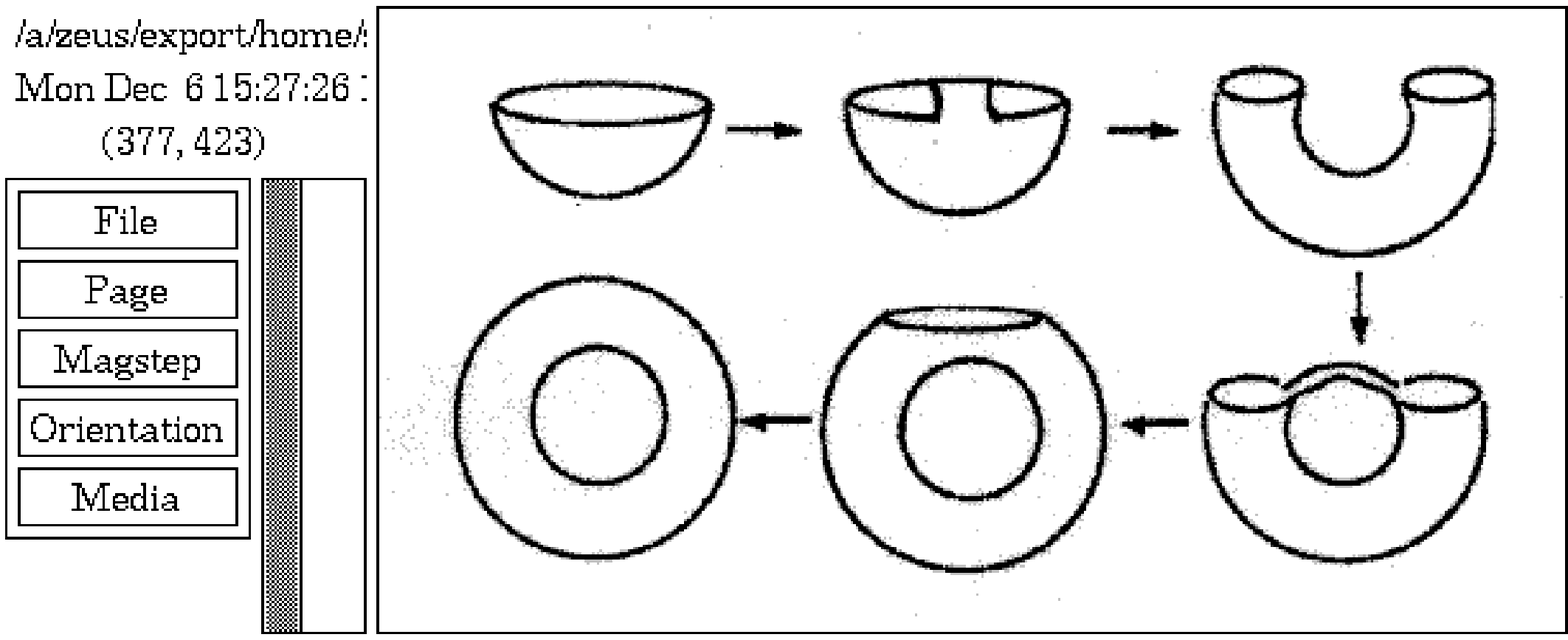,height=7cm,clip=true}} 
\caption{The building of a two-dimensional torus by attaching handles.
$(a)$ Attaching a 0-handle. $(b)$ attaching a 1-handle.
$(c)$ Attaching a 1-handle again. $(d)$ attaching a 2-handle to 
complete the torus. The symbol ``$\approx$'' means ``is diffeomorphic to''.} 
\label{fig_steps} 
\end{figure}

Let us see how this works in a simple example. Consider as our manifold $M$ a  
two-dimensional torus standing on a plane (think of a tyre in a ready-to-roll  
position), and define a function $f$ on it as the height of a point of $M$ 
above  
the floor level. If the $z$-axis is vertical, $f$ is the orthogonal projection 
of $M$ onto the $z$-axis. Such a function has four critical points, 
and the corresponding four  
critical levels of $f$, which will 
be denoted as $c_0,c_1,c_2,c_3$, respectively, 
are depicted in Fig.\ \ref{fig_torus}.
We can build our torus in separate steps: each  
step will correspond to the crossing of a critical level of $f$. The steps 
are  
pictorially described in Fig.\ \ref{fig_steps}. As long as $a < 0$, the 
manifold  
$M_a$ is empty. At $a = c_0 = 0$ we cross the first critical value,  
corresponding to a critical point of index 0. This means that we have to 
attach  
a 0-handle (a disk) to the empty set. Any $M_a$ with $0 < a < c_1$ is  
diffeomorphic to a disk, as we can see by cutting a torus at any height 
between  
0 and $c_1$ and throwing away the upper part. At $c_2$ we meet the second  
critical point, which now has index 1, so that we have to attach a 1-handle 
(a  
stripe) to the previous disk, obtaining a sort of a basket. Such a basket 
can be  
smoothly deformed into a U-shaped tube: in fact if we cut a torus at any 
height  
between $c_1$ and $c_2$ and we throw away the upper part, we get a U-shaped  
tube. The third critical point $c_2$ is again a point of index 1, so we have 
to  
glue another stripe to the tube. What we obtain can be smoothly deformed 
into a  
full torus with only the polar cap cut away from it. The last critical point 
has  
index 2, so that the crossing of it corresponds to the gluing of a 2-handle 
(a disk), which is just the polar cap we needed to complete the torus. 
 
\subsection {Topological invariants} 
 
Morse theory can be used also to give estimates, and  
sometimes to compute exactly, 
some topological invariants of our manifold $M$.  
For a   
(two-dimensional) surface, a very important topological invariant is the {\em  
genus} $g$, which equals the 
number of holes of the surface. The generalization  
to $n$ dimensions of the genus is given by the set of the {\em Betti numbers}  
$b_k(M)$, $0 \leq k \leq n$,  
which are the fundamental topological invariants of an 
$n$-dimensional manifold,
and completely describe its topology. For hyperspheres, all the Betti
numbers are zero.  Using the Betti numbers we can 
obtain another topological invariant, the {\em Euler characteristic} $\chi  
(M)$, which is nothing but the alternating sum of the $b_k$: 
\beq 
\chi (M) \equiv \sum_{k = 1}^n (-1)^k b_k (M)~. 
\label{chi} 
\eeq 
For (two-dimensional) surfaces, $\chi = 2 - 2g$ holds.  
 
Now let us consider a generic Morse function $f$ on $M$ and let us define the  
{\em Morse numbers} $\mu_k (M)$ as follows: $\mu_k$ is the total number of  
critical points of $f$ on $M$ which have index $k$. It turns out 
that the Morse  
numbers of a manifold are upper bounds of the Betti numbers, i.e., the 
following  
{\em (weak) Morse inequalities} hold: 
\beq 
b_k (M) \leq \mu_k (M) \qquad k = 0,\ldots,n~. 
\label{morse_ineq} 
\eeq 
Actually, a result stronger than Eqs.\ (\ref{morse_ineq}) holds, which states  
that alternate sums of two, three, four, ... subsequent Betti numbers are  
bounded from above by the alternate sums of the corresponding Morse numbers.  
Starting from this result one can prove the following identity 
\beq 
\chi (M) = \sum_{k = 1}^n (-1)^k \mu_k (M) ~, 
\eeq 
and this provides a way of computing exactly the Euler characterstic of a  
manifold once all the critical points of a Morse function are known. 
 
Among all the Morse functions on a manifold $M$, there is a special class  
(called {\em perfect} Morse functions) for which the Morse inequalities  
(\ref{morse_ineq}) hold as equalities. Perfect Morse functions characterize  
completely the topology of a manifold. It is possible to prove that the 
height  
function on the torus we 
considered above is a perfect Morse function \cite{Palais}. However,  
there are no simple general 
recipes to construct perfect Morse functions (this  
is actually an active area of research).

\section{Chaos in Hamiltonian dynamical systems} 
\label{app_chaos} 
For a long time the equations of Newtonian mechanics have been  
the paradigm of 
classical determinism.  Only quite recently has it been  
realized that 
``determinism'' and ``predictability'' are far from being  
the same concept, and 
that predictability also requires the {\em stability} of the  
solutions of the 
dynamical differential equations. Determinism implies that, once an
initial condition is given, the trajectory is uniquely determined for
all the forthcoming times; stability means that two initially close 
trajectories will remain close in the future (more precisely, their
distance will grow slower than a power of the time). If this is not
true, it becomes impossible to predict the evolution of a system 
even for very small times, as explained, for instance, by 
R.\ P.\ Feynman \cite{Feynman}:

\begin{quotation}
It is true classically that if we knew the position and the velocity of every 
particle in the world, or in a box of gas, we could predict exactly what would
happen. And therefore the classical world is deterministic. 
Suppose, however, that we have a finite accuracy and do not know {\em exactly}
where just one atom is, say to one part in a billion. Then as it goes along 
it hits another atom, and because we did not know the position better than 
one part in a billion, we find an even larger error in the position after the
collision. And that is amplified, of course, in the next collision, so that 
if we start with only a tiny error it rapidly magnifies to a very great 
uncertainty. To give an example: if water falls over a dam, it splashes. 
If we stand nearby, every now and then a drop will land on our nose.
This appears to be completely random, yet such a behavior would be predicted
by purely classical laws. The exact position of all the drops depends upon 
the precise wigglings of the water befor it goes over the dam. How?
the tiniest irregularities are magnified in falling, so that we get complete
randomness. Obviously, we cannot predict the position of the drops unless
we know the motion of the water {\em absolutely exactly}.

Speaking more precisely, given an arbitrary accuracy, no matter how precise,
one can find a time long enough that we cannot make predictions valid for that
long a time. Now the pint is that this length of time is not very large.
It is not that the time is millions of years if the accuracy is one part
in a billion. The time gose, in fact, only logarithimically with the error,
and it turns out that in only a very, very tiny time we lose all our 
information. If the accuracy is taken to be one part in billions and billions and billions --- no matter how many billions we wish, provided we do stop
somewhere --- then we can find a time less than the time it took to state
the accuracy --- after which we can no longer predict what is going to happen!
\end{quotation}

As long as nonlinear  
dynamical systems are 
considered, stability is the exception rather than the rule. 
Even if this relies --- at least from a conceptual  
point of view --- upon 
mathematical results which have been known since the end of  
the last century, 
its importance has only been completely realized with the  
aid of a new and 
powerful approach: numerical simulation.  The very complicated
structure of some trajectories 
which can arise in nonlinear dynamical systems  
was discovered by Poincar\'e \cite{Poincare} in the late XIX  
century, but the 
physics community became fully aware of the existence and of the meaning  
of these structures only they were visualized by computer
simulation in 
the work of H\'enon and Heiles \cite{HenonHeiles}.
 
The instability we are referring to is known as intrinsic  
stochasticity of the
dynamics, or ``deterministic chaos''.  These terms mean that  
the dynamics, being 
completely deterministic, yet exhibits some features that make  
it indistinguishable 
from a random process.  The characteristic feature of a  
chaotic system, which is 
at the basis of the unpredictability of its dynamics, is the  
sensitive 
(exponential) dependence on initial conditions:  the  
distance between two trajectories
which originate in very close-by points in phase space grows  
exponentially in time 
so that the system looses the memory of its initial conditions. Regular 
dynamics, i.e., quasiperiodic motion,  
is --- as far as conservative systems are  
considered --- a ``weak'' 
property, because it is destroyed by very small  
perturbations of the system.  On the contrary,
chaos is a strong property, because given a dynamical system 
where chaos is present, in many cases it will be present even
after the system has been subjected 
to significant perturbations \cite{Lichtenberg}.   
 
Here we recall briefly the main concepts of the theory of Hamiltonian 
dynamical systems which are 
necessary for the understanding of the material on chaos presented 
in this Report.  
The main goal of this Appendix is then 
to provide the reader with a definition of 
the Lyapunov exponents and of a motivation for the introduction of these 
quantities as a ``measure'' of chaos in a dynamical system. 
 
A very good introduction 
to the subject is given in Lichtenberg and Lieberman's  
classic book 
\cite{Lichtenberg}, and,  
at a more pedagogical level, in Tabor's \cite{Tabor} and Ott's textbooks 
\cite{Ott}. An interesting selection of reprints  
can be found in 
MacKay and Meiss \cite{MacKay}. We assume the reader is familiar, at
least at a basic level, with 
the concepts of {\em ergodicity} and {\em mixing}. A discussion on these topic would be far beyond the scope of the present Report; a good introduction on 
these topics can be found in any of the references just mentioned above.
 
\subsection{A simple example of chaotic dynamics: the perturbed pendulum} 

Throughout the paper we have been concerned with Hamiltonian dynamical systems
with a large number of degrees of freedom. However, the main features of 
chaos can be better appreciated starting with an example of a system with
only one degree of freedom, subjected to an external perturbation: the
forced pendulum. Although the behavior of many-degree-of-freedom is much more
complicated, nevertheless some of the essential features of chaos are already 
present in this simple example.  

The forced pendulum is a system obeying the follwing equation of motion:
\beq  
\ddot q + \sin q = \varepsilon \sin (q - \omega t)~.
\label{forced_pendulum}  
\eeq  
The phase space of the forced pendulum is three-dimensional, because in
addition to the coordinate $q$ and to the momentum $p = \dot q$,
one has to take into account also the time $t$, because the forcing term
on the right-hand side of Eq.\ (\ref{forced_pendulum}) depends explicitly
on time. The forced pendulum is, however, a Hamiltonian system, and since
the dimension of the phase space of 
autonomous Hamiltonian dynamical systems is $2N$,
where $N$ is the number of degrees of freedom, it is customary to refer 
to systems like the forced pendulum as to systems with 
``1.5 degrees of freedom''.

\begin{figure}
\centerline{\psfig{file=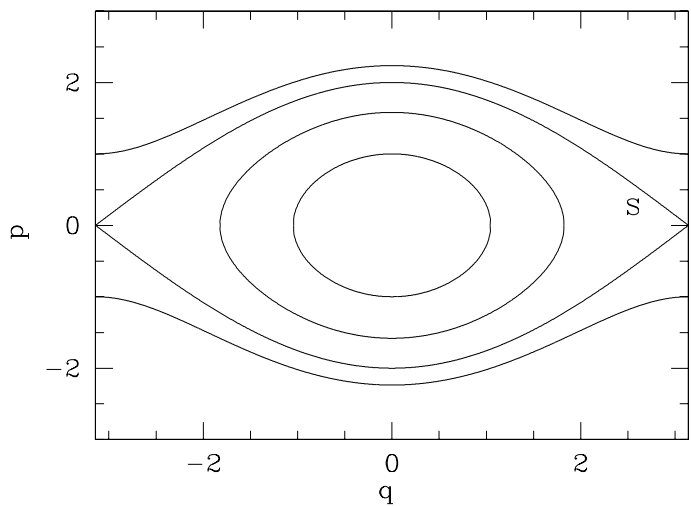,height=8cm,clip=true}} 
\caption{Phase-space trajectories of a simple pendulum. The closed curves
are the oscillations, the curves above and below the separatrix $S$ 
are clockwise 
and counterclockwise rotations, respectively.}
\label{fig_pendulum}
\end{figure}

As long as $\varepsilon = 0$, the system obeying Eq.\ (\ref{forced_pendulum})
is a simple pendulum, and its Hamiltonian 
\beq
{\cal H} = \frac{p^2}{2} - \cos  q
\eeq
is an integral of motion, so that its value, the energy $E$, is a constant of the motion and the system is {\em integrable}, as every  
one-degree-of-freedom 
autonomous Hamiltonian systems is. The word ``integrable'' is used here
in a wider sense than its immediate meaning ``such that the equations of
motion can be solved''; a Hamiltonian system is integrable when it has
a sufficiently large number of integrals of motion ($N$, for an autonomous
system with $N$ degrees of freedom), such that its trajectories do not 
explore the whole phase space, but are confined to lower-dimensional subsets
called {\em invariant tori}\footnote{The origin of the term is as follows: 
such subsets are {\em invariant} 
because if a trajectory starts on one of them, it 
remains there forever; they are called {\em tori}, 
because they are topologically
equivalent to multi-dimensional tori.}. When $N=1$, each invariant torus
coincides with a trajectory. Some of these are depicted in 
Fig.\ \ref{fig_pendulum}. We remark that there are two distinct classes
of trajectories: oscillations, which correspond to bounded motions, and
rotations, which are unbounded. the two classes are separated by a curve 
called the {\em separatrix}. The separatrix is the trajectory pursued by
the pendulum when it starts precisely 
at the unstable equilibrium point $(p, q) = 
(0,\pm \pi)$ with $E=1$, i.e., just the energy that is required to come back
to the same point (note that $q = \pi$ and $ q = -\pi$ 
must be identified). The motion on the separatrix requires an infinite 
amount of time.

But what happens if $\varepsilon \not = 0$? Once $\varepsilon  
\not = 0$, no matter 
how small, the system (\ref{forced_pendulum}) is no longer  
integrable, and the separatrix, which was a {\em unique} curve in the
$\varepsilon = 0$ case, splits into {\em two} distinct
invariant curves.
These curves must intersect transversally each other infinitely many times,
as Poincar\'e showed for the 
first time 
\cite{Poincare}.  These intersections are referred to as  
{\em homoclinic 
intersections},
and force the trajectories to fold themselves  
giving rise to a very complicated structure: in Poincar\'e's own words 
\cite{Poincare},

\begin{quotation}
these intersections form a sort of texture, or of a net 
whose meshes are infinitely tight; each of these two curves can never
intersect itself, but has to fold in a complicated way as to intersect all the 
meshes of the net an infinite number of times. One is amazed by the
complexity of this picture, which I do not even attempt to draw. 
\end{quotation}

\begin{figure}
\centerline{\psfig{file=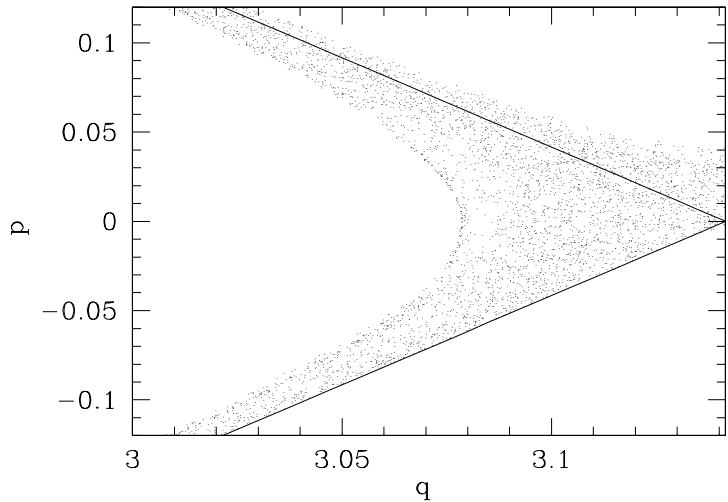,height=8cm,clip=true}} 
\caption{Section of the phase space of a perturbed pendulum,
showing the appearance of chaotic seas close to the separatrix of
the unperturbed system (the solid line in the figure). The dots are obtained
 from a single trajectory issuing from a point very close to the unperturbed separatrix.  The amplitude of the perturbation is $\varepsilon = 10^{-4}$.}
\label{fig_chao_sea}
\end{figure}

As a consequence
of the presence of these intersections, in a neighborhood
of the region in phase space which was occupied by the separatrix
in the integrable case, a so-called {\em chaotic sea} suddenly
appears. The chaotic sea is the region irregularly filled by dots 
in Fig.\ \ref{fig_chao_sea}, where a two-dimensional section\footnote{This
section has been obtained as a stroboscopic 
Poincar\'e section \protect\cite{Tabor}, so that each point on the plot corresponds to an intersection of a
trajectory of the system with the planes $t = 2n\pi/\omega$.}  of the
3-$d$ phase space of the system is shown.
If we now follow the evolution
of two intially close points in the chaotic sea, we see that their separation
grows {\em exponentially} in time, so that the dynamics in the
chaotic sea is unpredictable.

The appearence, in phase space, of irregular regions like the chaotic
sea could justify by itself the  
use of the term ``chaotic  
dynamics''. However, there are also other properties of the dynamics
described by Eq.\ (\ref{forced_pendulum}) which justify the use of
such a term.
For example, if we introduce a  
symbolic coding of the 
dynamics in which the symbol 0 is associated with each  
passage through the point 
$ q = 0$ with $\dot  q > 0$ and the symbol 1 to  
each passage through 
the same point with $\dot q < 0$, then given any bi-infinite sequence of 
zeros and ones, for example generated by coin tosses, this  
sequence corresponds 
to a real trajectory of the system (\ref{forced_pendulum}).   
The motion of the 
system, though deterministic, is thus indistinguishable from a random  
process. 

We can intuitively understand the origin of such a behaviour  
if we think that 
when the phase point is on a trajectory very close to the  
separatrix, an 
infinitesimal variation may {\em qualitatively} change the  
character of the 
motion (e.g.,  from oscillations to rotations).  
This is an example of the sensitive dependence  
on initial 
conditions that, in general, leads to the exponential separation
of initially close-by trajectories.
 
This example is extremely simple but contains the essential  
features of the 
problem.  In fact even the case of a Hamiltonian system with  
$N$ degrees of 
freedom can be treated in a similar way and shows analogous,  
though much more 
complicated, results. In that case, there is no need of an external forcing 
to get chaos, for as $N > 1$ an autonomous, nonlinear Hamiltonian
system is generically nonintegrable (the integrable systems being
a very small subset of all the possible systems).
However, even in the simple low-dimensional cases, by means of 
concepts like homoclinic intersections, it
is possible only to give a qualititative description of the onset 
of chaos, but a quantitative description of the stochastic regions
is impossible, i.e., it is impossible to compute {\em how fast}
two initially close-by points will separate. For $N$-dimensional
systems, the situation is obviously even worse: there exists a
method, which is a generalization of 
Poincar\'e's method, obtained  by
Mel'nikov \cite{Melnikov} and later by Arnol'd \cite{Arnold_diff}, 
which allows one to  
show the 
existence of homoclinic intersections near separatrices for very small  
perturbations even for large systems, but again
there is no possibility of describing quantitatively  
the stochastic regions.  

To obtain  
quantitative 
informations on chaotic dynamics we must introduce the concept  
of Lyapunov 
exponent. 
 
\subsection{Lyapunov exponents} 
\label{lyapunov} 
 
We now give a definition and an explanation of the Lyapunov exponents.
Our discussion will be aimed at showing how to define and compute
the Lyapunov exponents for a dynamical system which is defined by 
a system of ordinary differential equations, i.e., for a flow,
because Hamiltonian dynamical systems belong to this class. For a more
general discussion of Lyapunov exponents, see Eckmann and Ruelle
\cite{EckmannRuelle}.

From a physical point of view, given a trajectory of a dynamical system,
it is important to find answers to the
following questions: Is the trajectory chaotic? And if so, how strong 
chaos is, or how fast
do two initially close-by points separate in phase space, i.e., how long
should we wait until the system exhibits its chaotic feature?
The concept of Lyapunov exponents is introduced to answer these questions,
since
Lyapunov exponents are defined in order to provide an average measure 
of the rate of exponential divergence of nearby orbits in phase space, 
which is the distinctive feature of chaos. 

Lyapunov exponents are defined for a given trajectory of a  
dynamical system, and this allows us to give a definition of 
a chaotic trajectory  as follows:
a trajectory is said to be  
{\em chaotic} if its (largest) Lyapunov exponent is 
{\em positive}.

We now give a definition of the Lyapunov exponents. 
Let us consider a dynamical system whose 
trajectories in an $n$-dimensional  
phase space $M$ are the solutions of the following system of 
ordinary differential equations: 
\beq 
\begin{array}{ccc} 
\dot x_1 & = & X_1(x_1,\ldots,x_n)~,  \\
\vdots & & \vdots \\
\dot x_n & = & X_n(x_1,\ldots,x_n)~,
\end{array}
\label{dyn_system}
\eeq  
If we denote by $x(t) = (x_1(t),\ldots,x_n(t))$ a given trajectory
whose initial condition is $x(0)$, and by $y(t)$ another trajectory
which is initially close to $x(t)$, and we denote by $\xi(t)$
the vector  
\beq
\xi(t) = y(t) - x(t)~,
\label{def_xi}
\eeq
then the evolution of $\xi$ describes the separation of the two
trajectories in phase space. The vector $\xi$ is assumed to obey
the linearized equations of motion, because it is assumed to be
initially small. These equations are, as can be shown by 
inserting Eq.\ (\ref{def_xi}) into the equations of motion (\ref{dyn_system})
and expanding in a power series up to the linear terms, 
\beq 
\begin{array}{ccc} 
\dot \xi_1 & = & {\displaystyle  \sum_{j = 1}^n 
\left(\frac{\partial X_1}{\partial x_j}\right)_{x(t)}\xi_j }~,  \\
\vdots & & \vdots \\
\dot \xi_n & = & {\displaystyle  \sum_{j = 1}^n 
\left(\frac{\partial X_n}{\partial x_j}\right)_{x(t)}\xi_j }~,  \\
\end{array}
\label{tgdyn_system}
\eeq 
and are referred to as the {\em tangent dynamics equations}\footnote{This
notation follows from that the dynamics of the vector $\xi$ takes
place in the tangent space $T_{x(t)}M$ of the phase space $M$.}, which we
already wrote in the main text in the particular case of a 
standard Hamiltonian system (see Eq.\ (\ref{eq_dintang})). Note that
(\ref{tgdyn_system}) is a system of linear differential equations, whose
coefficients, however, depend on time. 
According to the definition (\ref{def_xi}), the norm 
$|\xi|$ of the vector $\xi$, i.e.,
\beq
|\xi(t) | = \left[ \sum_{i = 1}^n \xi_i^2(t) \right]^{1/2}~,
\eeq
measures the distance of the two trajectories as a function of $t$. If
the trajectory $x(t)$ is unstable, all its perturbations grow exponentially,
so that $|\xi(t)| \propto \exp(\lambda t)$. If the elements of the 
Jacobian matrix 
$\partial X_i/\partial x_j$, which are the coefficients of the
linear equations (\ref{tgdyn_system}), were either constant or periodic,
it would be possible to solve the system, but, since the Jacobian matrix
depends on the trajectory $x(t)$, its entries are in general neither
constant nor periodic, so that the rate of exponential divergence 
varies with time. Therefore, one introduces an asymptotic rate
of exponential growth of $\xi$ as the {\em
Lyapunov exponent} $\lambda$
\beq 
\lambda = \lim_{t\rightarrow\infty}\frac{1}{t} \log \frac{|\xi(t)|}
{|\xi(0)|}~,
\label{lambda_def}  
\eeq 
which measures the degree of 
instability of a trajectory:  if $\lambda$ is positive, the  
trajectory is 
unstable with a characteristic time $\lambda^{-1}$.  In  
principle $\lambda$ 
depends on both the initial values of $x$, $x(0)$, and of $\xi$, $\xi(0)$.  
However,  
Oselede\v{c} 
\cite{Oseledec} has shown that
the limit (\ref{lambda_def}) exists, is finite and can assume only
one of the $n$ values 
\beq  
\lambda_1 \leq \lambda_2 \leq \cdots \leq 
\lambda_n~.   
\eeq  
The set $\{\lambda_i\}$ is the called {\em Lyapunov spectrum}~.   
The 
exponent $\lambda$ defined in (\ref{lambda_def}) 
takes the $n$ different values of the  
spectrum as the initial 
condition $\xi$ in the tangent space $T_{x(0)}M$ is varied; the  
latter admits a 
decomposition in linear subspaces,  
\beq  
T_{x(0)} M = E_1 \oplus E_2 \oplus \cdots 
\oplus E_n~,  
\eeq  
and each $\lambda_i$ is associated with the corresponding 
subspace $E_i$, in that a vector $\xi(0) \in E_i$ will exponentially
grow with the exponent $\lambda_i$.  If there exists on the phase space $M$
a probability measure $\mu$, which is ergodic and invariant for the
dynamics on $M$, then the numbers $\lambda_i$ do {\em not} depend on
the initial condition $x(0)$, apart from a possible subset of
initial conditions of measure zero with respect to $\mu$.

In practice, the evolution of the  
norm of a tangent 
vector is sensitive only to the first --- the largest --- exponent,
because a generic initial vector $\xi(0)$ will have
a nonvanishing component in the $E_1$ subspace, so that the 
largest exponent $\lambda_1$ will always dominate in the long-time
limit:   
choosing $\xi$ at 
random with respect to a uniform distribution we have  
$\lambda = \lambda_1$ with 
probability one.  This means that Eq.\ (\ref{lambda_def}) provides
a practical definition for the {\em largest} Lyapunov exponent $\lambda_1$,
which we have always denoted simply by $\lambda$ in the main text.

Let us now apply the above  
to a standard 
Hamiltonian system, whose Hamiltonian is  of the form (\ref{H}); 
the dimension of the phase space is $n = 2N$, and
the equations of motion (\ref{dyn_system}) are now Hamilton's equations 
\beq 
\begin{array}{ccc} 
\dot q_1 & = & p_1~,  \\
\vdots & & \vdots \\
\dot q_N & = & p_N~, \\
\dot p_1 & = & {\displaystyle -\frac{\partial V}{\partial q_1} }~,  \\
\vdots & & \vdots \\
\dot p_N & = & {\displaystyle -\frac{\partial V}{\partial q_N} }~,
\end{array}
\label{H_dyn_system}
\eeq  
and also the linearized dynamics  
(\ref{tgdyn_system}) can be 
cast in the canonical form 
\beq 
\begin{array}{ccc} 
\dot \xi_1 & = & \xi_{N+1}~,  \\
\vdots & & \vdots \\
\dot \xi_n & = & \xi_{2N}~, \\
\dot \xi_{N+1} & = & {\displaystyle  -\sum_{j=1}^N 
\left(\frac{\partial^2 V}{\partial q_1 \partial q_j}\right)_{q(t)} \xi_j }~,  \\
\vdots & & \vdots \\
\dot \xi{2N} & = & {\displaystyle  -\sum_{j=1}^N 
\left(\frac{\partial^2 V}{\partial q_N \partial q_j}\right)_{q(t)} \xi_j }~.
\end{array}
\label{H_tgdyn_system}
\eeq  
This equation was already introduced as Eq.\ (\ref{eq_dintang}) in 
Sec.\ \ref{sec_geodyn}, and is 
usually referred to as the tangent dynamics equation for Hamiltonian
systems.
To measure the largest Lyapunov exponent $\lambda$ in a numerical simulation,
one integrates numerically both Eqs.\ (\ref{H_dyn_system}) and 
(\ref{H_tgdyn_system}), and then makes use of the definition 
(\ref{lambda_def}),
which can be rewritten, in this case, as 
\beq 
\lambda = \lim_{t \to \infty} \frac{1}{t} \log  
\frac{\left[\xi_1^2(t) +  
\cdots + \xi_N^2(t) + \dot\xi_1^2(t) +  
\cdots + \dot\xi_N^2(t)\right]^{1/2}} 
{\left[\xi_1^2(0) +  
\cdots + \xi_N^2(0) + \dot\xi_1^2(0) +  
\cdots + \dot\xi_N^2(0)\right]^{1/2}} ~,
\label{def_lambda_standard} 
\eeq 
where we have used that 
$\dot \xi_i = \xi_{i + N}$ (see Eq.\ (\ref{H_tgdyn_system})).
More precisely, in a numerical simulation one uses the  
discretized 
version of Eq.\ (\ref{def_lambda_standard}), i.e.,  
\beq 
\lambda = \lim_{m \to \infty} \frac{1}{m} \sum_{i=1}^m  
\frac{1}{\Delta t} 
\log\frac{|\xi(i\Delta t + \Delta t)|}{|\xi(i\Delta t)|}~,  \label{lambda_num}  
\eeq  
where, after a given number of time steps $\Delta t$, the value
of $|\xi|$ has to be renormalized to a fixed value, in order to  
avoid overflow \cite{BGS}.  

The definition (\ref{lambda_def}) does not allow one to measure
the other exponents of the Lyapunov spectrum. To measure
them, one has to observe that they can be related to the growth of
{\em volumes} in the tangent space. A two-dimensional area $V_2$ in the
tangent space, spanned by two linearly independent 
tangent vectors $\xi^{(1)}$ and $\xi^{(2)}$, will expand according to
\beq
V_2(t) \propto \exp[(\lambda_1 + \lambda_2) t]~,
\eeq
a three-dimensional volume, as
\beq
V_3(t) \propto \exp[(\lambda_1 + \lambda_2 + \lambda_3) t]~,
\eeq
and so on, so that, 
choosing $k\leq n$  
linearly independent and 
normalized vectors $\xi^{(1)},\xi^{(2)},\ldots,\xi^{(k)}\in T_x M$ we  
obtain  
\beq 
\lim_{t\rightarrow\infty}\frac{1}{t} \log  
|\xi^{(1)}(t)\wedge 
\xi^{(2)}(t) \wedge\cdots\wedge\ \xi^{(k)}(t)| =  
\sum_{i=1}^k 
\lambda_i~,  
\eeq  
Therefore the algorithm (\ref{lambda_num}) can be generalized  
to obtain an 
algorithm to compute numerically the whole Lyapunov spectrum  
\cite{BGS}.  However,
such a computation is very hard when the number $N$ is  
large. 

The sum of {\em all} the $n$ 
Lyapunov exponents in the Lyapunov spectrum, $\sum_{i = 1}^n \lambda_i$,
measures the expansion rate of $n$-volumes in phase space. Therefore,
for a Hamiltonian system,  
\beq
\sum_{i = 1}^{2N} \lambda_i = 0~,
\label{sumzero}
\eeq
because volumes in phase space are conserved. In addition, for Hamiltonian
systems a result stronger than (\ref{sumzero}) holds, i.e., there
is a symmetry in the Lyapunov spectrum such that
\beq
\lambda_i = - \lambda_{2N - i + 1}~.
\label{pairing}
\eeq
Equation (\ref{pairing}) for Hamiltonian systems
is a consequence of the symplectic structure
of the Hamilton's equations \cite{Ruffo}, however it has been recently
generalized to a class of non-Hamiltonian systems \cite{Dettmann}.

The numerical integration of the Eqs.\  (\ref{tgdyn_system}) and  
the consequent measure 
of $\lambda$ --- or of the spectrum $\{\lambda_i \}$ when it  
is possible in 
practice --- is the standard technique to characterize  
Hamiltonian chaotic 
dynamics.  An operative definition of a chaotic dynamical  
system can be stated 
as follows:  a system is chaotic if it has at least one  
positive and one
negative Lyapunov exponent.  In fact this ensures that the  
system shows
(almost everywhere with respect to the ergodic measure  
$\mu$ used to define the Lyapunov exponents) the 
distinctive features of chaos as described in the  
example of the 
forced pendulum. In fact, the presence of a positive exponent ensures
the presence of a exponential divergence of nearby orbits, and 
the presence of a negative one ensures that they also fold and mix in a very complicated way, so that
they can produce those structures we refrred to as ``chaotic seas''. 
However, as long as  
autonomous Hamiltonian 
systems are considered, the anti-symmetry of the spectrum 
(\ref{pairing}) ensures that the presence of a positive
exponent implies the presence of a negative one with the same
absolute value, so that a single (the largest)  
positive exponent is 
sufficient to have chaos; on the contrary, if the largest  
exponent vanishes the 
dynamics will be regular. These facts, together with that the largest
Lyapunov exponent 
$\lambda$ measures the smallest instability time scale, show how natural 
the use of the  
value of $\lambda$ is to 
measure chaos in such systems. 
 
It is important to specify with resepct to what invariant ergodic  
measure $\mu$  the 
Lyapunov exponents are defined:  this may be also a  
$\delta$-measure 
concentrated on a single trajectory, in which case we could speak  
of a chaotic trajectory 
rather than of a chaotic system.  In Hamiltonian  
systems with a large 
number of degrees of freedom we expect the microcanonical  
measure of the chaotic 
regions to be overwhelmingly larger than the measure of the  
regular regions; the  
existence of these regular regions  
is ensured --- at least as long as the system is  
not too far from an 
integrable limit --- by the Kol'mogorov-Arnol'd-Moser (KAM)  
theorem \cite{can_perturb_theory}. However, from a 
practical point of view the  
measure relevant for the 
definition of the Lyapunov exponent is indeed the  
microcanonical one.
Numerical experiments are 
in agreement with this expectation for large systems, since  
no relevant 
dependence of the Lyapunov exponent on the initial  
conditions has been detected, and this is the reason why
in the main text we have never referred explicitly to  
any dependence of $\lambda$ on $\mu$,  
treating the Lyapunov 
exponent as any other ``thermodynamic'' observable.   
Nevertheless for small 
systems (especially $N=2$ which is the best known case) the  
simulations show 
that the measure of the chaotic regions may be very small in  
a very large energy 
range, so that in that case one cannot speak of a truly  
chaotic system but only of 
a system in which chaotic and regular regions are  
simultaneously present (these 
systems are often referred to as {\em mixed} systems, as  
they are in between 
completely chaotic and regular ones). 
 
Since we are interested in large systems, up to the  
thermodynamic limit, a number of
questions naturally arises: what is the behavior of the  
Lyapunov exponents as 
$n$ increases;  does a thermodynamic limit exist for the  
Lyapunov spectrum, etc.
Numerical results \cite{LiviPolitiRuffo} 
have shown that as $n \rightarrow \infty$ the 
Lyapunov spectrum $\{\lambda_i\}$ appears indeed to converge to a  
well-behaved function  
\beq
\lambda(x) = \lim_{n\rightarrow\infty} \lambda_{xn}~.   
\eeq  
The function 
$\lambda(x)$ is a non-increasing function of $x \in [0,1]$.
Some rigorous work in this respect has been recently done by Sinai 
\cite{Sinai_spectrum}.
The existence of a limiting Lyapunov spectrum in the 
thermodynamic limit has many 
important consequences that we will not review here; a good  
discussion can be 
found in Ref. \cite{Ruffo}. We only want to remark  
here that the existence of a thermodynamic limit for the Lyapunov spectrum
implies that the largest Lyapunov exponent is  
expected to behave as an 
intensive quantity as $N$ increases.

\section{The stochastic oscillator equation} 
\label{app_sto} 
In the following we will briefly describe how to cope with the stochastic  
oscillator problem which we encountered in \S \ref{geom_formula}. The  
discussion closely follows Van Kampen (Ref.\ \cite{VanKampen}) where all 
the details can be found. 
 
A stochastic differential equation can be put in the general form 
\begin{equation} 
F(x,\dot x,\ddot x,\ldots,\Omega)=0, 
\label{a1} 
\end{equation} 
where $F$ is an assigned function and the variable $\Omega$ is a random 
process, defined by a mean, a standard deviation and an autocorrelation 
function. A function $~{\xi}(\Omega)$ is a solution of this equation, if 
$\forall \Omega$ $F(\xi (\Omega ),\Omega)=0$. 
If equation (\ref{a1}) is linear of order $n$, it is written as 
\begin{equation} 
\dot{{\bf u}}={\bf A}(t,\Omega){\bf u} 
\label{stoclin} 
\end{equation} 
where 
\beq
{\bf u} = \left( \begin{array}{c} u_1 \\ u_2 \\ u_3 \\ \vdots \\ u_n
\end{array} \right) = \left( \begin{array}{c} x \\ \dot x \\ \ddot x \\ 
\vdots \\ x^{(n)}
\end{array} \right) ~,
\eeq
and ${\bf A}$ is an $n\times n$ matrix whose  
elements $A_{\mu \nu}(t)$ depend randomly on time. 
 
For the purposes of our work, we are interested in the evolution of 
the quantities $u_\nu u_\mu$, 
rather than of the $u_\mu$'s themselves. 
The $u_{\mu \nu}$'s obey the differential
equation
\begin{equation} 
\frac{d}{dt}(u_{\nu}u_{\mu})=\sum_{k,\lambda}\tilde{A}_{\nu\mu,k\lambda}(t)( 
u_{k}u_{\lambda})~, 
\label{eq_stocmoments})
\end{equation} 
where 
\begin{equation} 
\tilde{A}_{\nu\mu,k\lambda}=A_{\nu 
k}\delta_{\mu\lambda}+\delta_{\nu k}A_{\mu \lambda}~. 
\label{Atilde} 
\end{equation} 
However, both Eq.\ (\ref{stoclin}) and (\ref{eq_stocmoments}) have 
exactly the same form and can be
solved using the same procedure, so that we will first
illustrate such a procedure in general. Therefore in the following formul\ae,
${\bf u}$ refers to a vector whose components are either the $u_\mu$'s
or the $u_{\mu \nu}$'s, and $\bf A$ denotes either the matrix  $\bf A$
in Eq.\ (\ref{stoclin}) or the matrix $\tilde{\bf A}$ whose
elements are given by Eq.\ (\ref{Atilde}), respectively. Then, we
will apply this procedure to the case of the stochastic harmonic oscillator.

Now, solving a linear stochastic differential equation means determining  
the evolution of the  
average of ${\bf u}(t)$, $\langle {\bf u}(t)\rangle$, where the average 
is carried over all the realizations of the process. 
Let us consider the matrix ${\bf A}$ as the sum 
\begin{equation} 
{\bf A}(t,\Omega)={\bf A}_{0}(t)+\alpha{\bf A}_{1}(t,\Omega) 
\label{decomp} 
\end{equation} 
where the first term is $\Omega$-independent and the second one is randomly 
fluctuating with zero mean. 
Let us also assume that ${\bf A}_{0}$ is time-independent. 
If the parameter $\alpha$ --- which determines the fluctuation amplitude --- 
is small we can treat Eq.\ (\ref{stoclin}) 
by means of a perturbation expansion. 
It is convenient to use the interaction representation, so that we put 
\begin{equation} 
{\bf u}(t)=\exp({\bf A}_{0}t){\bf v}(t) 
\end{equation} 
and 
\begin{equation} 
{\bf A}_{1}(t)=\exp({\bf A}_{0}t){\bf v}(t)\exp(-{\bf A}_{0}t)~. 
\end{equation} 
Formally one is then led to a Dyson expansion for the solution ${\bf v}(t)$.  
Then, going back to the previous variables and averaging, the second order 
approximation gives 
\begin{equation} 
\frac{d}{dt}\langle{\bf u}(t)\rangle=\{{\bf 
A}_{0}+\alpha^{2}\int^{+\infty}_{0}\langle{\bf A}_{1}(t)\exp({\bf 
A}_{0}\tau){\bf A}_{1}(t-\tau)\rangle\exp(-{\bf A}_{0}\tau)d\tau\}\langle{\bf 
 u}(t)\rangle~. 
\label{media2} 
\end{equation} 
 
Let us remark that, if the stochastic process $\Omega$ is Gaussian,  
Eq.\ (\ref{media2}) is more than a 
second order approximation: it is exact. 
In fact, the Dyson series can be written in compact form as 
\begin{equation} 
\langle{\bf u}(t)\rangle = T\left[  
\left\langle\exp\left(\int_{0}^{t}{\bf A}(t')dt'\right) 
\right\rangle \right] 
\langle {\bf u} (0)\rangle~, 
\label{tord} 
\end{equation} 
where $T[ \cdots]$ stands for a time-ordered product. 
According to Wick's procedure we can rewrite Eq.\  (\ref{tord}) as a cumulant 
expansion, and  
when the cumulants of higher than the second order vanish (as in the 
case of a Gaussian process) one can easily show that 
Eq.\  (\ref{media2}) is exact. 
 
We now apply this general approach to the case of interest 
for the main text, i.e., to the stochastic harmonic  
oscillator equation, which is the
the second-order linear stochastic differential equation given by
\beq 
\ddot x + \Omega(t)\, x = 0~,  
\eeq 
where $\Omega(t)$ is the random squared frequency, 
$\Omega =\Omega_{0}+\sigma^{\,}_{\Omega}\eta(t)$, where $\Omega_0$ is the 
mean of $\Omega(t)$, $\sigma_\Omega$ is 
the amplitude of the fluctuations, and $\eta(t)$  
is a stochastic process with zero mean. In this case,  
Eq.\ (\ref{stoclin}) has the  
form 
\begin{equation} 
\frac{d}{dt}\left( 
\begin{array}{c} 
x \\ 
\dot{x} 
\end{array}\right)=\left(\begin{array}{cc} 
0 & 1 \\ 
-\Omega & 0 \end{array}\right)\left(\begin{array}{c}x \\ 
\dot{x}\end{array}\right). 
\label{eqstocosmatrix}
\end{equation} 
In particular, we are interested in obtaining the averaged 
equation of motion for 
the second moments.  
Using Eqs.\ (\ref{Atilde}) and (\ref{eqstocosmatrix}), 
one finds that Eq.\ (\ref{eq_stocmoments}) becomes: 
\begin{equation} 
\frac{d}{dt}\left(\begin{array}{c} 
x^{2} \\ \dot{x}^{2} \\ x\dot{x} \end{array} \right) = \left( \begin{array}{ccc} 
0 & 0 & 2 \\ 
0 & 0 & -2\Omega \\ 
-\Omega & 1 & 0  \end{array} \right)\left(\begin{array}{c} x^{2} \\ 
\dot{x}^{2} \\ x\dot{x} \end{array} \right) 
={\bf A} 
\left(\begin{array}{c} x^{2} \\ 
\dot{x}^{2} \\ x\dot{x} \end{array} \right)~. 
\end{equation} 
Like in Eq.\  (\ref{decomp}), the matrix ${\bf A}$ splits into: 
\begin{equation} 
{\bf A}(t)={\bf A}_{0}+\sigma^{\,}_{\Omega}\eta(t){\bf A}_{1}= 
\left(\begin{array}{ccc} 
0 & 0 & 2 \\ 
0 & 0 & -2\Omega_{0} \\ 
-\Omega_{0} & 1 & 0 
\end{array}\right)+\sigma^{\,}_{\Omega}\eta(t)\left(\begin{array}{ccc} 
0 & 0 & 0 \\ 
0 & 0 & -2 \\ 
-1 & 0 & 0 \end{array}\right)~, 
\end{equation} 
so that the equation for the averages becomes 
\begin{equation} 
\frac{d}{dt} 
\left(\begin{array}{c} 
\langle x^{2}\rangle \\ \langle\dot{x}^{2}\rangle \\ \langle x\dot{x}\rangle 
\end{array}\right)=\{{\bf 
A}_{0}+\sigma^{2}_{\Omega}\int_{0}^{+\infty}\langle\eta(t)\eta(t-t') 
\rangle{\bf B}(t')dt' 
\} 
\left(\begin{array}{c} 
\langle x^{2}\rangle \\ \langle\dot{x}^{2}\rangle \\ \langle x\dot{x}\rangle 
\end{array}\right), 
\label{eqAB} 
\end{equation} 
where ${\bf B}(t)={\bf A}_{1}\exp({\bf A}_{0}t){\bf A}_{1} 
\exp(-{\bf A}_{0}t)$. 
 
When the process $\eta(t)$ is Gaussian and $\delta$-correlated,  
Eq.\ (\ref{eqAB}) is exact, and the integral can be computed explicitly: 
writing $\langle\eta(t)\eta(t-t')\rangle=\tau \delta(t')$, where 
$\tau$ is the correlation time scale of the random process, we obtain 
\begin{equation} 
\frac{d}{dt} 
\left(\begin{array}{c} 
\langle x^{2}\rangle \\ \langle\dot{x}^{2}\rangle \\ \langle x\dot{x}\rangle 
\end{array}\right)=\{{\bf 
A}_{0}+\frac{\sigma^{2}_{\Omega}\tau}{2} {\bf B}(0) 
\} 
\left(\begin{array}{c} 
\langle x^{2}\rangle \\ \langle\dot{x}^{2}\rangle \\ \langle x\dot{x}\rangle 
\end{array}\right)~. 
\end{equation} 
From the definition of ${\bf B}(t)$ it follows then that  
${\bf B}(0)={\bf A}_{1}^{2}$, and by an easy calculations we find  
\begin{equation} 
{\bf A}_{0}+\sigma^{2}_{\Omega}\tau {\bf A}_{1}^{2}=\left(\begin{array}{ccc} 
0 & 0 & 2 \\ 
\sigma^{2}_{\Omega}\tau & 0 & -2\Omega_{0} \\ 
-\Omega_{0} & 1 & 0 \end{array}\right) 
\end{equation} 
which is the result used in \S \ref{geom_formula}.

\end{document}